\documentclass[useAMS,usenatbib,babel,times]{mn2e}
\usepackage{multirow,balance,verbatim}
\usepackage[english,english]{babel}
\usepackage{amsmath,subfig}
\usepackage{amssymb,amsfonts,textcomp}
\usepackage{array}
\usepackage{supertabular}
\usepackage{hhline}
\usepackage{hyperref}
\usepackage[usenames]{color}
\hypersetup{dvips, colorlinks=true, linkcolor=black, citecolor=black, filecolor=black, urlcolor=black}
\usepackage[dvips]{graphicx}
\usepackage[squaren,Gray,cdot]{SIunits}
\def\gtrsim{\lower.5ex\hbox{$\; \buildrel > \over \sim \;$}}

\begin{document}

\author[Chisari et al.]{
  \parbox{\textwidth}{N. Chisari$^1$\thanks{elisa.chisari@physics.ox.ac.uk}, C. Laigle$^2$, S. Codis$^{3,2}$, Y. Dubois$^2$, J. Devriendt$^1$, \\
  L. Miller$^1$, K. Benabed$^2$ ,  A. Slyz$^1$,  R. Gavazzi$^2$, C. Pichon$^{2,4}$}
\vspace*{6pt}\\
\noindent
$^{1}$Department of Physics, University of Oxford, Keble Road, Oxford, OX1 3RH,UK.\\
$^{2}$Institut d'Astrophysique de Paris, CNRS \& UPMC, UMR 7095, 98 bis Boulevard Arago, 75014, Paris, France.\\
$^{3}$Canadian Institute for Theoretical Astrophysics, University of Toronto, 60 St. George Street, Toronto, ON M5S 3H8, Canada.\\
$^{4}$Korea Institute of Advanced Studies (KIAS) 85 Hoegiro, Dongdaemun-gu, Seoul, 02455, Republic of Korea.
}

\date{Accepted 2016 June 09. Received 2016 May 30; in original form 2016 February 26}

\title[Evolution of intrinsic alignments]{
Redshift and luminosity evolution of the \\intrinsic alignments of galaxies in Horizon-AGN}

\maketitle

\begin{abstract}
Intrinsic galaxy shape and angular momentum alignments can arise in cosmological large-scale structure due to tidal interactions or galaxy formation processes. Cosmological hydrodynamical simulations have recently come of age as a tool to study these alignments and their contamination to weak gravitational lensing. We probe the redshift and luminosity evolution of intrinsic alignments in Horizon-AGN between $z=0$ and $z=3$ for galaxies with an $r$-band absolute magnitude of $M_r\leq-20$. Alignments transition from being radial at low redshifts and high luminosities, dominated by the contribution of ellipticals, to being tangential at high redshift and low luminosities, where discs dominate the signal. This cannot be explained by the evolution of the fraction of ellipticals and discs alone: intrinsic evolution in the amplitude of alignments is necessary. The alignment amplitude of elliptical galaxies alone is smaller in amplitude by a factor of $\simeq 2$, but has similar luminosity and redshift evolution as in current observations and in the nonlinear tidal alignment model at projected separations of $\gtrsim 1$ Mpc. Alignments of discs are null in projection and consistent with current low redshift observations. The combination of the two populations yields an overall amplitude a factor of $\simeq 4$ lower than observed alignments of luminous red galaxies with a steeper luminosity dependence. The restriction on accurate galaxy shapes implies that the galaxy population in the simulation is complete only to $M_r\leq-20$. Higher resolution simulations will be necessary to avoid extrapolation of the intrinsic alignment predictions to the range of luminosities probed by future surveys. 
\end{abstract}

\begin{keywords}
cosmology: theory ---
gravitational lensing: weak --
large-scale structure of Universe ---
methods: numerical 
\end{keywords}

\section{Introduction}
\label{sec:intro}

The shapes and spins of galaxies are stretched and torqued by tides across the Universe, influenced by star formation and mergers, such that their intrinsic shapes and orientations are correlated over a large range of scales. These intrinsic alignments of galaxies constitute a contaminant to weak gravitational lensing, potentially biasing contraints on the evolution of dark energy from future surveys if unaccounted for \citep{Kirk10,Kirk12,Krause15}. Surveys such as {\it Euclid\,}\footnote{\url{http://sci.esa.int/euclid}}\!~\citep{Laureijs11}, the Large Synoptic Survey Telescope\footnote{\url{http://www.lsst.org}}\citep[LSST]{LSST} and WFIRST\footnote{\url{http://wfirst.gsfc.nasa.gov/}}~\citep{green11}, expect to put General Relativity to the test by measuring gravitational lensing distortions on galaxy shapes to quantify the evolution of the dark energy equation of state over the history of the Universe. Marginalization and mitigation schemes have been developed to suppress the impact of alignments on lensing \citep{King05,Zhang10,Joachimi08,Joachimi10,Joachimi11}, but the success of these techniques can be improved with prior knowledge of the strength, dependence of galaxy-type, redshift and luminosity evolution of the alignment signal.  

Recently, several studies \citep{Codis14,Tenneti15a,Tenneti15b,Chisari15,Velliscig15b} have been devoted to measuring intrinsic alignments in hydrodynamical cosmological simulations.  In \citet{Chisari15}, hereafter Paper I, we measured intrinsic alignments at $z=0.5$ in the Horizon-AGN simulation~\citep{Dubois14}, a hydrodynamical cosmological simulation of ($100 \, h^{-1}\, \rm Mpc$)$^3$  comoving volume. We showed that two alignment mechanisms are present for disc-like and spheroidal galaxies. The orientation of discs tends to be tangential about spheroidals, while those galaxies tend to point towards each other. We also presented projected correlation functions of galaxy shapes and the dark matter density field and model fits to those quantities. The strength of alignment for the highest luminosity galaxies was in agreement with predictions from other simulations \citep{Tenneti15a}. However, \citet{Tenneti15b} found no evidence for a disc-alignment mechanism different than for ellipticals in either the MassiveBlack II smoothed-particle-hydrodynamics simulation~\citep{Khandai15}, or in the moving-mesh Illustris simulation~\citep{Vogelsberger14}. The reason for this discrepancy remains currently unknown, but is possibly related to the implementation of different baryonic prescriptions in these simulations. A mass transition has been identified in the Horizon-AGN simulation \citep{Dubois14,Welker14}, whereby less massive halos transition from a parallel alignment between galaxy spins and the direction of the nearest filaments to a perpendicular alignment. This transition is consistent with recent observations \citep{tempel13},  numerous measurements in dark matter simulations \cite[e.g.][and references therein]{Codis12} and tidal torque theory \citep{Codis15b}, though it is not seen in the MassiveBlack II simulation \citep{Chen15}. 

In this work, we extend the study presented in Paper I by probing the redshift and luminosity dependence of the alignment signal in Horizon-AGN from $z=0$ to $z=3$ and for galaxies with $r$-band absolute magnitude $M_r\leq-20$. Taking this step is crucial for understanding the physical mechanisms that give rise to alignments, comparing to theoretical predictions and making forecasts of the intrinsic alignment contamination to weak lensing surveys. On the other hand, information on the redshift evolution of the alignment signal can also help us understand the discrepancies between different hydrodynamical simulations. 

This work is organised as follows. In section~\ref{sec:sim}, we give an overview of the properties of the Horizon-AGN simulation, including the galaxy catalogue and the computation of galaxy shapes. In section~\ref{sec:correl}, we define the correlation functions used to measure alignments in this work, along with the accompanying modelling. We present our results in section~\ref{sec:results}, followed by discussion in section~\ref{sec:discuss} and conclusions in section~\ref{sec:conclusion}. The appendices of this work include a discussion on the impact of more realistic galaxy colours (Appendix~\ref{app:dust}) and the presentation of the complete set of alignment correlations, which expands on the summary of the results presented in section~\ref{sec:results}.

\section{The Horizon-AGN simulation}
\label{sec:sim}

Horizon-AGN~\citep{Dubois14} is a cosmological hydrodynamical numerical simulation run using the adaptive mesh refinement code {\sc ramses}~\citep{teyssier02}. The cosmological box is $100 \, h^{-1}\, \rm Mpc$ in each dimension and cosmological parameters are matched to a $\Lambda$ Cold Dark Matter {\it WMAP7} cosmology~\citep{komatsuetal11}. For this cosmology, the total matter density is $\Omega_{\rm m}=0.272$, the dark energy density is $\Omega_\Lambda=0.728$, the amplitude of the matter power spectrum is $\sigma_8=0.81$, the baryon density is $\Omega_{\rm b}=0.045$, the Hubble constant adopted is $H_0=70.4 \, \rm km\,s^{-1}\,Mpc^{-1}$, and the power-law index of primordial fluctuations is $n_s=0.967$. Horizon-AGN uses $7$ levels of refinement, reaching a resolution of $\Delta x=1\, \rm kpc$ following a quasi-Lagrangian criterion. The volume is filled with $1024^3$ dark matter particles, corresponding to a dark matter mass resolution of $8\times 10^7 \, \rm M_\odot$.

Star formation follows a Schmidt law and is triggered wherever the hydrogen gas number density exceeds $n_0=0.1\,{\rm cm}^{-3}$ following a Poisson random process~\citep{rasera&teyssier06, dubois&teyssier08winds}. The stellar mass resolution is $M_{*, \rm res} \simeq 2\times 10^6 \, \rm M_\odot$.

Following~\cite{haardt&madau96}, heating from a uniform UV background is considered after $z_{\rm reion} = 10$, the reionization redshift. Gas cools via H and He cooling down to $10^4\, \rm K$, and the contribution from metals is accounted for \citep{sutherland&dopita93}. Metallicity is a passive variable of the gas, which changes according to the injection of gas ejecta from supernovae explosions and stellar winds~\citep[for details, see][]{Dubois14}.
Horizon-AGN includes feedback from Active Galactic Nuclei ~\citep[AGN,][]{duboisetal12agnmodel} assuming a Bondi-Hoyle-Lyttleton accretion rate $\dot M_{\rm BH}$ onto black holes capped at the Eddington rate, $\dot M_{\rm Edd}$. AGN feedback is modelled with two modes: the \emph{radio jet} mode, active when $\dot M_{\rm BH}/\dot M_{\rm Edd}< 0.01$, and the \emph{quasar heating} mode, operating otherwise (see~\citealp{duboisetal12agnmodel, Dubois14} for details).

Black hole relations to galaxy properties and AGN populations~\citep{volonterietal16}, as well as galaxy properties are reasonably well reproduced in Horizon-AGN (Kaviraj et al, in prep.; Dubois et al, in prep.). Thus, it is unique tool to infer alignments of galaxies in the cosmic web.

\subsection{Galaxy catalogue}
\label{sec:postprocess}

In each redshift snapshot of Horizon-AGN, we rely on the distribution of stellar particles to identify galaxies. The AdaptaHOP finder~\citep{aubertetal04} looks for the twenty nearest neighbours of each stellar particle and computes the local density around it. Only over-densities above a threshold of $\rho_{\rm t}=178$ times the average total matter density are considered further. The final mock catalogues solely include structures of $>50$ stellar particles. The mass of a galaxy identified by AdaptaHOP is the sum of the masses of its stellar particles.

We use stellar population models from~\cite{bruzual&charlot03} with a Salpeter initial mass function \citep{Salpeter55} to synthesize absolute AB magnitudes and rest-frame colours of mock galaxies. The contribution of each star particle to the flux per frequency depends on its mass, age and metallicity. We sum these contributions and filter them through $u$, $g$, $r$, and $i$ bands from the {\it Sloan Digital Sky Survey} \citep[SDSS,][]{Gunn06}. In the main body of this manuscript, we work with dust-free rest-frame fluxes (without accounting for redshifting). We assess the impact of more realistic flux and colour modelling in Appendix~\ref{app:dust}.

\subsection{Shapes and spins}
\label{sec:shape}

For each galaxy, its three dimensional shape is characterised by the {\it simple} inertia tensor,
\begin{equation}\label{eq:inerdef3d}
I_{ ij}=\frac{1}{M_*}\sum_{n=1}^{N} m^{(n)} x^{(n)}_{i} x^{(n)}_{j}\, .
\end{equation}
where $i,j=1,2,3$ correspond to the cartesian axes of the simulation box, $M_*=\sum_{n=1}^{N} m^{(n)}$ is the stellar mass, $m^{(n)}$ is the mass of the $n$-th stellar particle and ${\bf x}^{(n)}$, its position in the box with respect to the center of mass of the galaxy. We will also present results obtained by measuring galaxy shapes using the {\it reduced} inertia tensor, which is given by
\begin{equation}\label{eq:reduceddef}
{\tilde I}_{ ij}=\frac{1}{M_*}\sum_{n=1}^{N} m^{(n)} \frac{x^{(n)}_{i} x^{(n)}_{j}}{r_{n}^2},
\end{equation}
where $r_{n}=|| \mathbf{x}^{(n)} ||$ is the three dimensional distance of the stellar particle $n$ from the center of mass of the galaxy. Each inertia tensor is diagonalised to obtain the unit eigenvectors. The one with the smallest eigenvalue represents the orientation of the minor axis of the galaxy. 

We measure the projected shape of a galaxy from the projected inertia tensor, given by equation~(\ref{eq:inerdef3d}) restricted to $x$ and $y$ ($i,j=1,2$). This projected inertia tensor is diagonalised to obtain the unit eigenvectors corresponding to the smallest ($\lambda_1$) and largest ($\lambda_2$) eigenvalues, which represent respectively the semiminor and semimajor axis of the ellipsoid. 

The inclusion of weights $\propto r_n^{-2}$ in the contribution of each particle to the reduced inertia tensor is a closer representation of the observed shape of a galaxy as measured by shear estimation for weak gravitational lensing, compared to applying equal weights. Other works \citep{Tenneti14a} have explored the application of iterative algorithms to reduce the impact of the spherical symmetry imposed by $r_n^{-2}$ weights. In this work, we will only consider the simple and the reduced inertia tensor cases. We expect that an iterative procedure would produce results bracketed by those two choices.

The axis ratio of the galaxy, $q=b/a$, is related to the ratio of the semiminor to semimajor axes ($b=\sqrt{\lambda_1},a=\sqrt{\lambda_2}$). The complex ellipticity, typically used in weak lensing measurements, is given by
\begin{equation}
  (e_+,e_\times) = \frac{1-q^2}{1+q^2}[\cos(2\phi),\sin(2\phi)]\,,
  \label{eq:complexe}
\end{equation}
where $\phi$ is the orientation angle of the semimajor axis, $+$ indicates the radial component of the ellipticity and $\times$ is the $45\deg$-rotated component. By convention, we represent radial alignments as having negative $e_+$ (gravitational lensing yields a signal with the opposite sign). 

We also consider the orientation of the ``spin'' -- the intrinsic angular momentum of the galaxy --, obtained as
\begin{equation}
\label{eq:spindef}
  \mathbf{L} = \sum_{ n=1}^{N} m^{( n)} \mathbf{x}^{(n)} \times \mathbf{v}^{(n)} \, ,
\end{equation}
where $\mathbf{v}^{(n)}$ is the velocity of each particle relative to the center of mass. 

\begin{figure}
\includegraphics[width=0.45\textwidth]{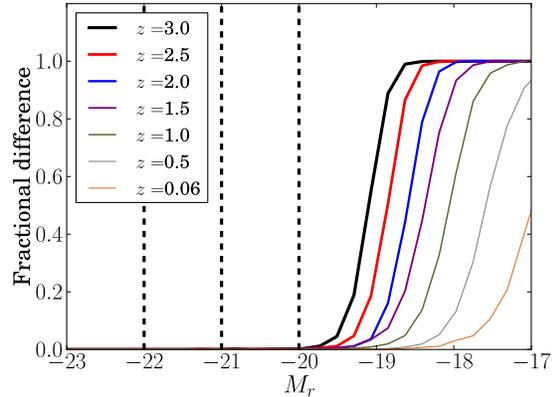}
\caption{The fractional difference in the distribution of $M_r$, rest-frame absolute magnitudes in the $r-$band, for galaxies with $>300$ particles and for the complete galaxy catalogue ($>50$ stellar particles), for different redshifts. The choice of absolute magnitude bins for the measurement of intrinsic alignments is indicated with the vertical dashed lines: $M_r\leq -22$, $-22<M_r\leq -21$ and $-21<M_r\leq -20$. A large fractional difference represents missing galaxies at that $M_r$ in the $>300$ particle sample. }
\label{fig:mrhist_frac}
\end{figure}

\subsection{Completeness}
\label{sec:complete}

Studies of intrinsic alignments in simulations rely on selecting a sample of galaxies with well-defined shapes. To this end, it is usual to impose cuts on the number of stellar particles required \citep{Tenneti14a,Tenneti15a,Chisari15,Tenneti15b,Velliscig15,Velliscig15b}. Naturally, this results in a lower limit in the stellar mass of galaxies for which the alignment signal is measured. Both in Paper I and in this work, we only consider galaxies with $>300$ stellar particles, which, given our resolution, results in an effective cut on stellar mass of $10^9$ M$_\odot$. In addition, this determines a lower luminosity threshold below which the sample of galaxies suffers from incompleteness. Fig.~\ref{fig:mrhist_frac} shows the fractional difference in the distribution of $r$-band absolute magnitudes for galaxies with $>300$ stellar particles and for all galaxies ($>50$ stellar particles) in Horizon-AGN and for different redshifts in the range $0.06<z<3$. A large fractional difference represents missing galaxies at that $M_r$ for the $>300$ particle sample. 

The completeness of the sample with measured shapes evolves with redshift such that at lower redshifts, we are complete to lower luminosities. The sample of galaxies with well-converged shapes becomes incomplete at $z>3$ for $M_r>-20$. We thus establish a minimum luminosity threshold of $M_r\leq-20$. To measure the intrinsic alignment signal as a function of luminosity, we divide the sample of galaxies at each redshift into three luminosity bins, whose boundaries are indicated by the vertical dashed lines: $M_r\leq-22$, $-22<M_r\leq -21$ and $-21<M_r\leq-20$. In Paper I, we found a non-monotonic dependence of intrinsic alignment amplitude with luminosity that originated in the alignment of low luminosity ellipticals. Due to incompleteness, we cannot probe this population over all of the redshift range of interest, $0<z<3$, and we discuss the implications in section~\ref{sec:discuss}.

\subsection{Substructure hierarchy}
\label{sec:hierarchy}

\begin{figure*}
  \includegraphics[width=0.32\textwidth,angle=270]{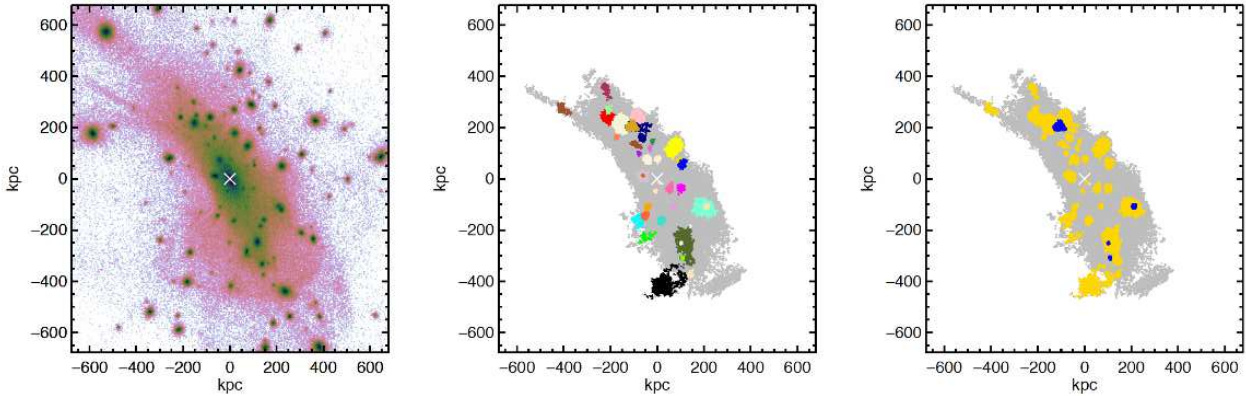}
  \caption{We show one of the most massive galaxies in Horizon-AGN at $z=0.06$ to illustrate the presence of different hierarchy galaxies as found by AdaptaHOP. The left panel shows the projected stellar density in a cube of $\sim 1.2$ Mpc a side. The middle panel shows the stellar particles of the central galaxy in grey and all other galaxies (higher in the hierarchy) in different colours. In the right panel, we show once more the central galaxy in grey, while the yellow points correspond to stellar particles of galaxies ranked second in the hierarchy (substructures), and blue points, to stellar particles of galaxies ranked third in the hierarchy (sub-substructures).}
  \label{fig:gal1}
\end{figure*}

AdaptaHOP identifies galaxies with different hierarchy levels, i.e., some galaxies are identified as substructures around a central galaxy. In Fig.~\ref{fig:gal1}, we show an example of substructures identified by AdaptaHOP around one of the most massive galaxies of Horizon-AGN at $z=0.06$. The left panel shows the distribution of stellar particles around a massive central galaxy projected along one of the directions of the box. The middle panel shows the stellar particles that belong to the central galaxy (in grey) and higher level galaxies around it, all of them with $>300$ stellar particles (different colours are used for each galaxy). Only four galaxies in this figure are third level of the hierarchy (substructure of substructure), and they are colour-coded in blue in the right panel, while second level galaxies are shown in yellow. Some satellites present in the left panel are classified as high level, and are not shown in the other panels. It as also worth noting that AdaptaHOP does not link stellar particles in the circum-galactic medium with any particular galaxy, and thus they do not contribute towards measuring galaxy ellipticities. Including them could potentially increase alignment amplitude, particularly those measured with the simple inertia tensor. However, the observability of stars in the circum-galactic medium depends on the surface brightness limit of future surveys.

Unless otherwise specified, we will consider all galaxies with more than $>300$ particles as contributors to the intrinsic alignment signal, regardless of their level in the hierarchy. However, we will discuss in section~\ref{sec:z0bis} the implications of this criterion. The second option we consider in that section is the case where all stellar particles belonging to higher level galaxies actually become part of the central galaxy. Notice that this is equivalent to exploring the impact of inefficient ``de-blending''. We do not attempt to mimic the exact performance of future surveys, but provide these cases as a qualitative estimate of the impact of substructure on intrinsic alignment correlations, especially at small separations. In practice, realistic image simulations with appropriate surface brightness cuts, point spread function convolution and detector systematics are required to match the observing conditions expected for {\it Euclid}, LSST and other weak lensing surveys.

\section{Correlation functions}
\label{sec:correl}

\subsection{Definitions}

As in Paper I, we define the correlation between the orientation of the unit eigenvectors corresponding to the minor axis of a galaxy, ${\hat {\bf u}}$ and the comoving separation vector between galaxies, ${\bf r}$, as 
\begin{equation}
\eta_e(r) = \langle |\hat{\bf r}\cdot \hat{\bf u}({\bf x}+{\bf r})|^2\rangle - {1}/{3}\,,
\label{eq:etaER}
\end{equation}
with $\hat{\bf r}={\bf r}/||\mathbf r||$.
A negative correlation is indicative of the minor axis pointing perpendicular to the separation vector, hence, of the galaxy being aligned radially towards the origin. A positive $\eta_e$ corresponds to a tendency for the separation vector and the minor axis of a galaxy to be parallel, or equivalently, for the galaxy to be elongated tangentially around the reference point. Analogously, we will refer to $\eta_s$ as the equivalent expression to equation~(\ref{eq:etaER}) when the orientation of a galaxy is determined from its spin
\begin{equation}
\eta_s(r) = \langle |\hat{\bf r}\cdot \hat{\bf L}({\bf x}+{\bf r})|^2\rangle - {1}/{3}\,,
\label{eq:etaSR}
\end{equation}
with $\hat{\bf L}={\bf L}/||\mathbf L||$.
We will also consider projected correlations, which are frequently used in observational studies to
measure intrinsic alignments of galaxies. The dataset of tracers of the density field is referred to as $D$; the set of galaxies with ellipticities, as $S_{+}$ (for the tangential ellipticity component of equation~\ref{eq:complexe}, and similarly for the $\times$ component); and we consider two sets of random points uniformly distributed in the simulation box, $R_S$ (matching the total number of galaxies with shapes) and $R_D$ (matching the number count in the density sample). From these data sets, we construct an estimator of the redshift-space correlation function of galaxy shapes and positions of density tracers, $\xi_{\delta +}(r_p,\Pi)$ as a function of projected separation, $r_p$, and along the line of sight, $\Pi$. This is given by
\begin{align}
\xi_{\delta +}(r_p,\Pi) &= \frac{S_+D}{R_SR_D}\,,\label{eq:xidp}\\
S_+D &= \sum_{(r_p,\Pi)} \frac{e_{+,j}}{2\mathcal{R}}\,\label{eq:splusD},
\end{align}
where $\mathcal{R}$ is the responsivity factor \citep{Bernstein02}, $\mathcal{R}=1-\langle e^2 \rangle$, $e_{+,j}$ is the tangential/radial component of the ellipticity vector of galaxy $j$, $\langle e^2 \rangle$ is the root mean square ellipticity per component, and the sum is over galaxy pairs in given bins of projected radius and line of sight distance. We project this correlation by projecting along one of the coordinate axes of the simulation box by integrating between $-\Pi_{\rm max}<\Pi<\Pi_{\rm max}$,
\begin{equation}
  w_{\delta+}(r_p) = \int_{-\Pi_{\rm max}}^{\Pi_{\rm max}} \textrm{d}\Pi\,\xi_{\delta +}(r_p,\Pi)\,,
  \label{eq:wdplusdef}
\end{equation}
where we take $\Pi_{\rm max}=\ell/2$, half the length of the simulation box. \citet{Tenneti15a} similarly adopt half of their simulation box for $\Pi_{\rm max}$. Observational constraints on alignments are typically obtained adopting $\Pi_{\rm max}=60\, h^{-1}\, \rm Mpc$ for spectroscopic samples \citep{Mandelbaum06,Hirata07} and no increase in the alignment signal is found from including information beyond this range. When the tracers of the density field are galaxies, we refer to the projected correlation functions of the two ellipticity components as $w_{g+}$ and $w_{g\times}$. Otherwise, to construct the density field, we subsample the DM particles adopting the same criterion as in Paper I and in \citet{Tenneti15a}; in Paper I, sub-per cent sampling of the DM particles guaranteed convergence to $10\%$ in the DM clustering correlation function. 

Alignments in three dimensions are typically measured to higher significance in the simulation than projected statistics (equation~\ref{eq:wdplusdef}). Projections lower the significance of the alignment signal because pairs with large separations along the line of sight (and low correlation in three dimensions) contribute weakly to the projected correlation. Moreover, weighting by ellipticity also dilutes the correlation found in three dimensions, as low ellipticity galaxies make a weak contribution even if their orientation vectors are aligned in three dimensions with the surrounding large-scale structure. Hence, it is useful to complement the measured projected signal with the three dimensional information on relative orientations accessible in the simulation.

Unless otherwise noted, the uncertainties in the projected correlation functions, including both the dispersion in the intrinsic shapes (``shape noise'') and cosmic variance of modes within the simulation box, are obtained from jackknife resampling \citep{Hirata04b,Mandelbaum06b}. We divide the simulation box into cubes of length $\ell/4$ in each dimension, which are removed one by one in each jackknife iteration. Due to the limited size of the box, it is expected that the error bars could be underestimated at large scales. In section~\ref{sec:results} and for computational reasons, we will consider the shape noise variance alone when the conclusions drawn are qualitative only. Notice that we found in Paper I that the overall uncertainty including cosmic variance can be a factor of $1-4$ greater than the bin variance at small scales ($<1\,h^{-1}$ Mpc), and up to an order of magnitude greater at the largest scales probed in this work ($20\,h^{-1}$ Mpc).

The effect of grid-locking arising from correlations of galaxy spins and shapes with the simulation grid, was studied in detail in Appendix A of Paper I, where we showed that position-shape and position-spin correlations are not affected by those systematics. While shape-grid or spin-grid correlations exist and depend on galaxy properties, there is no expected effect on the correlation of positions and shapes as galaxy separations are not grid-locked. We refer the reader to Paper I for more details on the jackknife resampling procedure and grid-locking contamination.

\subsection{Modelling}
\label{sec:model}

Alignments of luminous red galaxies are typically modelled assuming that the intrinsic component of the shape is proportional to the projected tidal field of the large-scale structure \citep{Catelan01}, $\gamma^I_{(+,\times)} \propto C_1 T_{(+,\times)}[\phi_p]$, where $C_1$ is a bias quantifying the response of a galaxy to the tidal field, $T_{(+,\times)}$ are the components of the projected tidal operator and $\phi_p$ is the primordial gravitational potential at the redshift of galaxy formation. While this model should only strictly be applied to elliptical galaxies and in the linear regime, several works have applied the tidal alignment model to the disc population as well \citep{WiggleZ,Heymans13}. Recently, \citet{Larsen15} validated this approach by showing that the scale-dependence of the tidal alignment model is a good approximation to that arising from a tidal torquing model for discs \citep{Catelan01,Mackey02,Hirata04}. Under these assumptions, the power spectrum of the density field and the $+$ component of galaxy shapes can be expressed as,
\begin{equation}
  P_{\delta +}({\bf k},z) = -A_I \frac{C_1\rho_{\rm crit}\Omega_m}{D(z)}\frac{k_x^2-k_y^2}{k^2}P_\delta({\bf k},z)\,,
  \label{eq:powerdi}
\end{equation}
where $\rho_{\rm crit}$ is the critical density of the Universe today, $D(z)$ is the growth function (normalized to unity at $z=0$), $k_x$ and $k_y$ are the components of the wavemode vector on the sky, and $P_\delta$ is the matter power spectrum. We adopt by convention a fixed value of $C_1\rho_{\rm crit}=0.0134$ \citep[from SuperCOSMOS measurements at low redshift]{Brown02} and we instead leave the $A_I$ parameter free. 
A simple nonlinear extension of the model replaces the linear matter power spectrum in equation~(\ref{eq:powerdi}) by its nonlinear analogue \citep[`NLA model']{Hirata07,Bridle07}, although there is evidence that this results in missing power at small scales \citep{Singh14,Blazek15}. In some works, a smoothing filter is applied to the NLA model to suppress the contribution of the tidal field within the typical scale of a halo \citep[e.g.][]{Catelan01,Hirata04,Blazek11,Chisari13}. In these cases, the smoothing scale is another free parameter of the model. We do not apply any smoothing kernel in this work.

The projected correlation function of galaxy intrinsic shapes and the density field is
\begin{align}
w_{\delta +}(r_p) &= -A_I \frac{C_1\rho_{\rm crit}\Omega_m }{\pi^2D(z)} \int_0^{\infty}dk_z \int_0^{\infty}dk_{\perp} \nonumber
\\
& \hskip -1cm \frac{k_\perp^3}{(k_{\perp}^2+k_z^2)k_z}P_\delta({\bf k},z)\sin(k_z\Pi_{\rm max})J_2(k_{\perp} r_p),
\label{eq:wplus_nla}
  \end{align}
where $r_p$ is the projected radial separation, $k_z$ is the Fourier mode component along the line of sight, $k_{\perp}$ is the component perpendicular to the line of sight and $J_2$ is the second order Bessel function of the first kind. Notice that the value of $\Pi_{\rm max}$ is incorporated into the modelling and thus $A_I$ should be insensitive to our particular choice for this parameter. The $\times$ component correlation is expected to be null and we consider it only as a test for systematics.

Observations have shown evidence of increased nonlinear power in alignments with respect to the predictions of the NLA model. This excess power was modelled by the addition of a one-halo contribution \citep{Schneider10,Singh14}. The contribution to the correlation function of the one-halo term is
\begin{equation}
w_{\delta +}^{1h} = \int \frac{dk_\perp}{2\pi} k_\perp P_{\delta,{\gamma^I}}^{1h}(k_\perp,z)J_0(k_\perp r_p)\,,
\end{equation}
where $P_{\delta,{\gamma^I}}^{1h}$ is the alignment one-halo power spectrum \citep{Schneider10} and $J_0$ is the zeroth order Bessel function of the first kind. A set of constants parametrize $P_{\delta,{\gamma^I}}^{1h}$, given by
\begin{equation}
P_{\delta,\gamma^I}^{1h}(k,z)=-a_h\frac{(k/p_1)^2}{1+(k/p_2)^{p_3}}\,,
\end{equation}
where $a_h$ is the halo model alignment amplitude. The parameters $p_1$, $p_2$, and $p_3$ are redshift-dependent and given by the following functional forms,
\begin{align}
p_1 &= q_{11} \exp(q_{12}z^{q_{13}}),\nonumber\\
p_2 &= q_{21} \exp(q_{22}z^{q_{23}}),\nonumber\\
p_3 &= q_{31} \exp(q_{32}z^{q_{33}}).
\label{eq:halopar}
\end{align}
We adopt the values of $q_{ij}$ given in Table 1 of \citet{Singh14}, obtained from fits to measured alignments of LRGs in SDSS. However, we will find the need to fit for some of them in section~\ref{sec:results}. Throughout this work, we will use the convention that radial alignments correspond to negative values of $w_{\delta+}$ and, as in Paper I, we will restrict the halo model fits to small scales: $r_p<0.8\, h^{-1}\, \rm Mpc$.

We model galaxy position-intrinsic shape correlations simply as $w_{g+}=b_gw_{\delta+}$, where $b_g$ is the linear galaxy bias. Note that any scale dependence of the galaxy bias is thus attributed to the alignment model through $w_{\delta+}$. As a result, fits to $A_I$ in the case of $w_{g+}$ and $w_{\delta+}$ are not directly comparable except at large scales, but modelling the small scale dependence of $b_g$ is beyond the scope of this work.

\section{Results}
\label{sec:results}

\begin{figure*}
  \includegraphics[width=0.33\textwidth]{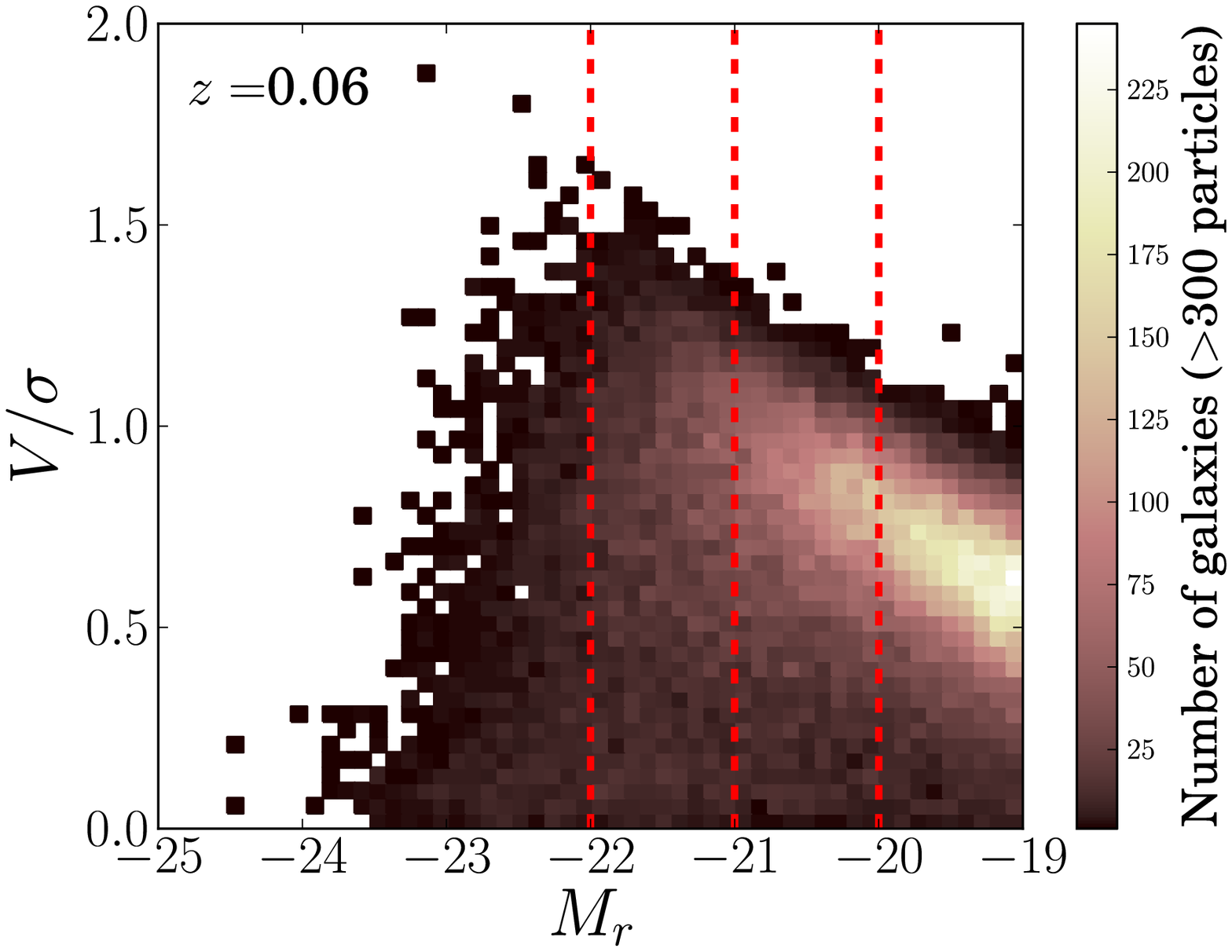}
  \includegraphics[width=0.33\textwidth]{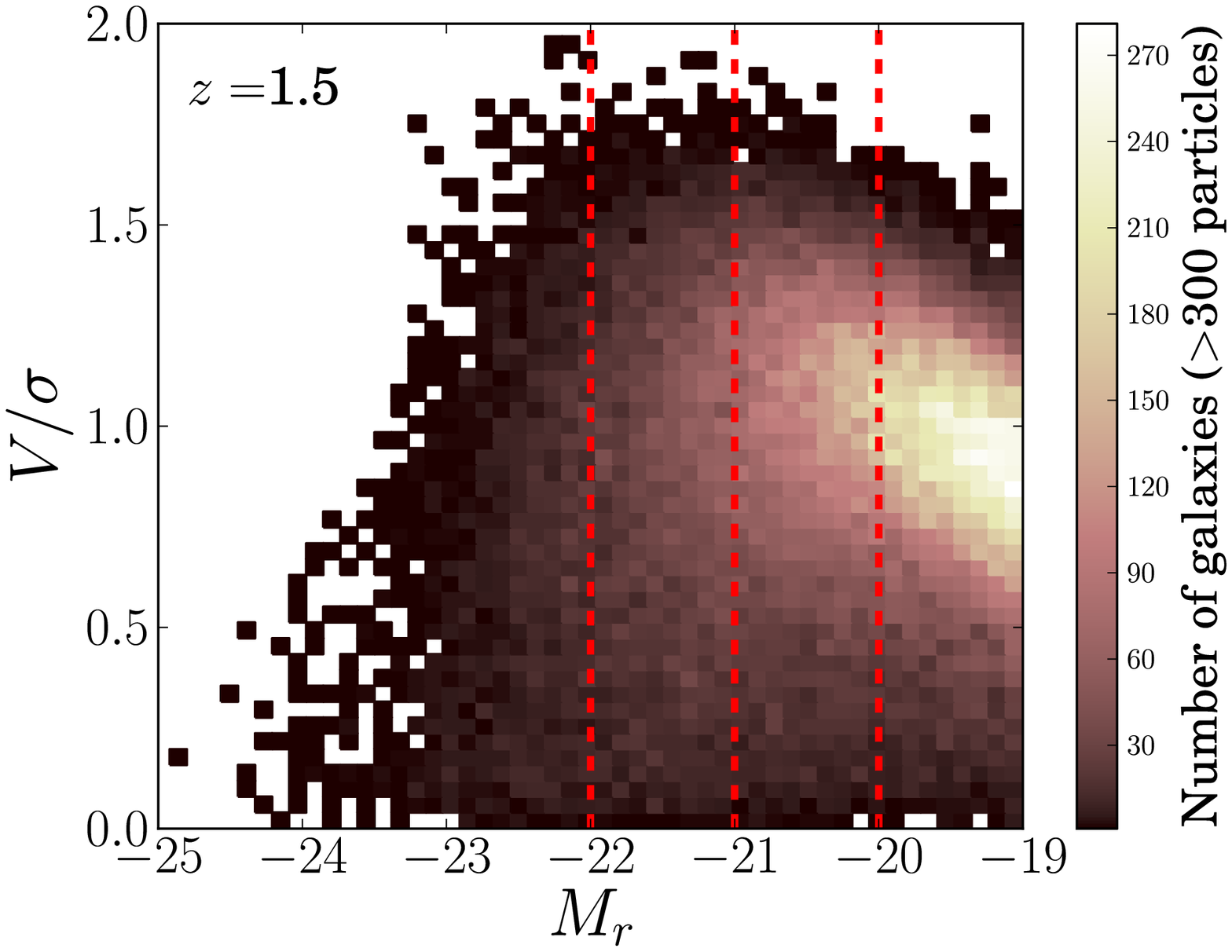}
  \includegraphics[width=0.33\textwidth]{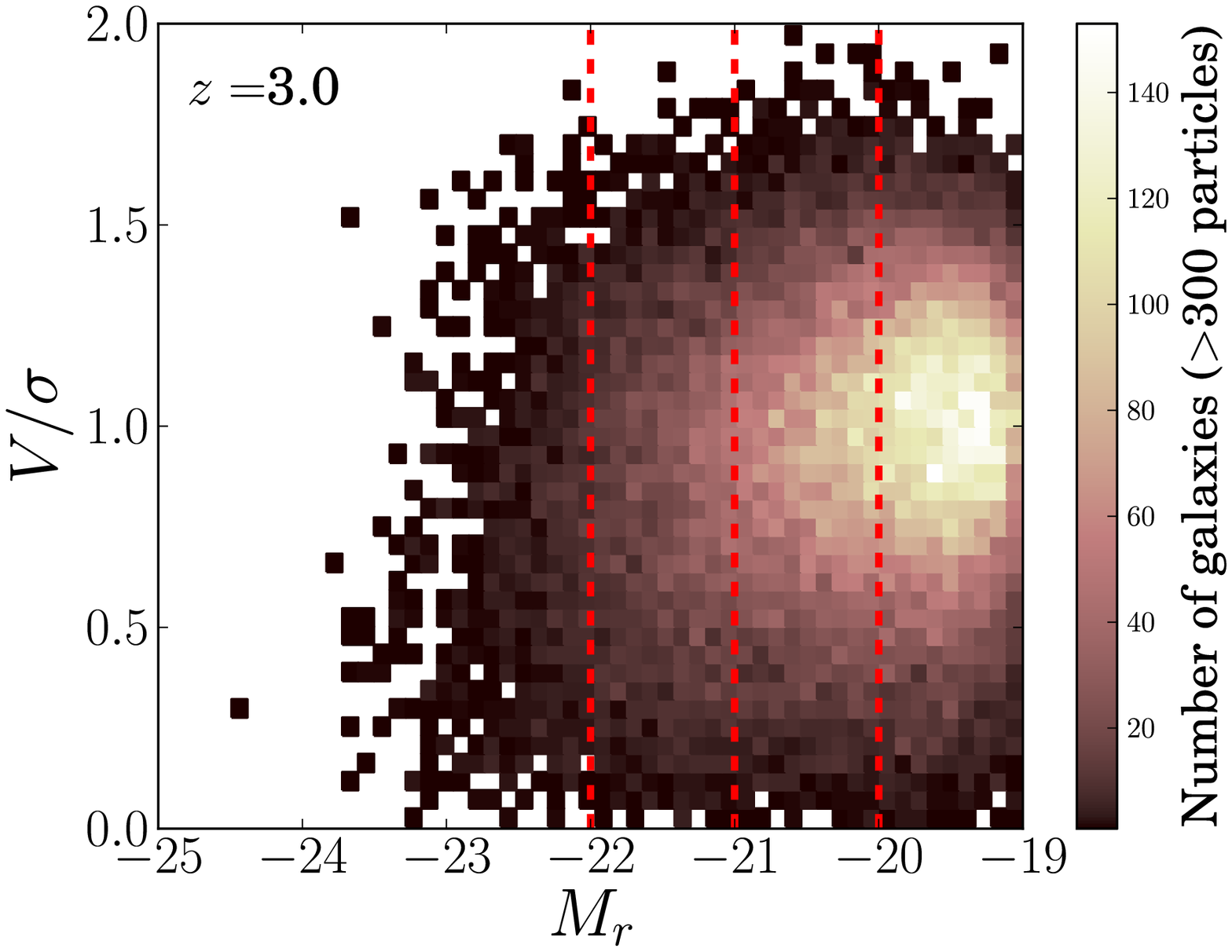}
  \caption{Distribution of $V/\sigma$ and $r$-band rest-frame absolute magnitude ($M_r$) at redshifts $z=0.06$ (left), $z=1.5$ (middle) and $z=3$ (right). The dashed vertical lines indicate the boundaries of the luminosity bins considered for the intrinsic alignment measurements in this work. Galaxies at higher redshift tend to have uniformly higher $V/\sigma$ at all luminosities, while the build up of the elliptical population at lower redshift results in a decrease of $V/\sigma$.}
  \label{fig:vsigmr}
\end{figure*}

We present results for the redshift and luminosity evolution of intrinsic alignments from Horizon-AGN in the redshift range: $0.06<z<3$, covering the range expected to be of relevance for the LSST survey, and exceeding the coverage needed for {\it Euclid}. Due to our selection of galaxies with $>300$ stellar particles, we apply a minimum threshold in luminosity of $M_r\leq-20$ (section~\ref{sec:complete}), above which the galaxy population is complete over the entire redshift range probed. We divide the galaxies into three luminosity bins: $-21<M_r\leq -20$, $-22<M_r\leq -21$ and $M_r\leq -22$. We work with rest-frame magnitudes that do not account for dust extinction; this makes our results more directly comparable to previous works. We gauge the impact of more realistic magnitudes (including dust, $K$-corrections, a Chabrier initial mass function and a metallicity correction) on the luminosity and redshift dependence of the alignment signal in Appendix~\ref{app:dust}.

\subsection{Elliptical and disc fractions}
\label{sec:efrac}

In Paper I, we identified two mechanisms for galaxy alignments at $z=0.5$ in Horizon-AGN. Discs have a tendency for tangential alignments around overdensities, while ellipticals tend to be elongated pointing towards them. In that work, we used a threshold of tangential velocity to velocity dispersion, $V/\sigma>0.55$, to define discs. Here, $V$ is the average of the tangential velocities of stars, and $\sigma^2=(\sigma_r^2+\sigma_t^2+\sigma_z^2)/3$, where $\sigma_r$, $\sigma_t$, and $\sigma_r$ are the dispersions around the average value of the radial, tangential, and vertical velocity components respectively in the cylindrical coordinates defined by the $z$-vertical spin axis of the galaxy.

The choice of this threshold was based on requiring that $2/3$ of the galaxy population fell into the disc category. Moreover, \citet{Dubois14} found that galaxies below $V/\sigma=0.6$ displayed a different alignment trend with respect to their nearest filament than galaxies above that threshold at $z=1.2$ in Horizon-AGN. We adopt once more the $V/\sigma=0.55$ threshold in this section to differentiate the two populations. The distribution of galaxies in the $V/\sigma-M_r$ plane for redshifts $z=\{0.06,1.5,3\}$ is shown in Fig.~\ref{fig:vsigmr}, and the boundaries of the different luminosity bins are indicated by dashed vertical lines. The highest luminosity bin has a larger contribution from elliptical galaxies and the mean $V/\sigma$ decreases towards low redshifts. Lower luminosity bins typically have larger contributions from disc-like galaxies, although we also find that a population of low luminosity ellipticals builds up towards lower redshift.

\begin{figure}
  \includegraphics[width=0.495\textwidth]{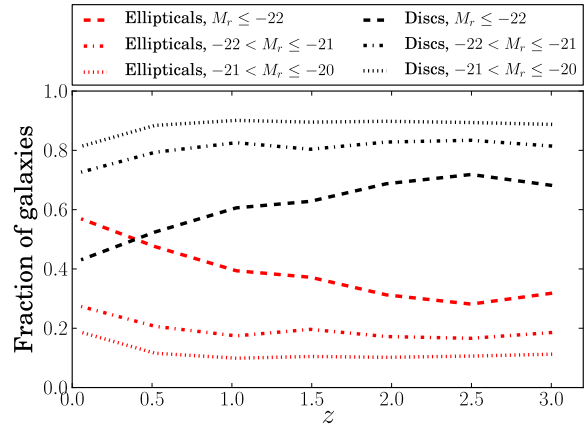}
  \caption{Fraction of discs and ellipticals in Horizon-AGN in each luminosity bin as a function of redshift. We adopt a cut in $V/\sigma=0.55$ to separate the two populations. The elliptical fraction increases towards low redshift, particularly for the highest luminosity bin. The total comoving number density of ellipticals with $M_r\leq-20$ ranges between $48 \times 10^{-4}\,h^3\,{\rm Mpc}^3$ and $26 \times 10^{-4}\,h^3\,{\rm Mpc}^3$ from $z=0$ to $z=3$.}
  \label{fig:rfrac}
\end{figure}

We measure the fraction of ellipticals and discs as a function of redshift and luminosity in the simulation. This provides complementary information to the intrinsic alignment signal per se. In Fig.~\ref{fig:rfrac}, we show the fraction of discs and ellipticals in each luminosity bin as a function of redshift. The elliptical fraction increases towards low redshift due to the build up of the red sequence. Their fraction is most significant at high luminosities and decreases almost monotonically with luminosity. However, Fig.~\ref{fig:vsigmr} shows the appearance of a population of ellipticals with $M_r>-20$ towards $z=0.06$. This population is beyond our completeness limit at higher redshift and is not studied here. The non-monotonic mass-dependence of the alignment signal measured in Paper I at $z=0.5$ was in fact attributed to low luminosity ellipticals. We will discuss their contribution further in section~\ref{sec:discuss}.

The comoving number density of ellipticals with $M_r\leq-20$ is approximately $48 \times 10^{-4}\,h^{-3}$ Mpc$^{-3}$ at $z=0.06$ and decreases with redshift to $\simeq 26 \times 10^{-4}\,h^{-3}$ Mpc$^{-3}$ at $z=3$. In comparison, the LOWZ sample of the SDSS-III survey \citep{Singh14} has an average comoving number density of $\bar{n}\sim 3\times 10^{-4}\,h^{-3}$ Mpc$^{-3}$ between $0.16<z<0.36$ and an average rest-frame $r$-band magnitude between $-21.95 < \langle M_r\rangle < -21.65$ in that redshift range. Selecting ellipticals within that magnitude range in Horizon-AGN, we obtain $\bar{n}\sim 5\times 10^{-4}\,h^{-3}$ Mpc$^{-3}$ at a comparable redshift. We do not expect to match exactly the selection effects of the LOWZ sample, since we do not include colour cuts or fibre collision effects, which can play a relevant role. However, it is encouraging to find similar order of magnitude counts by dynamical and luminosity selection.

\subsection{Spin and shape correlations}

In this section, we present three dimensional correlation functions of the shapes and spins, and projected correlation function of projected shapes (see section~\ref{sec:correl}) with respect to the position of the DM particles and the galaxies in the simulation box. We complement these measurements by measuring distributions of angles between the orientation of the minor and spin axes (not shown in the figures). 
%
%
To assess the significance of each alignment measurement, projected and three-dimensional, we use the full covariance matrix obtained through the jackknife procedure. We perform least-squares fits to the projected correlation functions of alignments, $w_{g+}$ and $w_{\delta +}$, using the NLA model and the halo model introduced in section~\ref{sec:model}. We apply least squares minimization using the diagonal of the covariance matrix to obtain constraints on the preferred parameters and their $1\sigma$ uncertainty. Note that the use of the diagonal of the covariance in the fits usually increases the strength of alignments compared to using the full covariance; in this sense, if the goal is to establish upper limits to alignment contamination to future weak lensing surveys, we are conservative in our estimate of alignment amplitudes. The covariance between adjacent bins can reach up to $\sim 70\%$ level on large scales.

\subsubsection{Relative orientations at $z=0$}
\label{sec:z0}

\begin{figure}
  \includegraphics[width=0.49\textwidth]{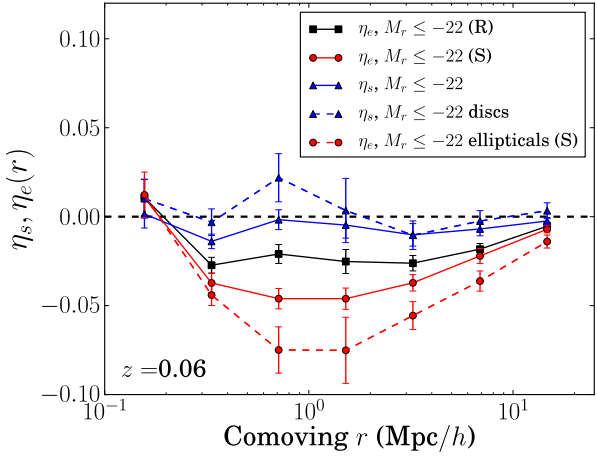}
  \caption{$\eta_e(r)$ and $\eta_s(r)$ at $z=0.06$ for galaxies with $M_r\leq -22$, obtained from orientations around DM particles. Orientations as defined from the simple inertia tensor are represented as circles; from the reduced inertia tensor, as squares; and for the spin, as triangles. The dashed red line represents $\eta_e(r)$ for ellipticals using the simple inertia tensor as the shape tracer. The dashed blue line corresponds to $\eta_s(r)$ for discs. The plotted error bars are obtained by means of the jackknife procedure. Ellipticals dominate the alignment signal, with their minor axes pointing tangentially around DM overdensities (corresponding to a negative $\eta_e$).}
  \label{fig:jack3d761}
\end{figure}

We find the strongest alignment signal at low redshift and we discuss those results separately in this section. At $z=0.06$, the alignment signal is the strongest for the most luminous galaxies ($M_r\leq-22$). The $\eta_e$ and $\eta_s$ statistics for this population are shown in Fig.~\ref{fig:jack3d761}. The negative trend in that panel indicates that both the spin and the minor axis of these galaxies are pointing tangentially around DM overdensities. This is equivalent to galaxies having their major axes laying parallel to the direction of the separation vector towards DM overdensities. This tendency coincides qualitatively with observational results for shape alignments of luminous red galaxies \citep{Mandelbaum06,Hirata07,Okumura09,Joachimi11,Singh14}. The amplitude of the signal depends on the observable: the reduced inertia tensor enhances the contribution of stellar particles closer to the center, and results in rounder shapes and lower alignment correlation than the simple inertia tensor. Nevertheless, both types of shape correlations are detected at $>99.99\%$ confidence level (C.L. hereafter). Other authors \citep{Singh15} have attributed the stronger alignment signal of the simple inertia tensor as coming from a more efficient twisting of the outer isophotes of galaxies in the direction of the tidal field.

For this population, we also find a trend for the spin to be pointing perpendicular to the separation vector at $99.7\%$ C.L. From the distributions of relative angles between spin and minor axes, we have confirmed that galaxies tend to have their spin and minor axes aligned with each other; although the level of correlation depends on $V/\sigma$ and luminosity. The median misalignment angle of the spin and minor axis is high for elliptical galaxies, ranging between $\sim 15-30\deg$, with lower values for lower luminosities and the reduced inertia tensor. For discs, the spin and the minor axis are better correlated, with a median misalignment of $\sim 3-6\deg$ depending on luminosity and shape estimator. As for ellipticals, the misalignment decreases for lower luminosities and the reduced inertia tensor.

\begin{figure}
  \includegraphics[width=0.49\textwidth]{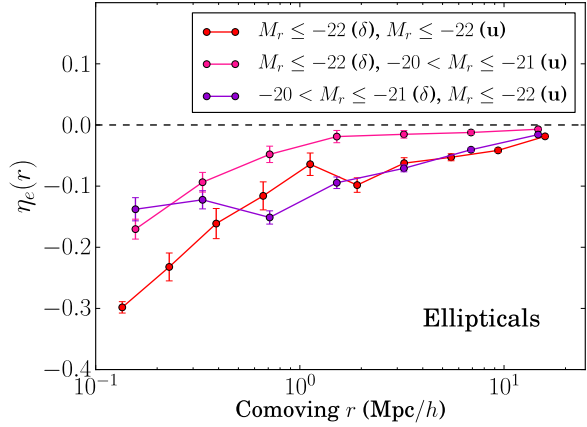}
  \caption{$\eta_e(r)$ for ellipticals around ellipticals in the different luminosity bins. The red line indicates the alignment of the minor axes of luminous ($M_r\leq-22$) ellipticals around other luminous ellipticals; the pink line corresponds to the orientation of low luminosity ($-21<M_r\leq-20$) ellipticals around luminous ellipticals; and the violet line, to those of luminous ellipticals around low luminosity ellipticals. A negative sign for $\eta_e$ corresponds to a tangential orientation of the minor axis of the shape (${\bf u}$) tracer around the density ($\delta$) tracer. The plotted error bars correspond to the bin variance alone; while this underestimates the true uncertainty, this figure is only shown for a qualitative purpose only.}
  \label{fig:threemodes}
\end{figure}
\begin{figure}
  \includegraphics[width=0.49\textwidth]{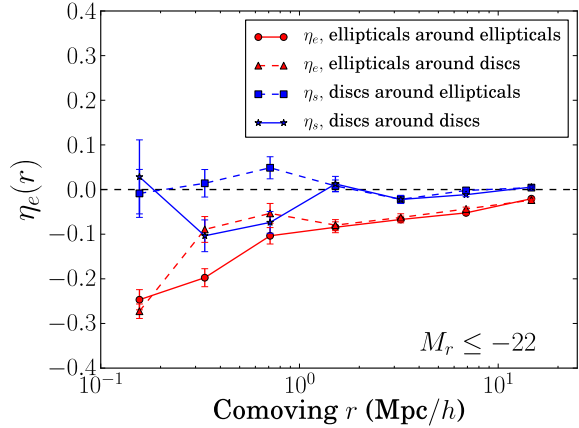}
  \caption{$\eta_e(r)$ for ellipticals and $\eta_s$ for discs with $M_r\leq -22$ at $z=0.06$, calculated around ellipticals and discs. We use the simple inertia tensor as shape tracer for the ellipticals. Black circles correspond to ellipticals around ellipticals; blue triangles, to ellipticals around discs; red squares, to discs around ellipticals and magenta stars, to discs around discs. Ellipticals are elongated towards other ellipticals and towards overdensities of discs. The plotted error bars correspond to the bin variance alone; while this underestimates the true uncertainty, this figure is only shown for a qualitative purpose only.}
  \label{fig:cross761}
\end{figure}

Splitting the population of $M_r\leq-22$ galaxies into discs (blue dashed representing their spin in Fig.~\ref{fig:jack3d761}) and ellipticals (red dashed representing the minor axis orientation), we find that the radial alignment signal with respect to the DM field is, overall, due to the elliptical population. It is also clear that the radial alignment of ellipticals is a decreasing function of luminosity for the simple inertia tensor. When the reduced inertia tensor is used, the highest luminosity sample of ellipticals shows a clear enhancement of alignment, while lower luminosity galaxies tend to show comparable alignment at this redshift. The overall amplitude and significance of the alignment signal  of ellipticals is reduced when averaged together with the disc population. The complete set of $\eta_e$ and $\eta_s$ correlations as a function of luminosity, redshift and $V/\sigma$ can be found in Fig.~\ref{fig:align3d} of Appendix~\ref{app:allcorr}.

Alignment trends of low and intermediate luminosity galaxies at this redshift retain a $>3\sigma$ significance for measurements performed with the simple inertia tensor. This suggests that there is an interplay between the fraction of ellipticals and the strength of alignment in each luminosity bin to yield the alignment signal of all galaxies, ellipticals and discs. As ellipticals are pressure-supported systems, their spin is a noisy quantity, and it is thus expected that the spin alignment signal would be smaller than the shape alignment for this population. On the other hand, there is no evidence for alignment of discs at this redshift.

\begin{figure}
  \includegraphics[width=0.49\textwidth]{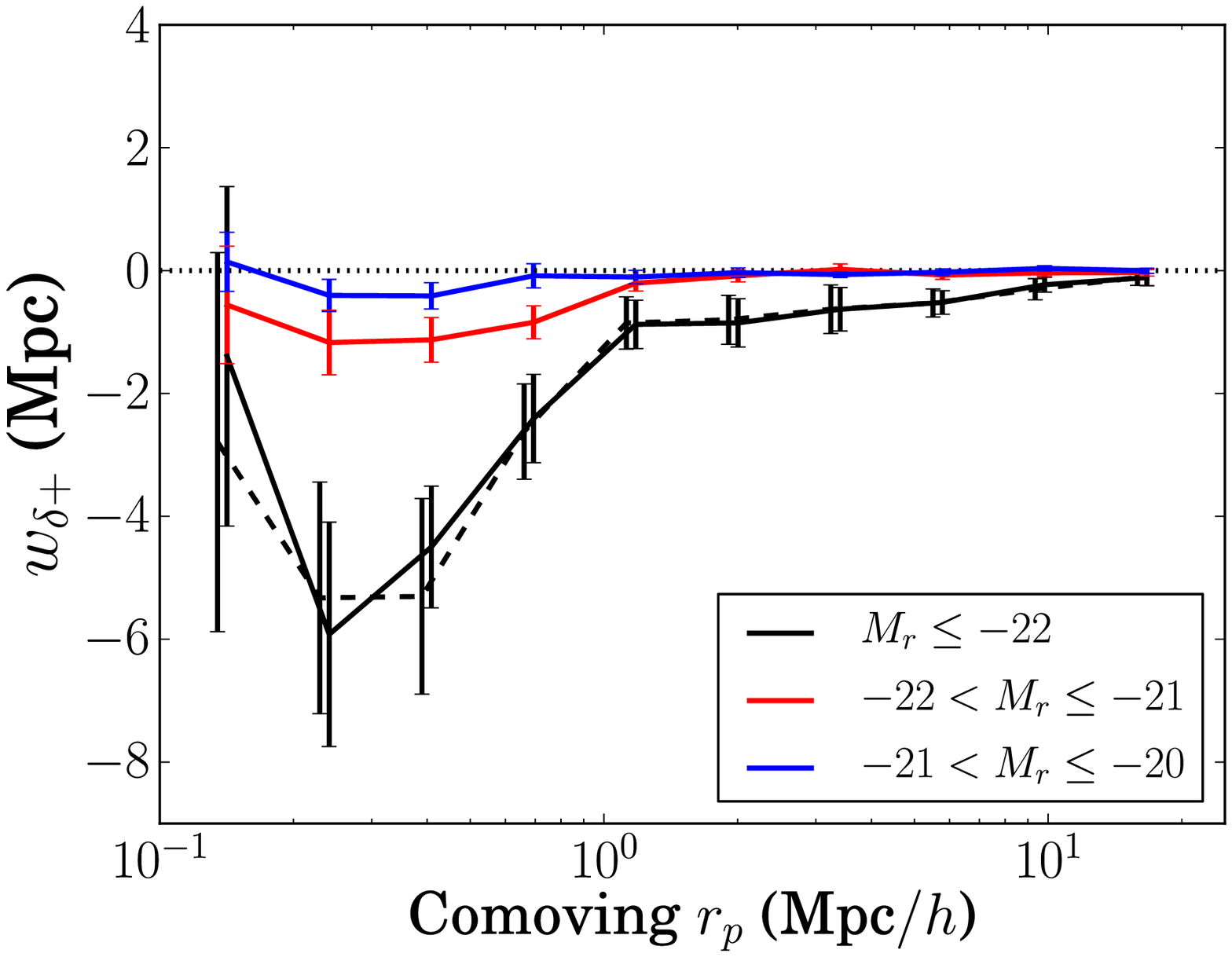}
  \includegraphics[width=0.49\textwidth]{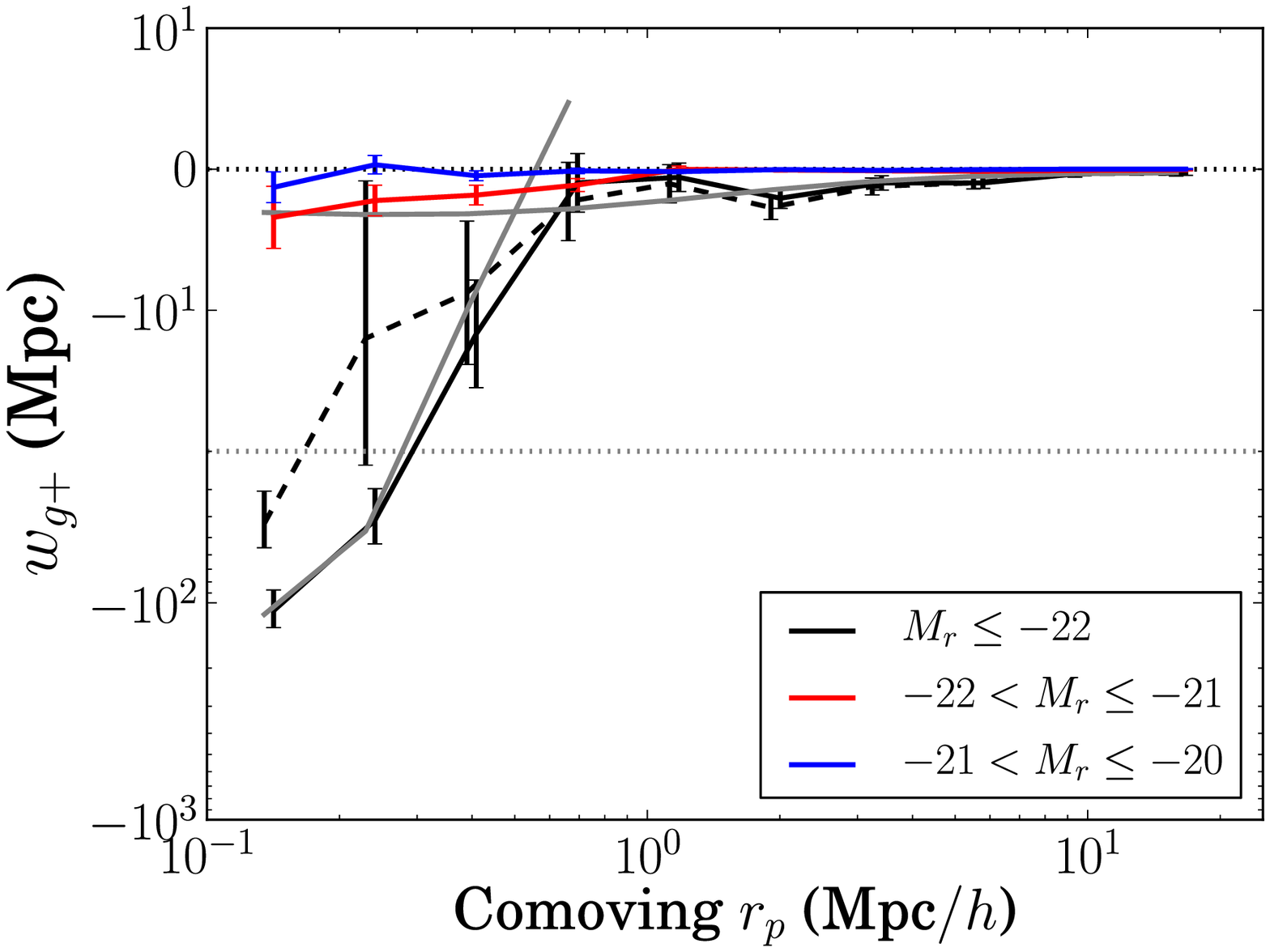}
  \caption{Projected correlation functions of alignments at $z=0.06$. The two panels show $w_{\delta+}$ for the simple inertia tensor (top) and $w_{g+}$ using the simple inertia tensor (bottom). The colours indicate the different magnitude bins: $M_r\leq -22$ (black), $-22<M_r\leq -21$ (red) and $-21<M_r\leq -20$ (blue); and the components (NLA+halo model) of the best fit (grey) to the highest luminosity bin results. Note the change between logarithmic scale and linear scale in the $y$-axis at $w_{g+}=-20$, indicated by a dotted black line. The solid lines indicate the scenario where substructures with $>300$ particles are identified as galaxies. The dashed lines correspond to the case where the stellar particles of substructures are considered as part of the corresponding central galaxy. The two procedures mostly yield consistent results within the error bars for $w_{\delta +}$; some departures are seen between the two measurements at small scales in the case of $w_{g+}$.}
  \label{fig:proj761}
\end{figure}

Correlations of galaxy orientations around {\it galaxies} give additional information on the dependence of alignments with environment. We split the galaxy population in each luminosity bin into discs and ellipticals and we measure relative orientations between them and also with respect to discs and ellipticals of different luminosities. We find that ellipticals present the strongest alignment around other ellipticals of the same luminosity. The alignment decreases as the shape tracer and/or the density tracer decrease in luminosity. These suggests that there are several components to the alignment signal (see extended discussion in Appendix~\ref{app:allcorr}) that can be connected to previous works. First, luminous galaxies point towards each other, as observed for LRGs \citep{Mandelbaum06,Hirata07,Okumura09,Joachimi11,Singh14}. Second, the correlation persists if low luminosity galaxies are used as tracers of the density field, indicating that they live preferentially in the direction of the semimajor axis of the central \citep{Binggeli82,Mandelbaum06,Welker15}. These low luminosity satellites are also radially elongated pointing towards the central, more luminous, galaxy \citep{Singh15}. All three modes can readily be identified in Fig.~\ref{fig:threemodes}. We also find an alignment signal of ellipticals around discs suggesting, as in Paper I, that ellipticals tend to orient their major axes towards overdensities of discs, i.e., filaments. An example of these two modes of alignments is seen in Fig.~\ref{fig:cross761} for high luminosities. 

\subsubsection{Projected correlations at $z=0$}
\label{sec:z0bis}

The top panel of Fig.~\ref{fig:proj761} shows the projected correlation function of the density field and intrinsic shapes, $w_{\delta +}$, for the simple inertia tensor. The most luminous galaxies have a stronger radial alignment signal than other galaxies. The signal persists to large scales, but there is a marked enhancement on scales $\lesssim 1\, h^{-1}\, \rm Mpc$. As inferred from $\eta_e$, the strength of alignment is monotonically decreasing with luminosity. Moreover, the use of the reduced inertia tensor reduces the significance of $w_{\delta +}$. In this case, we find evidence of significant alignment only for the most luminous galaxies. The $w_{\delta\times}$ component is consistent with null at $2\sigma$ C.L. 

The bottom panel of Fig.~\ref{fig:proj761} shows the projected correlation between the galaxy distribution and the intrinsic shapes, $w_{g+}$, at $z=0.06$ for the three luminosity bins for the simple inertia tensor. The results are similar, with a smaller amplitude, when the reduced inertia tensor is used. We have verified that $w_{g\times}$ is consistent with null for all luminosity bins considered. The $w_{g+}$ alignment at small scales ($r_p \lesssim 1$ Mpc$/h$) is much stronger than for $w_{\delta+}$, and increasing the number of DM particles used to trace the density field has no significant impact on the scale-dependence of $w_{\delta +}$. This difference between $w_{g+}$ and $w_{\delta +}$ can be attributed to an anisotropic distribution of satellites in the direction of the major axis of the central galaxy \citep{Welker15}. If the DM halos are more spherical than the distribution of satellites around the central galaxy, $w_{\delta +}$ will not capture this anisotropy, in contrast with $w_{g+}$. To test this last hypothesis, we have verified that a random perturbation of galaxy positions (but keeping the position of shape tracers fixed), which isotropizes their distribution up to $150$ kpc$/h$, can have a dramatic impact in the alignment amplitude of $w_{g+}$ in the first radial bin, making it consistent with null. Furthermore, neither miscentering of the DM distribution with respect to the galaxies, nor scale-dependence of the clustering bias can be responsible for such a suppression in $w_{\delta +}$ at small scales. To verify this, we compared the clustering of $M_r\leq-22$ galaxies at $z=0$ with their position correlation with DM particles. While there is evidence of a decreasing bias at small scales, $w_{\delta g}$ is monotonically increasing, and does not evidence the same small-scale nulling as $w_{\delta +}$.

We also present, in Fig.~\ref{fig:proj761}, a test for the impact of substructure identification, as discussed in section~\ref{sec:postprocess}. In this figure, the solid lines correspond to treating AdaptaHOP substructures as separate galaxies, which is our fiducial approach. The dashed lines indicate the projected correlations obtained when the stellar particles of galaxies identifies as substructure are considered as part of their host central galaxy. Our results are fully consistent within our error bars for both shape measurements. Although not shown, we also find that $\eta_e$ and $\eta_s$ are even more robust, suggesting that orientations are less affected than projected ellipticities. Overall, the details of how substructure is identified in the simulations do not change $w_{\delta +}$ significantly. Excluding galaxies classified as substructure altogether decreases the amplitude of the correlation slightly for the simple inertia tensor, although the change is still within error bars. In the case of $w_{g+}$, the identification of galaxies as substructure, rather than being part of the central, results in an increased correlation function at small separations $\lesssim 500\, h^{-1}\, \rm kpc$ due to the excess number of pairs. Thus, the details of the galaxy finder can potentially play a role in constraining the halo model of alignments on such scales.


Measurements of $w_{\delta +}$ and $w_{g+}$ for the highest luminosity galaxies at $z=0$ reject the null hypothesis at $\geq 3\sigma$ C.L. For these, neither the NLA model nor the NLA+halo model are a good fit. In the case of $w_{\delta+}$, the halo model tends to overestimate the power on the first bin; while for $w_{g+}$, the scale-dependence is not steep enough. The downturn of $w_{\delta +}$ at the smallest radii cannot be reproduced by relaxing the parameters of the halo model provided by \citet{Schneider10} because this model relies on the assumption that the intrinsic shear of satellite galaxies in a spherical halo is invariant with separation to the center of the halo. On the other hand, it is possible to fit the scale-dependence of $w_{g+}$ if we allow $p_2$ (equation~\ref{eq:halopar}) to vary. The preferred parameters and their $1\sigma$ uncertainty are: $p_2=1.74\pm0.17$ (compared to the fiducial value of $p_2\simeq 0.7$), $b_ga_h=0.048\pm 0.013$ and $b_gA_I=8.8\pm 1.9$ for the simple inertia tensor; $p_2=1.86\pm0.22$, $b_ga_h=0.0168\pm 0.0057$ and $b_gA_I=3.2\pm1.0$ for the reduced inertia tensor case. However, notice that there are only $4$ points at small scales and $2$ free parameters for the halo model, and that these points have correlations of up to $\sim50\%$. Nevertheless, the best-fit to the simple inertia tensor case is shown as a gray line in the bottom panel of Fig.~\ref{fig:proj761}. 

Alternatively, a fit of the NLA model to $w_{g+}$ only at large scales (defined as $0.8<r_p<20$ Mpc$/h$) yields $b_gA_I=8.7\pm 2.1$ for the simple inertia tensor, and $b_gA_I=3.1\pm1.1$ for the reduced inertia tensor. These results are very similar to those quoted in the previous paragraph, suggesting there is little covariance between the measurement on small and large scales. For $w_{\delta+}$, we obtain $A_I=4.98\pm 0.79$ for the simple inertia tensor, and $A_I=1.64\pm0.47$ for the reduced inertia tensor, implying that the bias of $M_r\leq -22$ galaxies at this redshift is $b_g\simeq 2$. If we restrict to the elliptical population with $M_r\leq-22$, we obtain a $30\%$ enhancement of the alignment amplitude at large scales: $A_I=7.38\pm0.87$ for the simple inertia tensor, and $A_I=2.20\pm0.49$ for the reduced inertia tensor. 

\subsubsection{Evolution of relative orientations}
\label{sec:align3d}

In this section, we focus on the evolution of $\eta_s$ and $\eta_e$ alone; projected correlation functions and model fits are discussed in the next section. The complete set of $\eta_s$ and $\eta_e$ measurements as a function of redshift, luminosity and $V/\sigma$ is presented in Appendix~\ref{app:allcorr}; we hereby summarise our main findings.

\begin{figure}
  \includegraphics[width=0.45\textwidth]{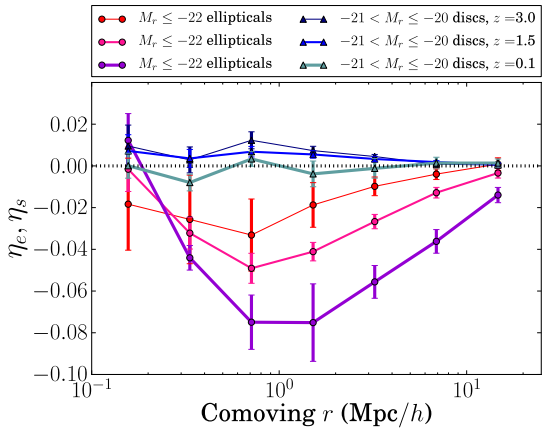}
  \caption{$\eta_s$ and $\eta_e$ around DM particles for high luminosity elliptical galaxies (circles) and low luminosity spirals (triangles) at redshifts $z=0.06,1.5,3$ (from thick to thin). For the triangles, the orientation is determined from the spin. The circles correspond to orientations derived from the minor axis of the simple inertia tensor. The dashed black line corresponds to no alignment. A positive correlation indicates that the minor axis or spin vector is pointing radially towards the DM particles.}
  \label{fig:align3d_small}
\end{figure}

The alignments of ellipticals decrease in strength with decreasing luminosity and increasing redshift. Fig.~\ref{fig:align3d_small} shows this evolution. These trends are a consequence of an intrinsic evolution of the alignment strength for this population. At $z=3$, we still observe radial alignments for the most luminous ellipticals. We also find a significant trend for both shape and spin {\it tangential} alignments of low luminosity discs at $>99.99\%$ C.L. The alignments of the overall population of discs and ellipticals, which are shown in Appendix~\ref{app:allcorr}, are dominated by the luminous ellipticals at low redshift and the low luminosity discs at high redshift. They tend to decrease with decreasing luminosity and increasing redshift as a consequence of intrinsic evolution in the alignment amplitude of ellipticals, and also due to increased fraction of discs. Eventually, at $z=3$, the overall signal changes sign as a consequence of the tangential alignment of discs.

At low redshift and high luminosities, we found different amplitudes for the alignment signal measured from the simple and reduced inertia tensor. The reduced inertia tensor yields rounder shapes than the simple case and the alignment signal is suppressed. This is a consequence of the signal being dominated by the alignments of ellipticals, for which the reduced inertia tensor produces a lower alignment strength. At high redshift and low luminosities, the overall alignment signal is dominated by the contribution of discs. As a consequence, the simple inertia tensor is an equally good tracer of alignments as the reduced inertia tensor or the spin. The orientation of the discs is equally well determined in all three cases, with median misalignment angles between the minor axis and the spin direction of $\sim 3-6 \deg$ across the whole redshift range probed.

In Paper I, we had found a tangential alignment of discs around ellipticals. It is clear from Figs.~\ref{fig:vsigmr} and \ref{fig:rfrac} that discs dominate the population of galaxies in the low luminosity bin, and their relative fraction increases towards high redshift. It is interesting to note that these tangential alignments are not observed when we measure $\eta_s$ of low luminosity discs around other galaxies with $M_r\leq -20$ at that redshift, instead of measuring it around DM particles. This suggests that the tangential alignment has its origin in the alignment of discs around lower density regions of the density field.

Other authors \citep{Tenneti15b} have found no evidence for tangential disc alignments in hydrodynamical cosmological simulations such as MassiveBlack II and Illustris. Our results suggest that a direct comparison of alignment mechanisms at high redshift ($z\sim 3$) across the different simulations could help elucidate the discrepancies between simulation techniques and/or the impact of baryonic physics. On the other hand, in Appendix~\ref{app:dust}, we show that the lowest luminosity bin where this signal is detected is beyond the magnitude limit of LSST ($m_r\sim 26$); and we will see in the next section that, in addition, disc alignments are highly suppressed in projection.

\subsubsection{Evolution of projected correlations}
\label{sec:resproj}

The redshift evolution of the projected correlations, $w_{\delta +}$ and $w_{g+}$, can inform us about the potential contamination of alignments to future lensing surveys. We measure these correlations as a function of redshift and luminosity in Horizon-AGN. 

\begin{figure*}
  \includegraphics[width=0.32\textwidth]{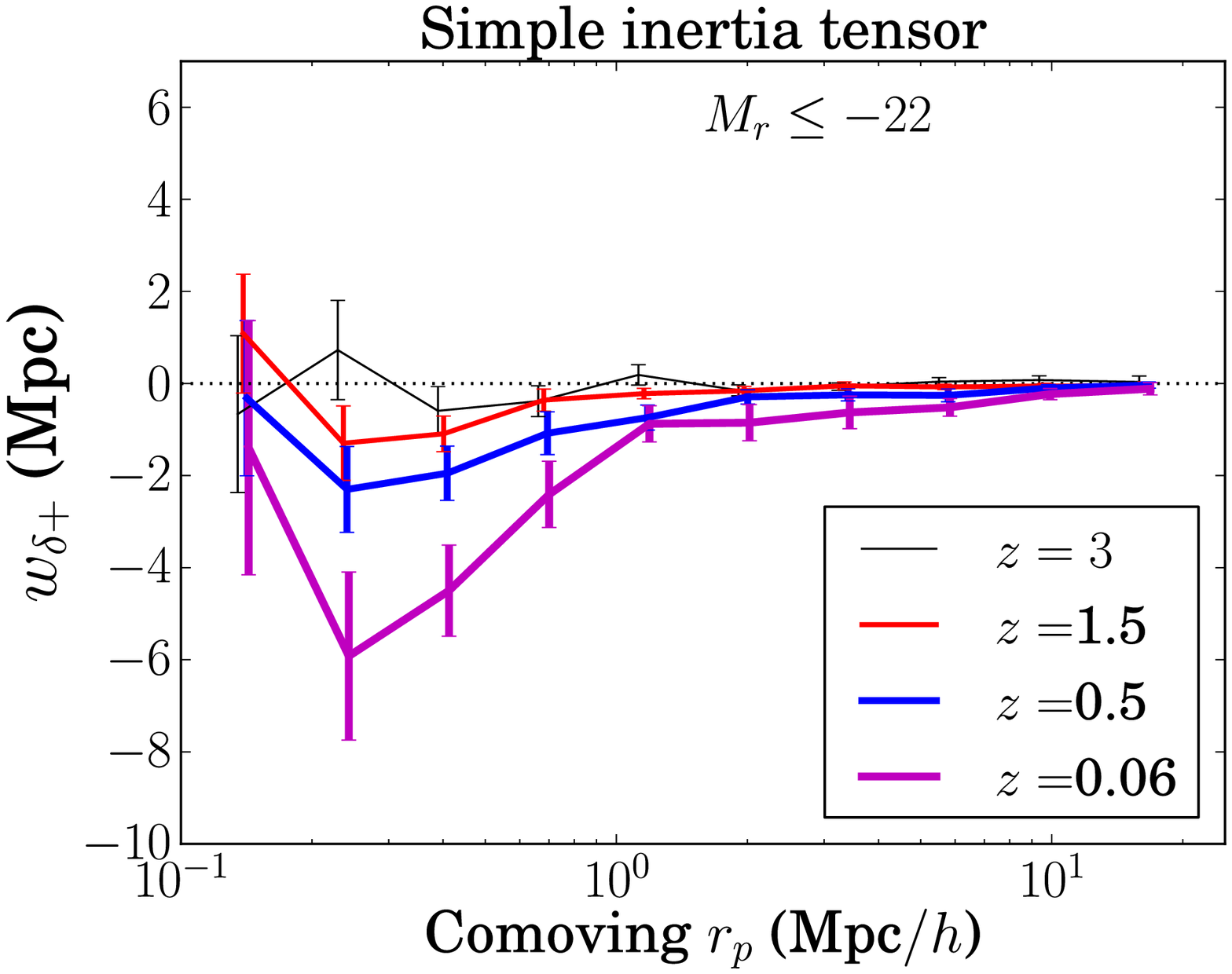}
  \includegraphics[width=0.32\textwidth]{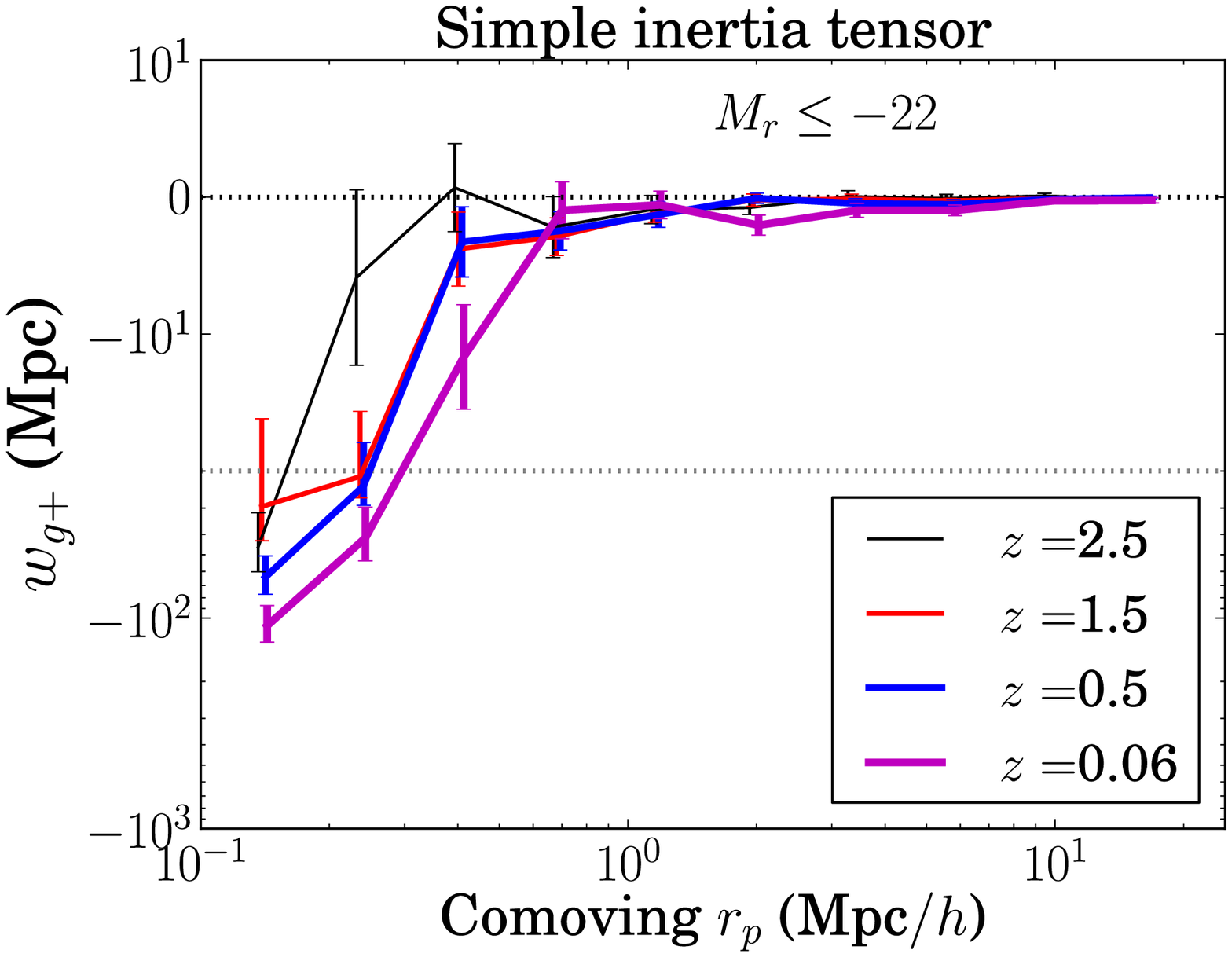}
  \includegraphics[width=0.32\textwidth]{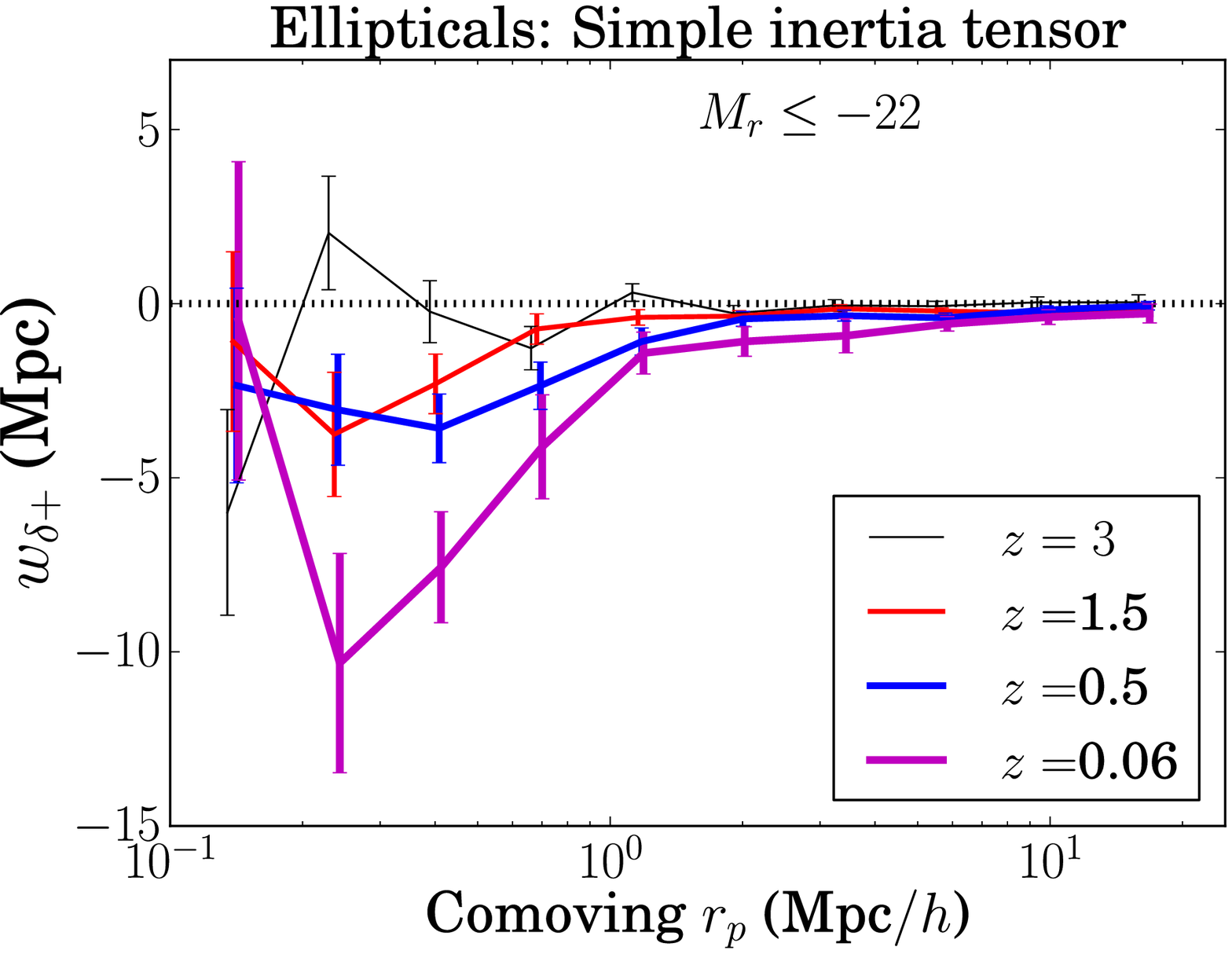}
  \caption{Redshift evolution of the projected correlation function $w_{\delta +}$ for all galaxies (left panel), for ellipticals (right panel) and of $w_{g+}$ for all luminous galaxies (middle panel) in the high luminosity bin. In all cases, the ellipticity is obtained from the simple inertia tensor. Selected redshifts are shown, but all redshifts available are used in the model fits. The error bars shown correspond to the jackknife. Note that the alignments of ellipticals are stronger than when the whole population is considered.}
  \label{fig:wproj}
\end{figure*}

For $w_{\delta +}$ and consistently with the results presented in section~\ref{sec:align3d}, we find a decrease of the amplitude of projected alignments of luminous galaxies with redshift. We show this evolution in Fig.~\ref{fig:wproj}. Galaxies at intermediate and low luminosities display lower amplitude alignment signals and are not shown here. In particular, low luminosity, high redshift discs do not show significant evidence for tangential alignments in projection. We also measure $w_{\delta\times}$ as a test for systematics. We find this to be consistent with null at the $3\sigma$ C.L. for all redshifts and luminosities considered.

We find that $w_{g+}$ correlations among galaxies in the same luminosity range retain a $>2\sigma$ significance for luminous galaxies ($M_r\leq -22$) from $z=0.06$ to $z=1.5$ for the simple inertia tensor (middle panel of Fig.~\ref{fig:wproj}), and only up to $z=0.5$ for the reduced inertia tensor. The $w_{g+}$ measurement is consistent with null for all redshifts at intermediate luminosities. We find an indication of a potential signal at $z=3$ for low luminosities, rejecting the null hypothesis at $97\%$ C.L. for both simple and reduced inertia tensor. $w_{g\times}$ is always consistent with null at the $2\sigma$ level, except for reduced inertia tensor measurement of the lowest luminosity bin at $z=1.5$; although it remains within the $3\sigma$ C.L. in this case. Examining the cross-correlations of galaxy positions and shapes across luminosity bins, we find qualitatively similar results for these correlations as for $\eta_e$ in section~\ref{sec:align3d}. Alignment amplitudes increase with the luminosity of the density and/or shape tracer. The complete set of relevant $w_{g+}$ cross-correlations is presented in Appendix~\ref{app:allcorr}.

With the goal of comparing the alignment amplitude obtained with Horizon-AGN with current observations of LRGs, we measure $w_{\delta +}$ selecting ellipticals alone in the simulation. The evolution of this correlation with redshift for the most luminous ellipticals is shown in the right panel of Fig.~\ref{fig:wproj}. The trends with redshift and luminosity are similar to those measured for the complete galaxy population, but with an increased amplitude. Isolating the elliptical galaxies at each redshift enhances the radial alignment signal. 

\begin{figure*}
  \includegraphics[width=0.4\textwidth]{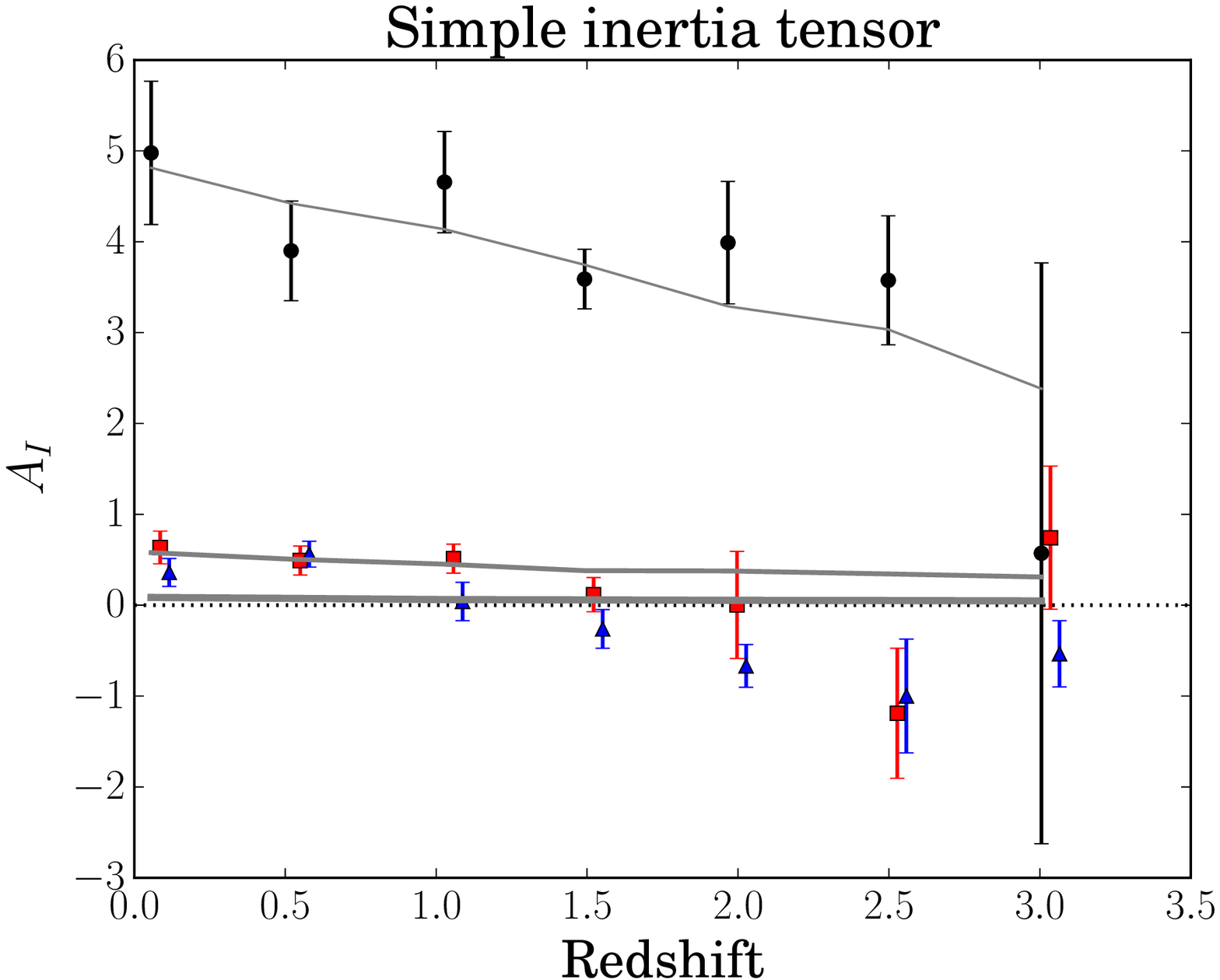}
  \includegraphics[width=0.4\textwidth]{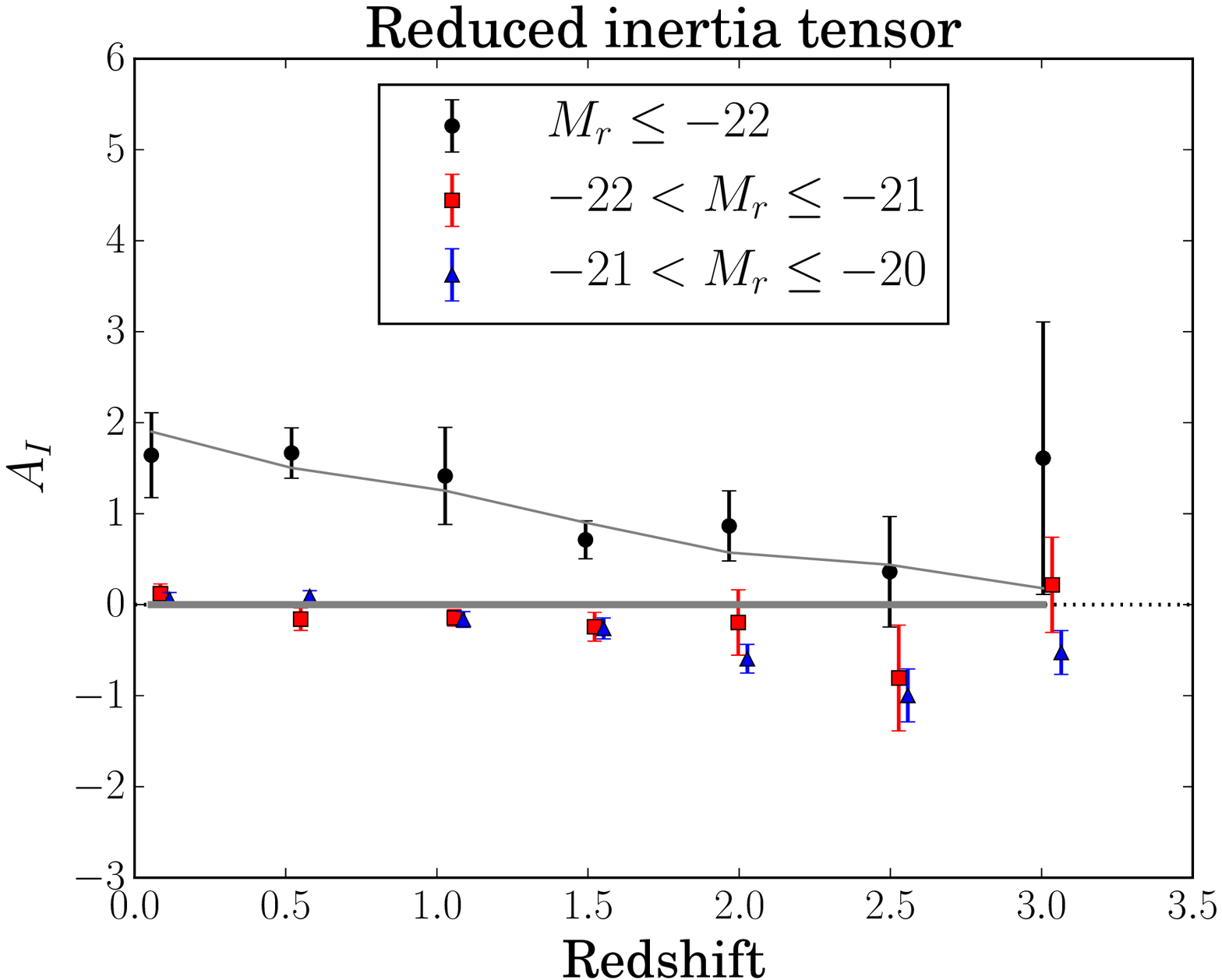}
  \includegraphics[width=0.4\textwidth]{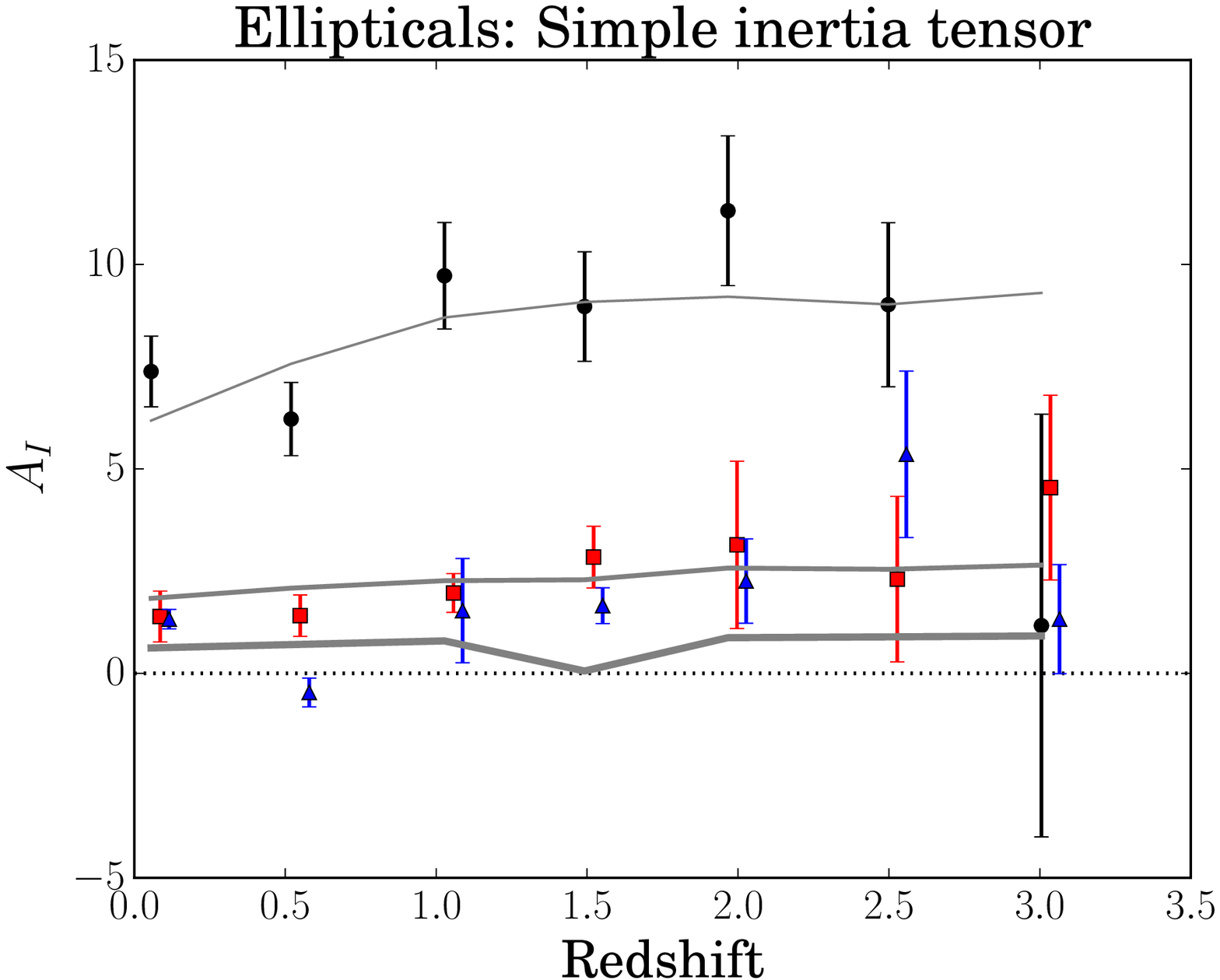}
  \includegraphics[width=0.4\textwidth]{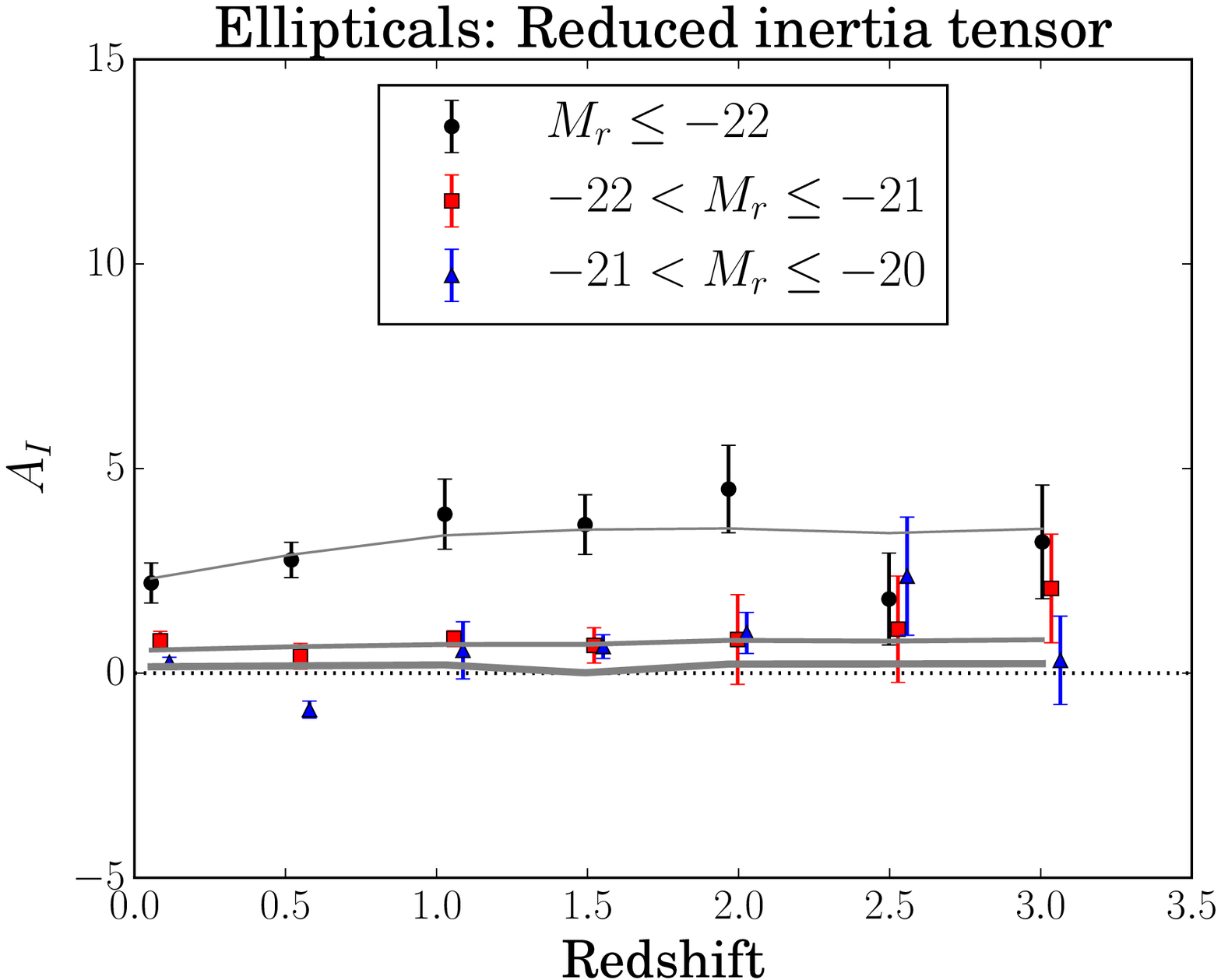}
  \includegraphics[width=0.4\textwidth]{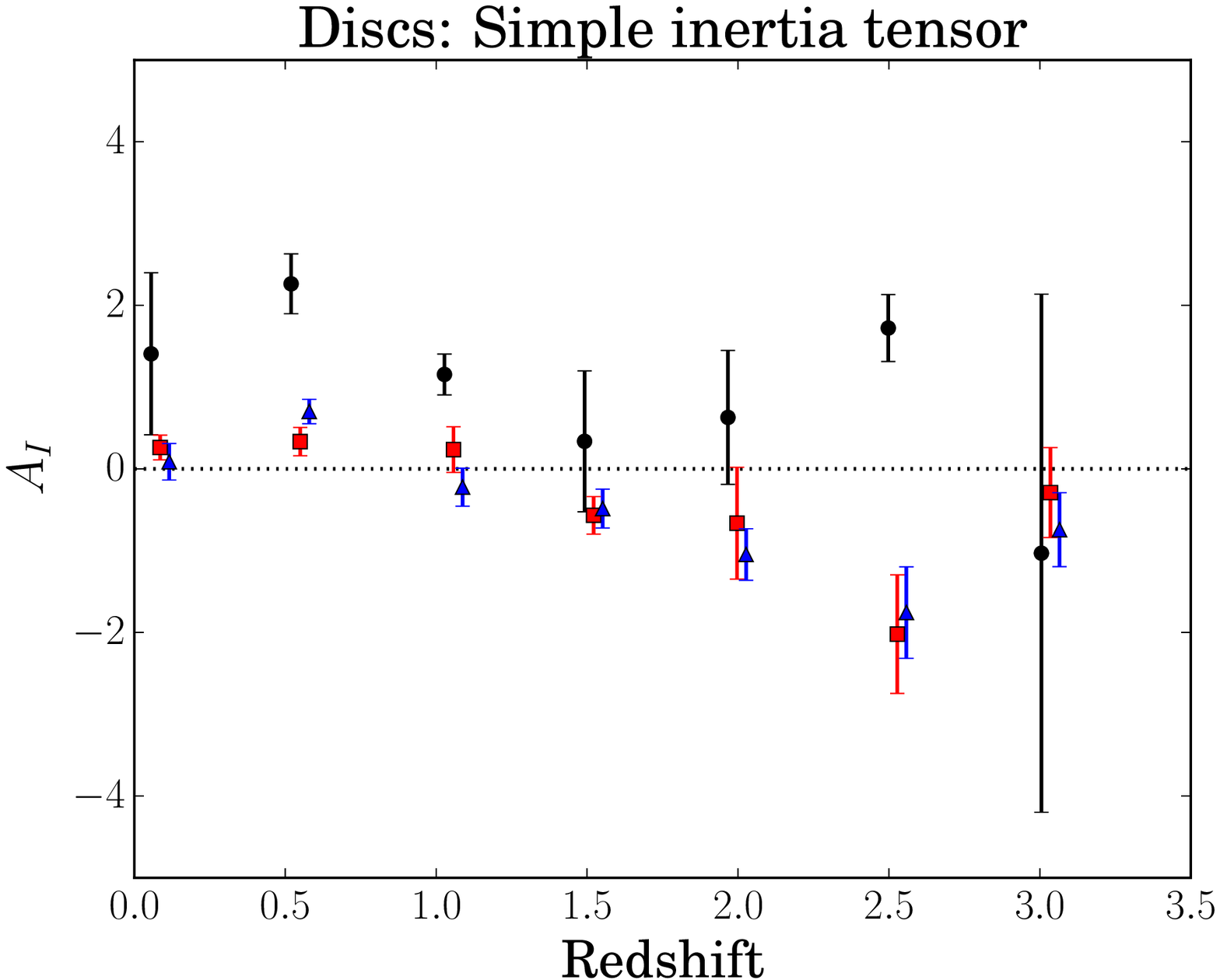}
  \includegraphics[width=0.4\textwidth]{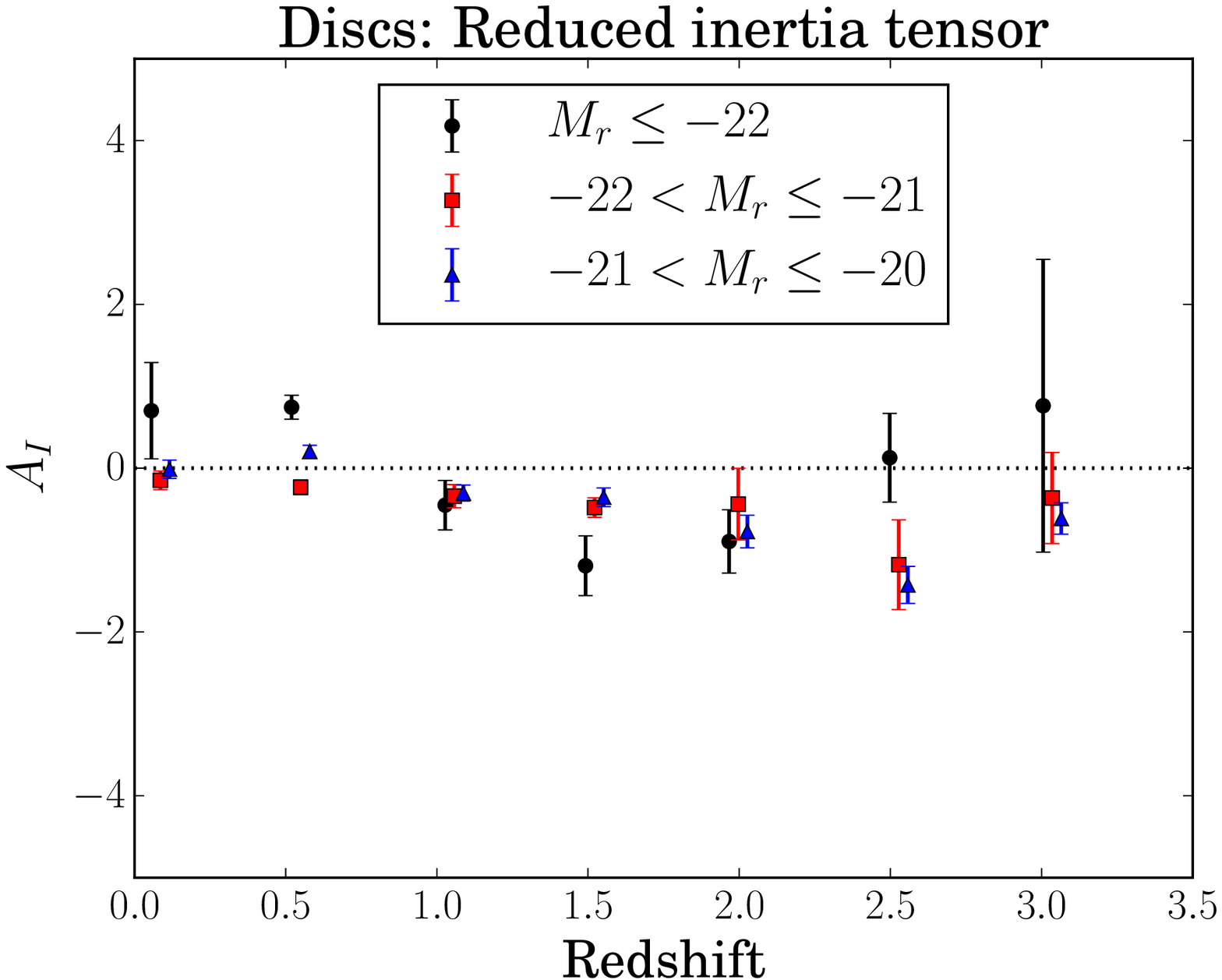}
  \caption{Evolution of $A_I$ as a function of redshift and luminosity. The top left panel corresponds to fits to $w_{\delta +}$ for all galaxies when shapes are measured from the projected simple inertia tensor; the top right panel, with the projected reduced inertia tensor. The middle row corresponds to constraints obtain for elliptical galaxies alone. Black circles represent the highest luminosity bin ($M_r\leq -22$), red squares indicate intermediate luminosities ($-22<M_r\leq -21$) and blue triangles, low luminosities ($-21<M_r\leq -20$).
    Analogously, the bottom row shows results for discs alone.
    All error bars correspond to $68\%$ confidence level constraints on $A_I$. The grey lines connect the points corresponding to the best-fit values. Notice that the functional form being fit, equation~(\ref{eq:ailz}), is a function of both $z$ and $\langle L_r\rangle/L_0$, which is not necessarily a smooth function of redshift.}
  \label{fig:aifit}
\end{figure*}

As discussed in section~\ref{sec:z0bis}, the scale-dependence of the halo model does not reproduce the steepness of $w_{g+}$ nor the downturn at small scales of $w_{\delta+}$ at $z=0.06$. We found an alternative fit to $w_{g+}$ by letting the $p_2$ parameter free. However, this value of $p_2$ is not suitable to describe the scale-dependence of $w_{g+}$ at higher redshifts. To constrain the linear amplitude of alignment, we consider fitting the NLA model only at large scales ($r_p>0.8 \, h^{-1}\, \rm Mpc$) from $w_{\delta+}$. While this method avoids the problem of the missing power in the model at small scales, at the same time, it underestimates the contamination of alignments to lensing due to this missing power. The constraint on $A_I$ as a function of redshift and luminosity are presented in the top row of Fig.~\ref{fig:aifit} for simple (left panel) and reduced (right panel) inertia tensor. 


Previous works have adopted the following parametrization of $A_I$ as a function of redshift and mean luminosity of LRGs \citep{Joachimi11,Tenneti15a},
\begin{equation}
  A_I = A \left(\frac{\langle L_r\rangle}{L_0}\right)^{\alpha_L}(1+z)^{\alpha_z}\,,
  \label{eq:ailz}
\end{equation}
and we obtain the following constraints from our measurements of $w_{\delta +}$ (for discs and ellipticals jointly) for the simple inertia tensor case: $A=1.72\pm0.34$, $\alpha_z=-0.41\pm0.25$ and $\alpha_L=2.17\pm0.28$. The parameters $A$ and $\alpha_z$ are correlated at the $0.72$ level; $A$ and $\alpha_L$, at the $0.37$ level; and redshift and luminosity dependence, at the $0.25$ level. In the case of the reduced inertia tensor, we find alignments to be fully consistent with null ($A_I=0.037\pm0.37$), and completely correlated with $\alpha_L$. Notice that for the simple inertia tensor, the redshift evolution is consistent with the tidal alignment model at $2\sigma$ and we find evidence of significant luminosity evolution.

This parametrization restricts $A_I$ to not change sign. However, we have seen in section~\ref{sec:align3d} that alignments indeed transition from radial to tangential as we go to high redshifts and low luminosities. While this trend is suppressed in projection, for the high luminosity, low redshift galaxies we find both $w_{g+}$ and $w_{\delta+}$ to be different from null with $>99.7\%$ confidence. Hence, the functional form of equation~(\ref{eq:ailz}) is not expected to provide a good fit for the overall population of discs and ellipticals. In the top panels of Fig.~ \ref{fig:aifit}, we see $A_I$ drop below zero at high redshift and low luminosities. 

We also fit equation~(\ref{eq:ailz}) to $w_{\delta +}$ of elliptical galaxies. In this case, we find an increased amplitude compared to the results obtained for the overall galaxy population. The best fit parameters for the simple inertia tensor case are:  $A=3.09\pm0.56$, $\alpha_z=0.31\pm0.25$ and $\alpha_L=1.16\pm0.17$. In this case, $A$ and $\alpha_z$ have a $0.8$ correlation coefficient; $A$ and $\alpha_L$, a $0.23$ correlation coefficient; and the redshift and luminosity evolution parameters are poorly correlated. We obtain a smaller amplitude for the reduced inertia tensor:  $A=1.02\pm0.25$, $\alpha_z=0.32\pm0.31$ and $\alpha_L=1.38\pm0.26$. The luminosity dependence of the signal is less steep for ellipticals than for the mix of ellipticals and discs, potentially due to the contribution of either null or tangential alignments of discs at low luminosities. This luminosity dependence is also mostly driven by the $M_r\leq-22$ population of ellipticals, which show a marked enhancement in their alignment in comparison to lower luminosity galaxies in Figure~\ref{fig:aifit}. We fit the NLA model at large scales to $w_{\delta +}$ for discs alone, despite the overall poor significance of the measurements. The typical amplitudes obtained are shown in the bottom panels of Fig.~\ref{fig:aifit}. The significance here is artificially enhanced by not taking into account off-diagonal elements of the covariance in the fit. In this sense, we are conservative in determining alignment amplitudes.

  In an effort to bring predictions closer to observations, we have found that the addition of dust obscuration and the redshifting of galaxy spectral energy distributions steepens the luminosity dependence of the alignment signal found in this work (see Appendix~\ref{app:dust}). Moreover, including surface brightness cuts \citep{Puchwein13} is likely to yield rounder galaxy shapes, reducing the alignment amplitude. 

\section{Discussion}
\label{sec:discuss}

Galaxy alignments have been detected in observations of the shapes of LRGs to high significance by \citet{Mandelbaum06,Hirata07,Joachimi11,Singh14}. \citet{Hirata07} used the LRG sample in SDSS ($0.15<z<0.35$) and the 2dF-SDSS LRG and QSO (2SLAQ) Survey ($0.4<z<0.8$) to constrain the redshift, scale and luminosity scaling of the cross-spectrum of lensing and alignments. That work found a $> 3\sigma$ detection of alignments on scales up to $60\, h^{-1}\, \rm Mpc$ at low redshift, but the higher redshift measurement was only $2\sigma$ significant. \citet{Joachimi11} extended these measurements to a combination of MegaZ-LRG \citep{Collister07} and SDSS, with a redshift coverage up to $z\sim0.7$ and higher number densities. As a result, they constrained equation~(\ref{eq:ailz}) fitting the NLA model at $r_p> 6\, h^{-1}\, \rm Mpc$. Their best-fit parameters are: $A=5.76^{+0.60}_{-0.62}$, $\alpha_L=1.13^{+0.25}_{-0.20}$ and $\alpha_z$ in agreement with the NLA model. Recently, \citet{Singh14} used the LOWZ sample of the BOSS survey to improve the constraints on alignments at low redshift profitting from a factor of $3$ increase in the number density of LRGs. These authors found that the NLA model can be extended to scales down to $r_p\sim 4\, h^{-1}\, \rm Mpc$, but they also found an excess amplitude of alignments with respect to the NLA model at small scales ($r_p<1.5\, h^{-1}\, \rm Mpc$), which they fit using the halo model \citep{Schneider10} and letting both $a_h$ and $q_{21}$ free. Their best-fit values are: $A_I=5.1\pm0.4$, $a_h=0.014\pm0.004$ and $q_{21}=1.1\pm0.1$ (other $q_{ij}$ parameters are given in their Table 1). Their fits to equation~(\ref{eq:ailz}) result in the following best-fit parameters: $A=4.9\pm 0.6$, $\alpha_L=1.3\pm0.27$ (under the assumption that $\alpha_z=0$). Their results and those of \citet{Joachimi11} are in good agreement.

Using the Horizon-AGN galaxy catalogue between $0.06<z<3$ and for $M_r\leq -20$, we find that the NLA model provides a good fit to $w_{g+}$ and $w_{\delta +}$ at $r_p>0.8\, h^{-1}\, \rm Mpc$, allowing us to model smaller scales than in current observations. Below this scale, we find a significant excess alignment in $w_{g+}$, particularly at low redshift and high luminosities. The halo model with the parameters from Table 1 of \citet{Singh14} is not a good fit for this excess signal. We find it necessary to let the value of $p_2$ free to fit the halo model to the data in this case, steepening the scale-dependence of the resulting $w_{g+}$. We also find that this value of $p_2$ cannot reproduce the scale-dependence of $w_{g+}$ at higher redshifts and lower luminosities. Attributing the change in value of $p_2$ in this bin to a change in $q_{22}$ or $q_{21}$ (see equation~(\ref{eq:halopar})) does not ease the tension at higher redshift. Either more parameters should be varied or their functional forms should be modified. 

We detect an increase of $w_{g+}$ and $w_{\delta +}$ with growing luminosity of the shape tracers. The luminosity dependence of the clustering bias also produces a variation of $w_{g+}$ with luminosity of the density tracer. \citet{Blazek15} developed a standard perturbation theory approach to modelling the intrinsic alignment power spectra. Their modelling indicates that the luminosity dependence of $A_I$ arises from the weighting of the tidal field by the density of the shape tracers. Our results are in qualitative agreement with their predictions. However, in Paper I, we had noted a non-monotonic dependence of radial alignments around dark matter particles with the mass of the shape tracer. Given that mass and luminosity are strongly correlated, we similarly expect that at low redshift, the alignment signal can be non-monotonic with luminosity. In this work, the functional form presented in equation~(\ref{eq:ailz}) provides a sufficiently good description of the measured correlations at large scales. We attribute this to the fact that we are only probing galaxy alignments in the range $M_r\leq-20$. At lower luminosities, a population of dwarf ellipticals emerges at low redshift (see Fig.~\ref{fig:vsigmr}) which could also carry a radial alignment. Their number counts are considerable; at $z=0.06$, ellipticals in the range $-20<M_r\leq-19$ double the number density of ellipticals overall. As we note in Appendix~\ref{app:dust}, higher resolution simulations are needed to probe lower luminosity alignments within the magnitude limit of LSST at low redshift and to determine whether our findings on the luminosity dependence of alignments can be extrapolated to that regime.

We compare our results for the redshift and luminosity evolution of $A_I$ to those obtained by \citet{Joachimi11}. Note that our fits to $w_{\delta+}$ and $A_I(L_r,z)$ were performed on scales $r_p>0.8$ Mpc$/h$ and thus neglect the small scale power in alignments, which is a relevant source of contamination to cosmic shear measurements. Our fits cannot be extrapolated to such small scales, where we further find evidence of deviations from the NLA model. When considering all galaxies with converged shapes regardless of their morphology, we obtain an alignment amplitude that is suppressed with respect to the measurements of \citet{Joachimi11}. This is expected due to the contribution of discs, for which we find only marginal evidence for alignment in projected statistics. We find constraints on late-type galaxy alignments from \citet{Heymans13} at $z<1.3$ consistent with our results for $A_I$ for discs and with the observational constraints for blue galaxies from \citet{WiggleZ} from SDSS and the WiggleZ samples at low redshift. We emphasize that we have not aimed to model the details of their sample selection. If we restrict to ellipticals alone and using the simple inertia tensor, we find a lower value of $A$ by a factor of approximately $2$, but good agreement between our results for $\alpha_L$ and $\alpha_z$ in comparison to the \citet{Joachimi11} and \citet{Singh14} constraints for LRGs\footnote{Note that matching the different choice of normalization of the redshift factor in equation~(\ref{eq:ailz}), increases our alignment amplitude by $\sim 8\%$ for elliptical galaxies; this is a small factor compared to the uncertainty in $A$.}. The results obtained using the reduced inertia tensor give a further lower alignment amplitude by another factor of $\simeq 2$.

  In observational works, observed galaxy shapes are measured up-weighting the pixels closer to the galaxy center proportionally to $r_p^{-2}$. Our results using the reduced inertia tensor would thus be expected to be more comparable to observational measurements of alignments. However, \citet{Tenneti15a} found evidence in the MassiveBlack II hydrodynamical simulation that the reduced inertia tensor tends to produce overly round shapes. Observational results by \citet{Singh15} confirmed that $A_I$ indeed depends on the shape estimator adopted. Shape estimators that put more weight on the outer isophotes of galaxies result in an enhanced alignment amplitude by $\sim 40\%$. Moreover, the details of survey selection on the LRG sample, including colour selection and fiber collisions, for example, could play a role in determining the alignment amplitude. In this work, we have implemented a selection cut on ellipticals based on the dynamical properties of galaxies in the simulation, and this results in a slightly higher comoving number density of objects than is found in \citet{Singh14}. Nevertheless, the fact that the overall  redshift and luminosity evolution of the alignment signal is in good agreement with observations is encouraging.  

Other authors have explored the alignment signal in smoothed-particle-hydrodynamics and moving-mesh simulations \citep{Tenneti15a,Velliscig15,Velliscig15b,Tenneti15b}. \citet{Tenneti15a} studied the redshift evolution of the $\eta_e$ and $w_{g+}$ statistics for galaxies of different stellar masses in the Massive-Black II simulation in the redshift range $0.06<z<1$. There are several discrepancies between our work and theirs. They found that the $\eta_e$ correlation decreased towards lower redshift; and they did not detect any significant evolution of $w_{g+}$ with redshift, possibly as a consequence of evolution of the overall shape distribution of the galaxies with redshift. Contrary to their results, and assuming that stellar mass and luminosity are strongly correlated, we find {\it increased} alignments from $\eta_e$, $w_{\delta +}$ and $w_{g+}$ with decreasing redshift, and this trend is more evident at small scales. \citet{Tenneti15a} also found a shallower scale dependence of $w_{g+}$ compared to $w_{\delta +}$, while we find the inverse trend.

As mentioned above, a direct comparison between observations and simulations is challenging due to the details of sample selection. At first glance, it seems clear that the amplitude of alignment varies for different simulations. \citet{Tenneti15a} fit equation~(\ref{eq:ailz}) without placing colour or morphology cuts for their sample of galaxies. Their measurements, which rely on the iterative reduced inertia tensor and probe magnitudes of $M_r<-18$, yield $A=6.7\pm 1.7$, $\alpha_L=0.47\pm0.08$ and $\alpha_z=0.5\pm0.5$. In comparison, without dynamical selection, we find $A$ fully consistent with zero if we use the reduced inertia tensor. For the simple inertia tensor, on the other hand, we find a {\it stronger} luminosity dependence of $A_I$ and an amplitude that is a factor of $\simeq 4$ smaller than in Massive Black II. \citet{Singh14} compared the results from \citet{Tenneti15a} to the best-fit model obtained from LOWZ alignment measurements and found that MassiveBlack II overpredicts the alignment strength by a factor of $\sim 2$. Results from the cosmo-OWLS simulation are qualitatively similar. \citet{Velliscig15b} showed that shape-position alignment signal from cosmo-OWLS overpredicts the LOWZ alignments when all stars in a galaxy are used to trace its shape; while the agreement is better if only stars within the half-mass radius are used. Alignment measurements at $z=0.06$ from the Illustris moving-mesh simulation are lower in amplitude than for MassiveBlack II by a factor $1.5-2$ at $1\, h^{-1}\, \rm Mpc$ \citep{Tenneti15b}, and could potentially be in better agreement with Horizon-AGN, but their luminosity dependence remains to be probed.

In Paper I, we found evidence of {\it tangential} alignments of disc galaxies, i.e., with their plane of rotation oriented tangentially around DM overdensities. Tidal torque theory predicts the existence of a transition mass below which halos orient their spins parallel to the filament in which they are embedded \citep{Codis15b}. Above the transition mass, their orientation becomes perpendicular to the filament. Such trends have been confirmed for both the orientation of DM halos \citep[][and references therein]{Codis12} and of galaxies \citep{Dubois14,Codis15}. The transition mass has further been related to the merger history and the dynamical properties of galaxies. Galaxies growing through mergers become redder and pressure-supported as they overcome the transition mass \citep{Dubois14}. We attributed the tangential alignment signal of Paper I to disc galaxies in filaments which lie below the transition mass. The alignment of galaxies above the transition mass is better evidenced through their shapes, as the increased number of mergers tends to decorrelate shape from spin. In this case, they tend to be elongated pointing towards overdensities, partly due to the \citet{Binggeli82} effect, as discussed in Appendix~\ref{app:allcorr}. \citet{Welker15} found that, in addition to the large scale alignment with the filamentary distribution of matter in the cosmic web, there is a small scale component to the alignments arising from the settling of satellites in the galactic plane of the central galaxy. This is particularly relevant for galaxies above the transition mass, whose galactic plane is also aligned with the filament. Similar conclusions were reached by \citet{Velliscig15b} in their analysis of intrinsic alignments in the EAGLE and cosmo-OWLS hydrodynamical simulations.

In this work, we found evidence in section~\ref{sec:align3d} of a transition from radial alignments at low redshifts and high luminosities to tangential alignments at high redshifts and low luminosities. This is consistent with the findings of \citet{Dubois14}, who found that spin alignments decrease with cosmic time due to mergers and the quenching of cold flows and star formation. On the other hand, \citet{Tenneti15b} do not find any evidence for {\it tangential} disc alignments either in MassiveBlack II or in Illustris on scales $r_p>0.1\, h^{-1}\, \rm Mpc$ at $z=0.6$, in contrast with our Paper I results. Hence, while the agreement between high luminosity alignments in Horizon-AGN and Illustris might be better than with MassiveBlack II, we do not expect this to hold as a function of morphology. \citet{Velliscig15b} find qualitatively similar results to \citet{Tenneti15b}: a reduced amplitude of alignment for galaxies selected to be more spherical (i.e., discs), but no evidence for a transition to tangential alignments for this population. This resuts are based on the EAGLE and cosmo-OWLS simulations, which apply similar numerical methods to MassiveBlack II. Our finding in section~\ref{sec:align3d} of significant tangential alignments at high redshift and low luminosities suggests that a more direct comparison between simulations might be possible in this regime. On the other hand, observational results on the amplitude of alignment of blue galaxies suggest this is consistent with null at $z<1.3$ \citep{WiggleZ,Heymans13}. Our measurements indeed suggest that the tangential alignment signal of disc galaxies is only marginally present in projection, even at $z=3$, where the $\eta_e(r)$ statistic is most significant. However, within current constraints, contamination from disc galaxies to weak lensing observables could still be significant \citep{Chisari15b}.

\section{Conclusions}
\label{sec:conclusion}

We have studied the redshift and luminosity evolution of alignments of galaxies, as traced by their stellar particles, in the Horizon-AGN simulation. The main result of this paper is the identification of a transition from radial, elliptical-dominated, alignments at low redshifts and high luminosities to tangential, disc-dominated, alignments at high redshift and low luminosities. The evolution of the mixture of populations in the luminosity bins considered is a consequence of these trends and of the slow evolution of the fraction of discs and ellipticals in each luminosity bin.
We also reached the following conclusions:
\begin{itemize}
\item The alignment signal is strengthened at low redshifts and high luminosities. At $z=0.06$, we have determined that $w_{g+}$ carries a much stronger alignment signal than $w_{\delta+}$; which is related to the anisotropic distribution of satellites in the one-halo regime. Alignments decrease with luminosity and can remain significant at small scales even at low luminosities. The fiducial halo model \citep{Schneider10} is not a good fit to the alignment with DM particles or galaxies at this redshift; but the measurements can be modelled by relaxing one of the parameters governing the scale-dependence of the model, i.e., with a steeper decrease of intrinsic shear with projected radius inside the one-halo regime. 
\item The detected transition between tangential disc alignments at low luminosity and high redshift to radial alignments of ellipticals at high luminosity and low redshift is accompanied by a decorrelation between minor axes and spin orientations. We attribute this to the result that the alignment signal of ellipticals is better evidenced through the simple inertia tensor, but not through the spin or reduced inertia tensor. On the contrary, at high redshift and low luminosity, the galaxy population is mostly dominated by discs and the orientation of spins and minor axes, regardless of the shape estimator, coincide.
\item The fraction of discs at low luminosities remains mostly constant throughout the redshift range probed, while the alignment amplitude is clearly evolving. We therefore conclude that the intrinsic evolution in the fraction of discs and ellipticals cannot be fully responsible for the evolution of the alignment signal. We have confirmed that the amplitude of alignment itself for each of these populations evolves with redshift.
\item Projecting along the line of sight and weighting by ellipticity tend to dilute the alignment signal measured from orientations. The NLA model provides a good template for fitting the projected DM density-ellipticity correlations at large scales, $r_p>0.8\, h^{-1}\, \rm Mpc$. At smaller separations, we find evidence of significant excess power that cannot be modelled with the fiducial parameters of the halo model; nor with the best-fit parameters obtained at $z=0.06$. We thus have constrained the luminosity and redshift dependence of the alignment strength, $A_I(L_r,z)$, from the large scale measurement of $w_{\delta +}$ alone. We find $A_I$ for elliptical galaxies in Horizon-AGN to be a factor of $\simeq 2$ smaller than observational results for LRGs when measured from the simple inertia tensor. Note that this is not representative of small scale alignments, and that our $w_{g+}$ measurements indicate a strong contribution of the one-halo term, particularly at high luminosities. The alignment amplitude of discs from the projected measurements is in agreement with current observational constraints. 
\item We have identified several components to the alignment signal: 1) very luminous galaxies are radially oriented towards other luminous galaxies; 2) low luminosity galaxies cluster preferentially in the direction of the semimajor axis of the central \citep{Binggeli82,Mandelbaum06,Welker15}; 3) low luminosity galaxies are also elongated towards high luminosity centrals.
\item Finally, we conclude that the comparison between different simulation techniques and baryonic physics prescriptions is likely to be more effective at high redshift, where we find alignments to be dominated by discs in Horizon-AGN. In particular, tangential alignments in Horizon-AGN, although predicted by tidal torque theory, have not been found in other hydrodynamical cosmological simulations. We expect that a more detailed comparison between them will be the topic of future work.
\end{itemize}

\section*{Acknowledgements}
This work has made use  of the HPC resources of CINES (Jade and Occigen supercomputer) under the time allocations 2013047012, 2014047012 and 2015047012 made by GENCI. This work is partially supported by the Spin(e) grants {ANR-13-BS05-0005} (\url{http://cosmicorigin.org}) of the French {\sl Agence Nationale de la Recherche} and by the ILP LABEX (under reference ANR-10-LABX-63 and ANR-11-IDEX-0004-02). Part of the analysis of the simulation was performed on the DiRAC facility jointly funded by STFC, BIS and the University of Oxford. NEC acknowledges support from a Beecroft fellowship. LM is supported by STFC grant ST/N000919/1. We thank  S. Rouberol for running  smoothly the {\tt Horizon} cluster for us. 

\bibliographystyle{mn2e}
\bibliography{author}

\appendix

\section{Impact of dust, metallicity and $K$-corrections on galaxy colours and luminosities}
\label{app:dust}

\begin{figure}
  \includegraphics[width=0.5\textwidth]{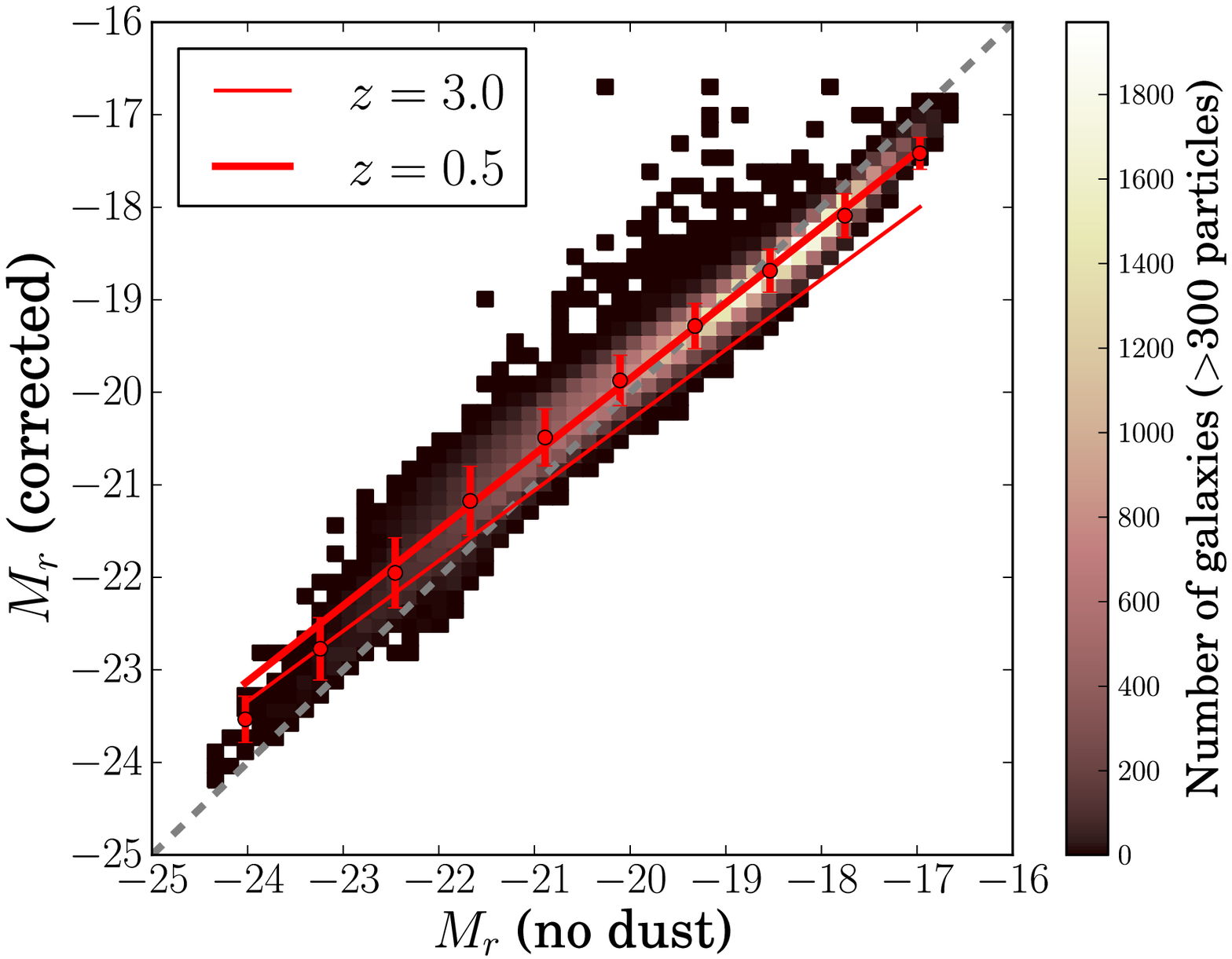}
  \caption{Comparison between dust-free galaxy absolute magnitudes in the $r$-band and the dust-corrected, metallicity-boosted, magnitudes at $z=0.5$ in Horizon-AGN. The red points indicate the mean and $1\sigma$ dispersion in bins of dust-free $M_r$. The thick red line indicates the best linear fit to the underlying galaxy population at $z=0.5$; and the thin red line, at $z=3$. The dashed grey line is the identity, plotted for reference. With the dust and metallicity corrections, galaxies tend to be dimmer at the bright end, where the dust-correction is dominant. At the dim end, galaxies can became brighter on average than the dust-free $M_r$ due to the impact of the metallicity correction.}
  \label{fig:mrmr519}
\end{figure}

\begin{figure*}
  \includegraphics[width=0.32\textwidth]{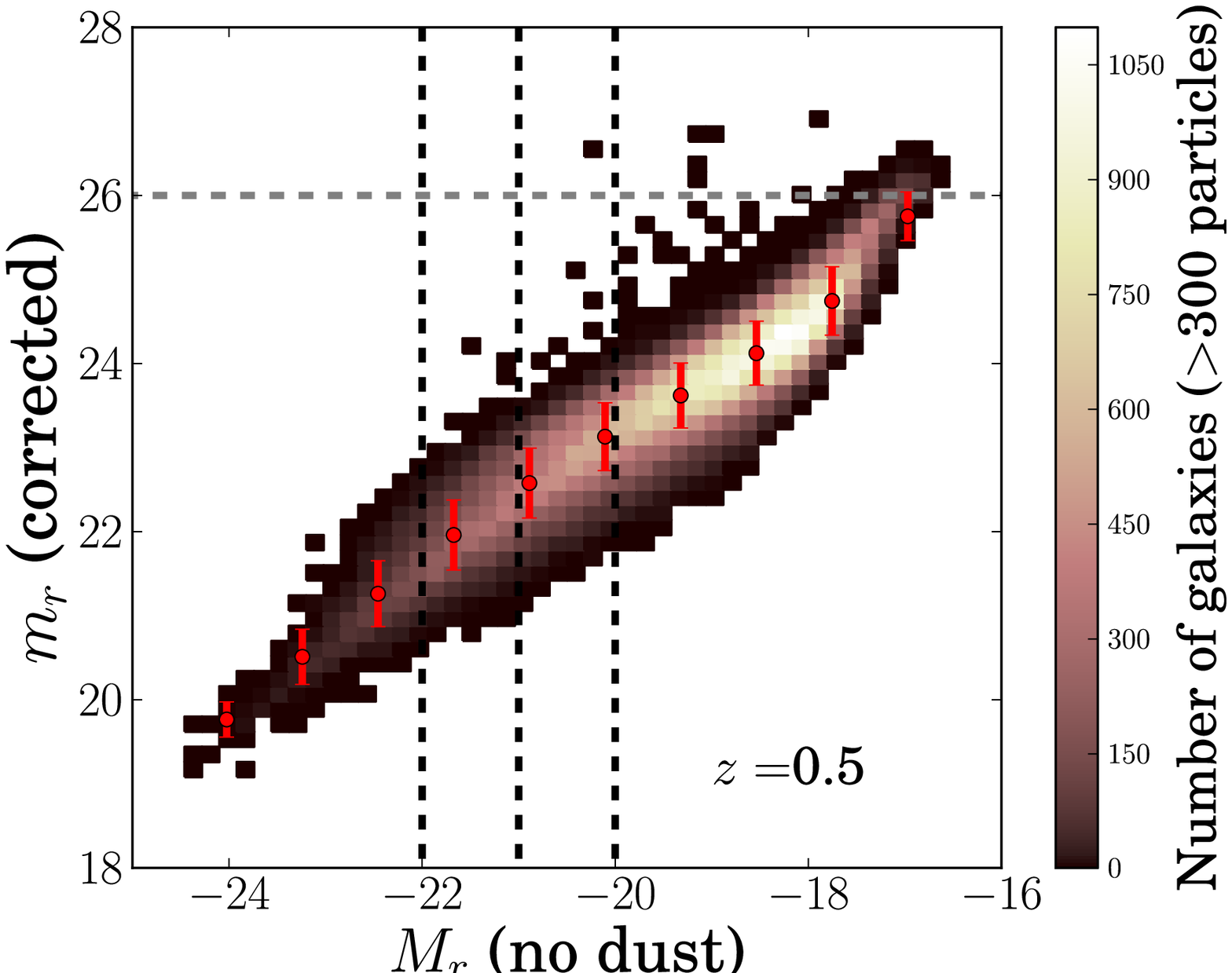}
  \includegraphics[width=0.32\textwidth]{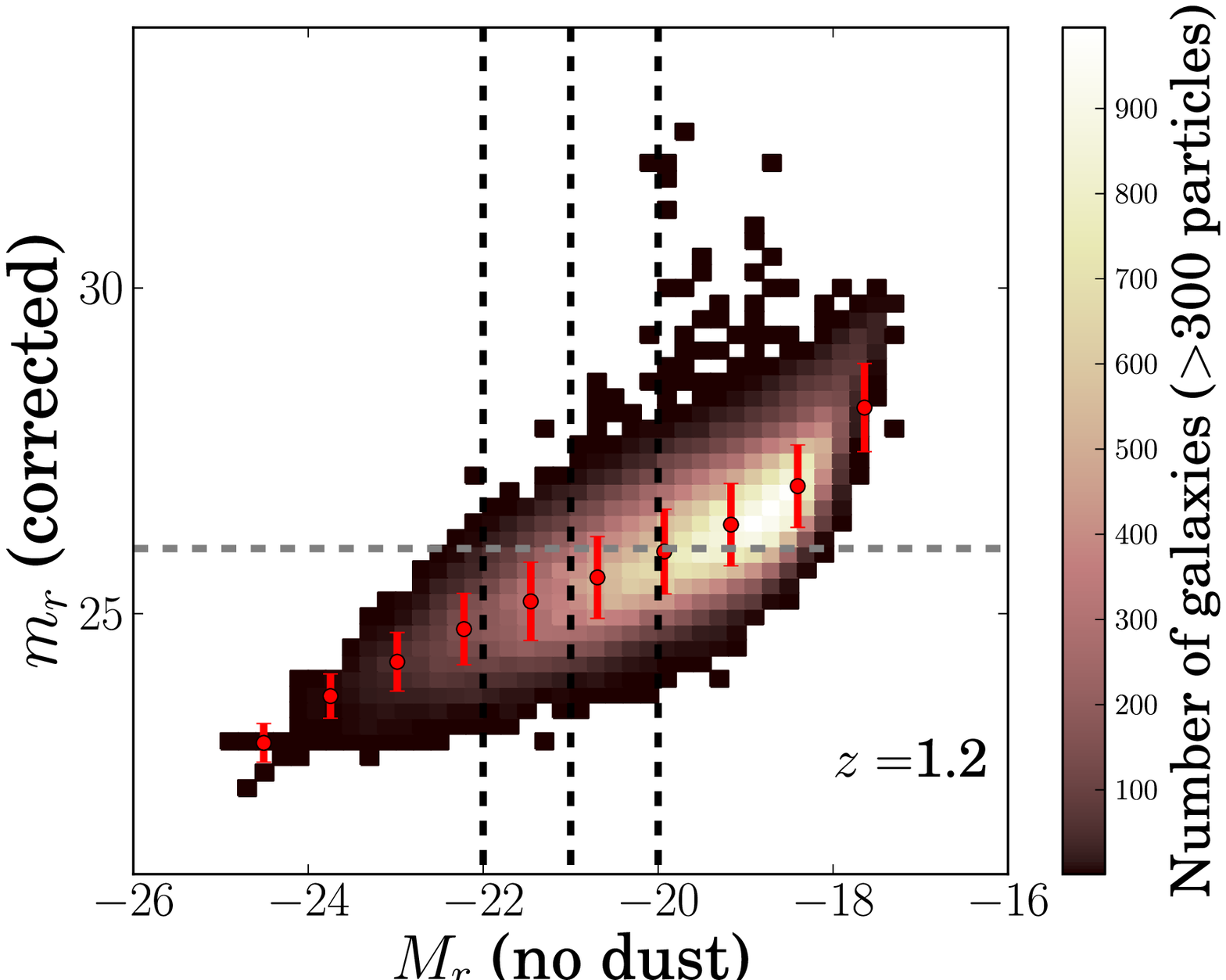}
  \includegraphics[width=0.32\textwidth]{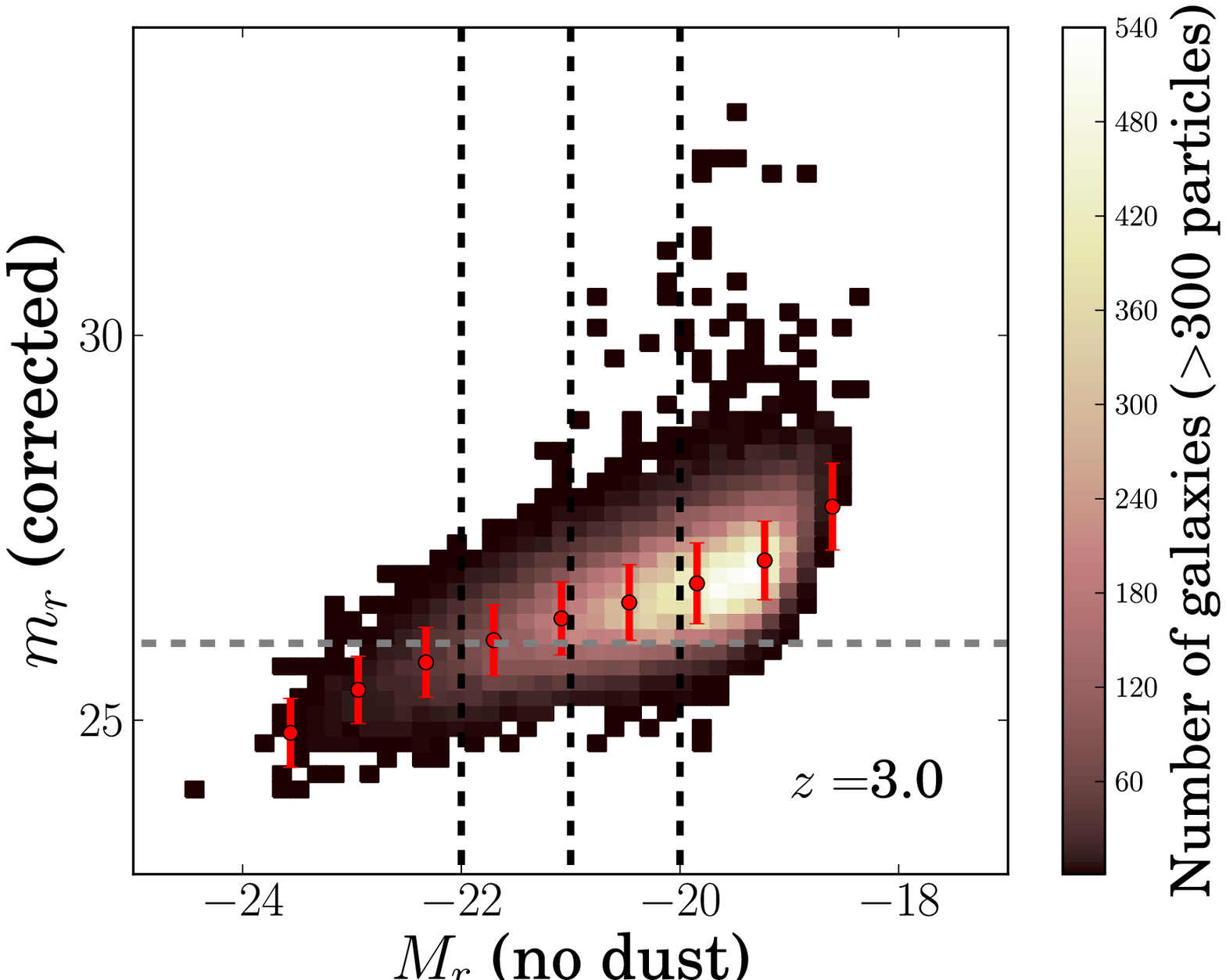}
  \caption{Dust-corrected and metallicity boosted apparent magnitudes of galaxies as a function of the dust-free absolute magnitudes in $r$-band for redshifts $z=0.5$ (left), $z=1.2$ (middle) and $z=3$ (right). The grey dashed line indicates the expected magnitude limit for LSST and the points indicate the median relation and one standard deviation. Vertical dashed black lines indicate the thresholds of the luminosity bins adopted for measuring intrinsic alignment correlations from $z=0$ to $z=3$: $M_r\leq -22$, $-22<M_r\leq -21$ and $-21<M_r\leq -20$.}
  \label{fig:apprdust}
\end{figure*}

The results presented in the main body of this manuscript rely on rest-frame dust-free absolute magnitudes of the simulated galaxies. In this appendix, we explore the impact of dust extinction and stellar metallicity on galaxy absolute magnitudes, and we include $K$-corrections in the modelling of galaxy apparent magnitudes. The overall comparison of galaxy properties in Horizon-AGN (i.e., galaxy colours, the luminosity function and stellar mass function) as a function of redshift is addressed in Kaviraj et al. (in preparation). We expect the corrected colours to have various effects on our fiducial measurements of alignments. First, they can result in a different evolution of $\langle L_r\rangle$ as a function of redshift and can affect the luminosity dependence of $A_I$. Second, dust extinction and $K$-corrections can have an impact on apparent magnitudes, pushing more galaxies over the magnitude limit of a given survey.

We compute the galaxy apparent and absolute magnitudes using the single stellar population (SSP) models from \cite{Bruzual03} and a Chabrier initial mass function. The spectral energy distribution (SED) is sampled at $1221$ wavelengths between $91$ and $16\times10^{5}$ \angstrom\,  for $6$ values of metallicity (from $10^{-4}$ to $0.05$) and $221$ values of age from 0 to $2\times10^{10}$ years. We assume that each star particle is an individual SSP and we compute its contribution to the total SED by logarithmically interpolating in metallicity and age. Due to an intrinsic suppression in metallicity in Horizon-AGN compared to observations, both gas and stars metallicity have been boosted by a factor empirically derived by fitting by the mass-metallicity relation with observations \citep{Maiolino08,Kewley08}. Hydrodynamical cosmological simulations usually need such a metallicity enhancement because the limited mass resolution effectively results in metals not being created fast enough. This boost factor, $\beta_{\rm metals}$, is redshift dependent and approximated by second-degree polynomial: $\beta_{\rm metals} = 4.08 - 0.21z- 0.11z^{2}$. 

To take into account dust attenuation, we extract the gas density and metallicity in a cube around each galaxy. The dust mass is assumed to scale with the mass of metals in the gas, with a dust-to-metal ratio of 0.4 \citep{Dwek98,Jonsson06,Smith15}. This allows us to compute the column density of dust, which we then connect to the optical depth along the line of sight (one axis of the box) to each star in the galaxy. We use the $R=3.1$ Milky Way dust grain model by \cite{Weingartner01}. We assume that the gas is transparent beyond one virial radius of the galaxy. Inclination-dependent dust extinction is naturally included, since we model the dust distribution from the gas, and the attenuation of each star particle is individually computed. The total dust-attenuated SED is the sum of the contribution of all star particles and passed through the relevant SDSS filters. To compute apparent magnitudes, we directly shift the galaxy spectrum before passing it through the filter. Notice that this means that we are thus taking into account the effect of $K+e$-corrections in addition to the dust attenuation and metallicity renormalisation. In comparison, alignment measurements from observations typically correct for the former.

Fig.~\ref{fig:mrmr519} shows an example of the impact of the improved luminosity modelling on absolute magnitudes in the $r$-band. The solid grey lines are the best linear relation fits to the population of galaxies at $z=0.5$ and $z=3$. On average, dust- and metallicity-corrected absolute magnitudes tend to be dimmer at the bright end of dust-free $M_r$, and brighter at the dim end. At the bright end, the dust extinction dominates, while the metallicity correction tends to enhance luminosities at the dim end. For the range of luminosities used in the alignment measurements presented in this work, we are mostly interested in the impact of dust extinction at the bright end, where $M_r\leq-20$. Overall, the reduction of the parameter space of $r$-band luminosities implies a higher value of alignment amplitude and a steeper dependence with luminosity. For example, the impact of these corrections on the fitting parameters in equation~(\ref{eq:ailz}) is a slight increase in alignment amplitude ($\simeq 70\%$) and a steeper luminosity dependence (an increase in $\alpha_L$ of $20\%$) at $z=0.5$. The effect is less significant towards higher redshift. At $z=3$, $A$ increases by $\simeq10\%$ and $\alpha_L$, by $\simeq30\%$. 

Consider now the `gold' sample of galaxies with weak lensing shapes envisaged to be used in the LSST survey \citep{LSST} as an example. This sample will be approximately magnitude limited to $m_r\sim26$. The apparent magnitudes in $r$-band at three different redshifts are shown in Fig.~\ref{fig:apprdust}. We have indicated the expected apparent magnitude limit of LSST with a horizontal dashed grey line. At $z=0.5$ (left panel), all galaxies in the three luminosity bins considered lie within the magnitude threshold of LSST. As the redshift increases, a larger fraction of dimmer galaxies exceed the magnitude limit. As a consequence, the contribution to the intrinsic alignment signal from high luminosity galaxies is enhanced with respect to the lower luminosity bins. In particular, the tentative tangential alignment signal detected for the $-21<M_r\leq-20$ bin at $z=3$ is beyond LSST reach.

Notice that galaxies with $M_r>-20$ lie within the magnitude limit of LSST at $z=0.5$. These galaxies were not included in the intrinsic alignment measurement presented in the main body of this manuscript because the adopted cut in the number of stellar particles implies that the sample is incomplete above this absolute magnitude. As a consequence, alignment measurements from hydrodynamical simulations require extrapolation to low luminosities at low redshifts to quantify alignment contamination to LSST, for example. Higher resolution is needed to avoid this extrapolation. Admittedly, our requirement that $M_r\leq-20$ to guarantee completeness could perhaps be relaxed if we restrict to lower redshifts in the alignment measurement. However, this would be insufficient since we could barely extend the alignment measurement to $M_r<-19$ if we restricted to $z<0.8$, well below the median redshift of LSST sources.

\section{Alignments as a function of redshift, luminosity and galaxy dynamics}
\label{app:allcorr}

\begin{figure*}
  \includegraphics[width=0.32\textwidth]{figs_paper2/dmcross3d_761_magr22_lge1_jack.eps}
  \includegraphics[width=0.32\textwidth]{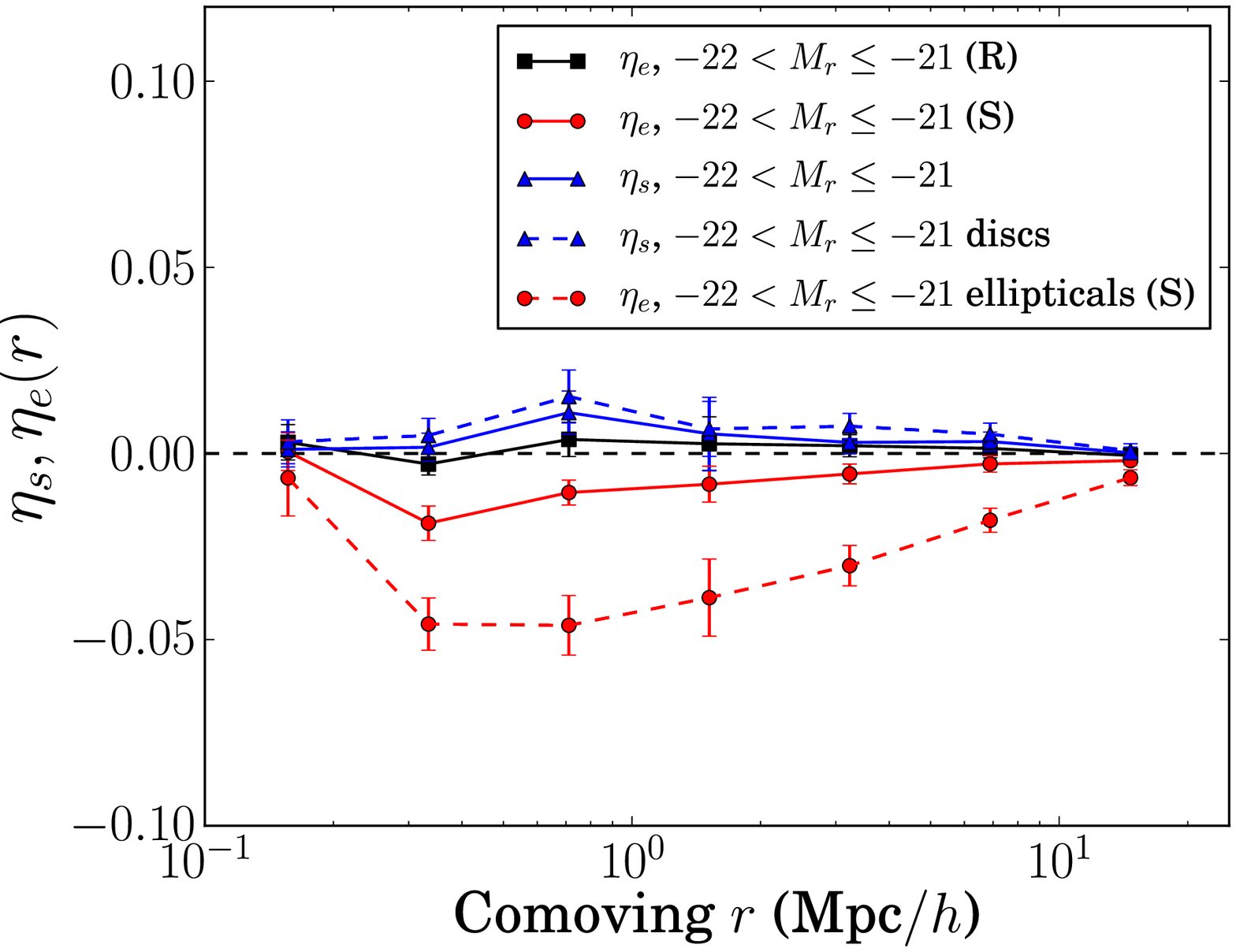}
  \includegraphics[width=0.32\textwidth]{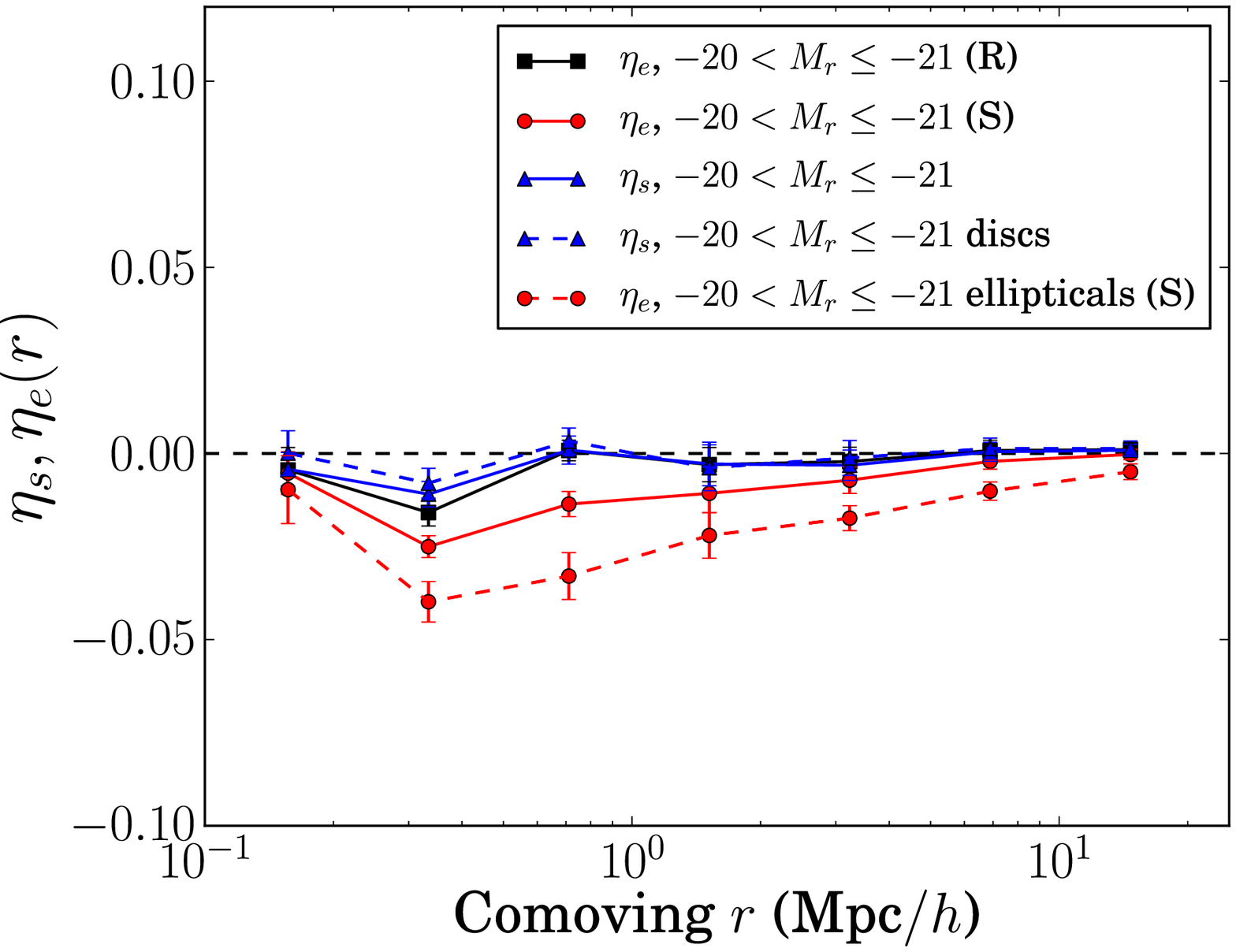}
  \includegraphics[width=0.32\textwidth]{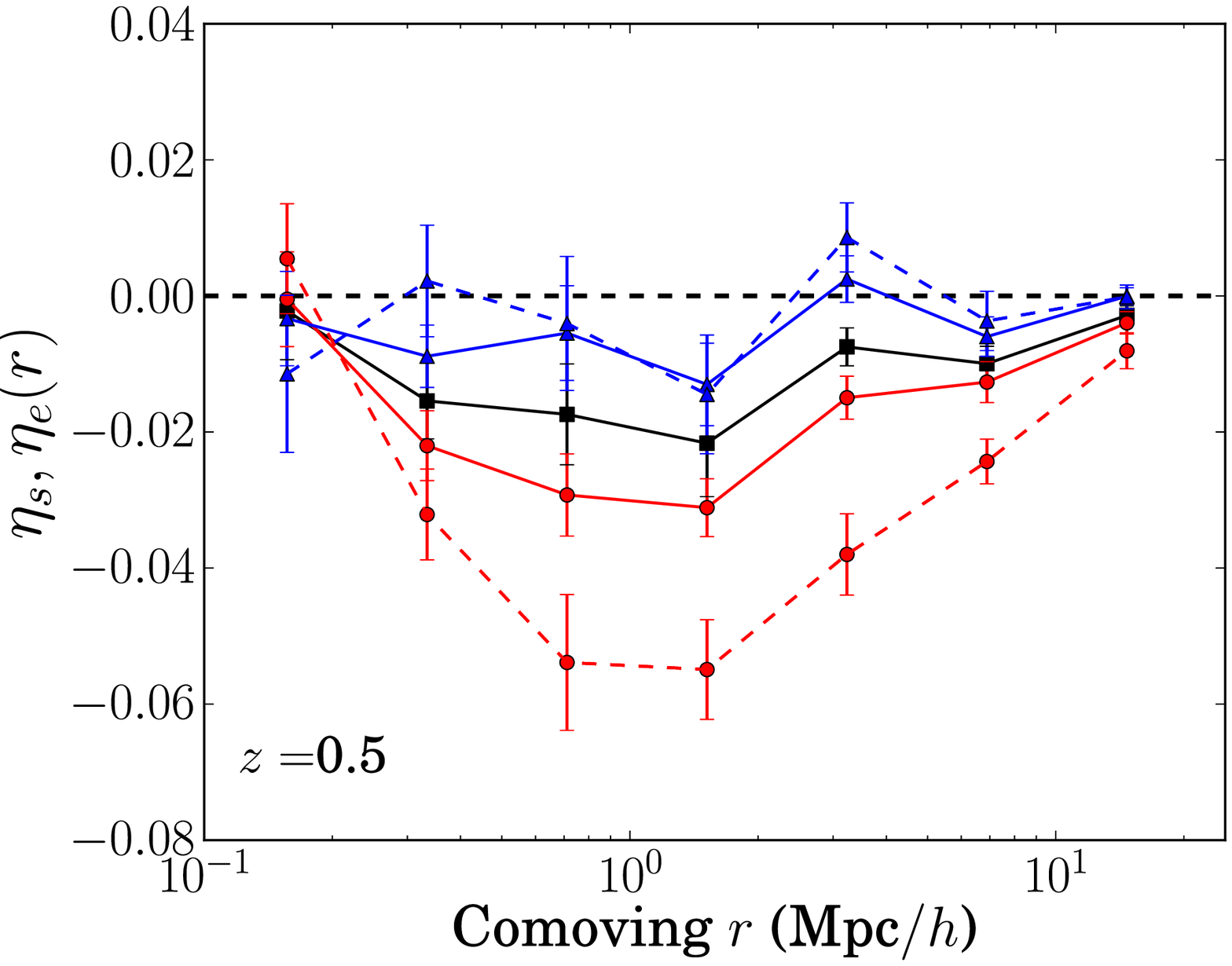}
  \includegraphics[width=0.32\textwidth]{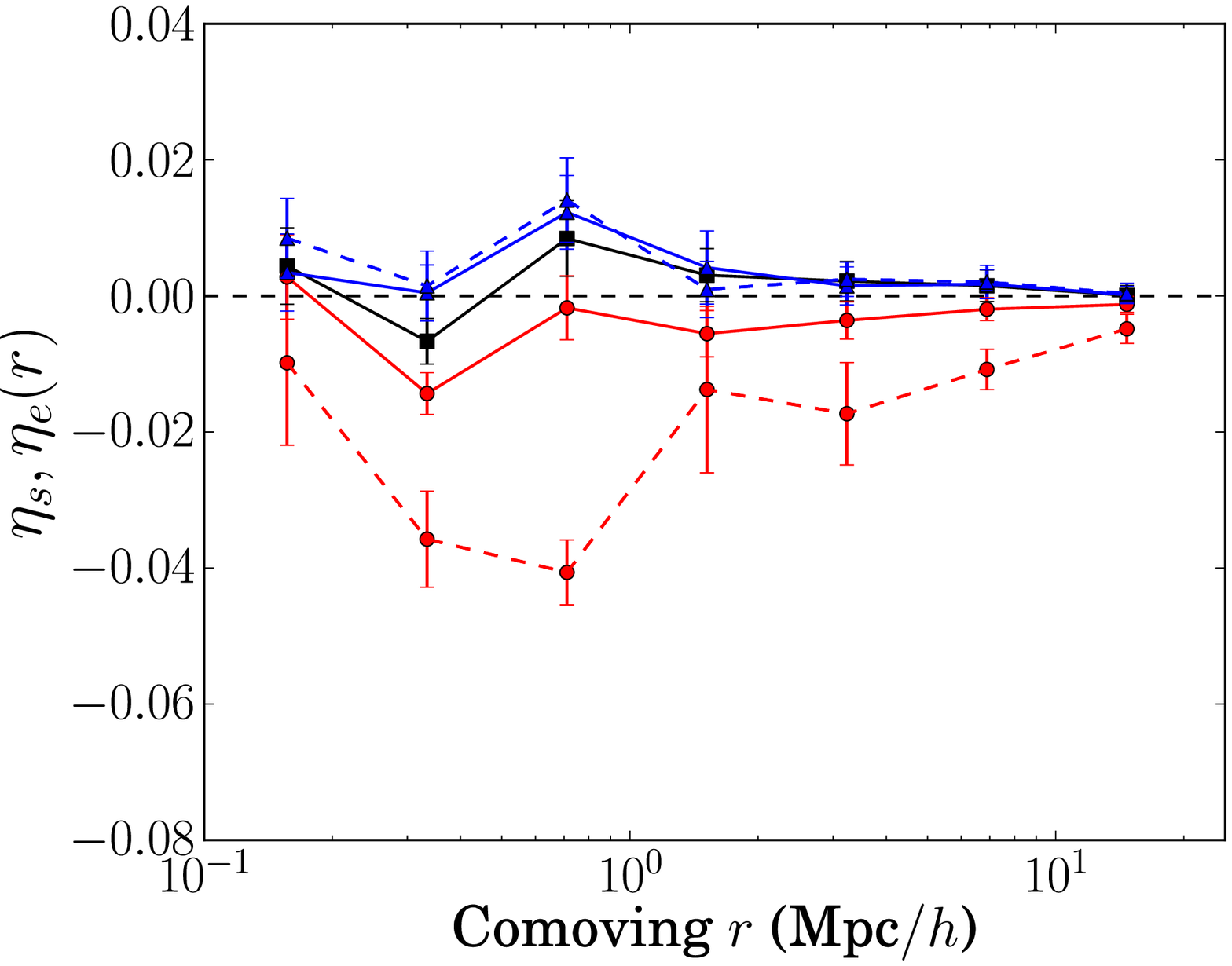}
  \includegraphics[width=0.32\textwidth]{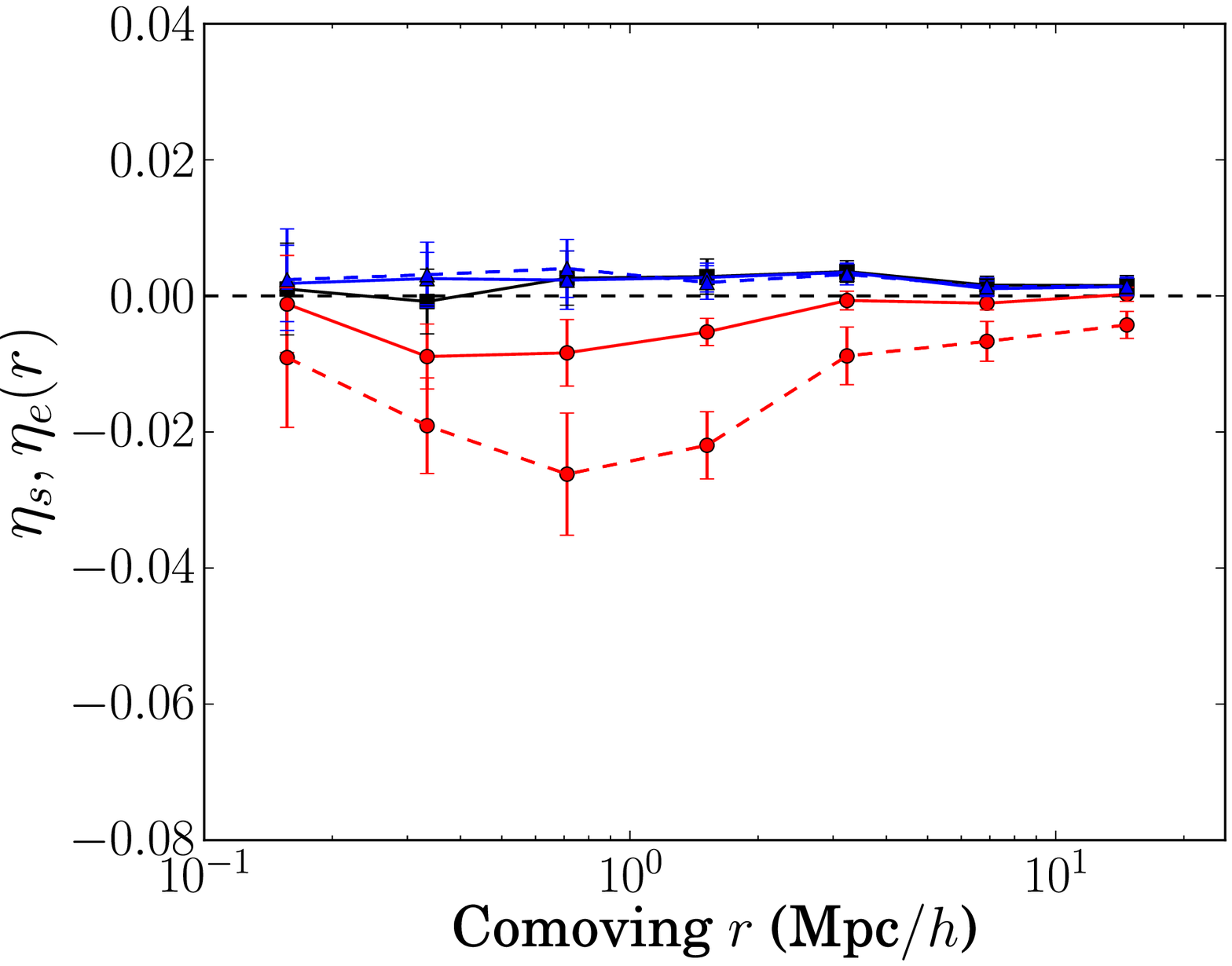}
  \includegraphics[width=0.32\textwidth]{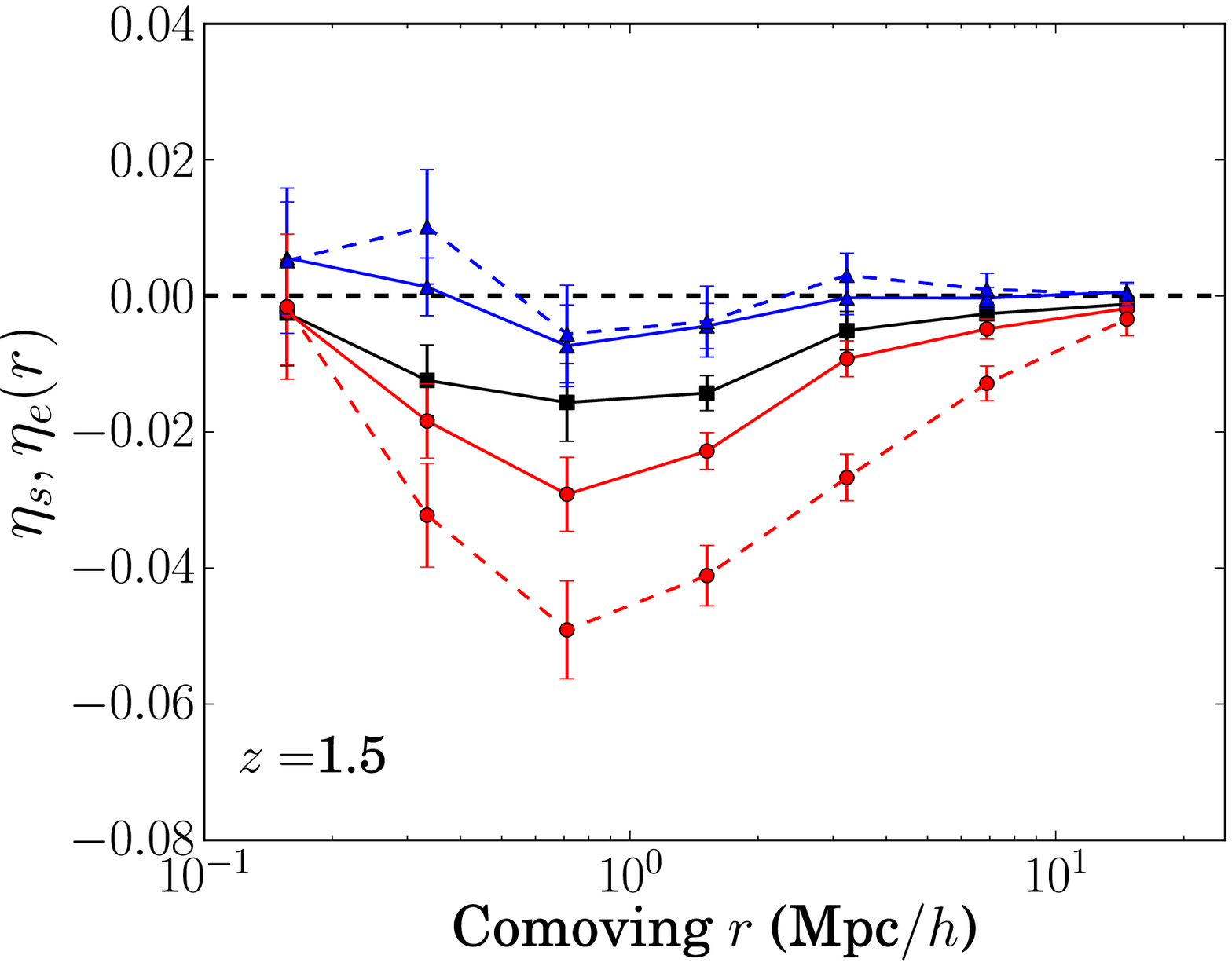}
  \includegraphics[width=0.32\textwidth]{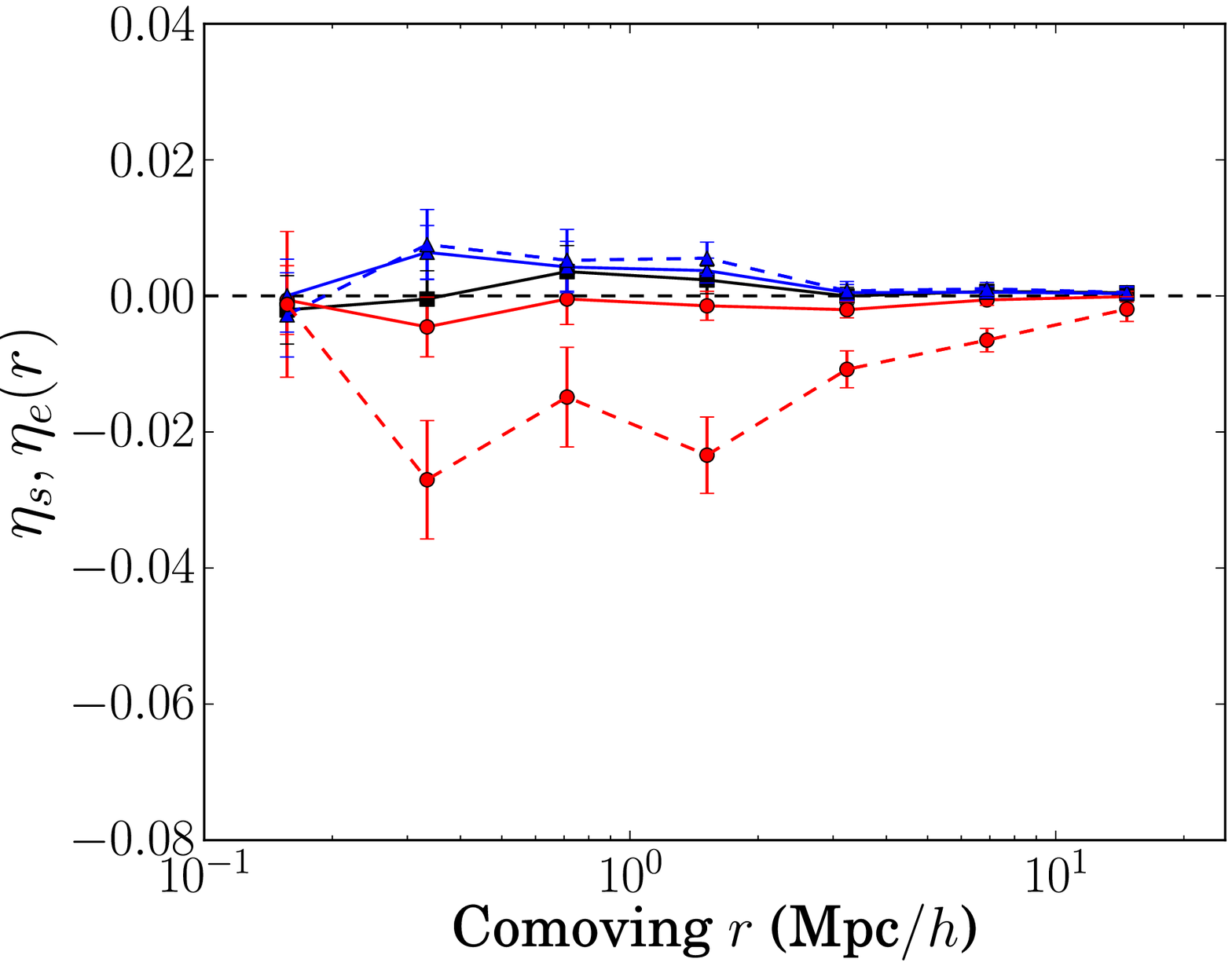}
  \includegraphics[width=0.32\textwidth]{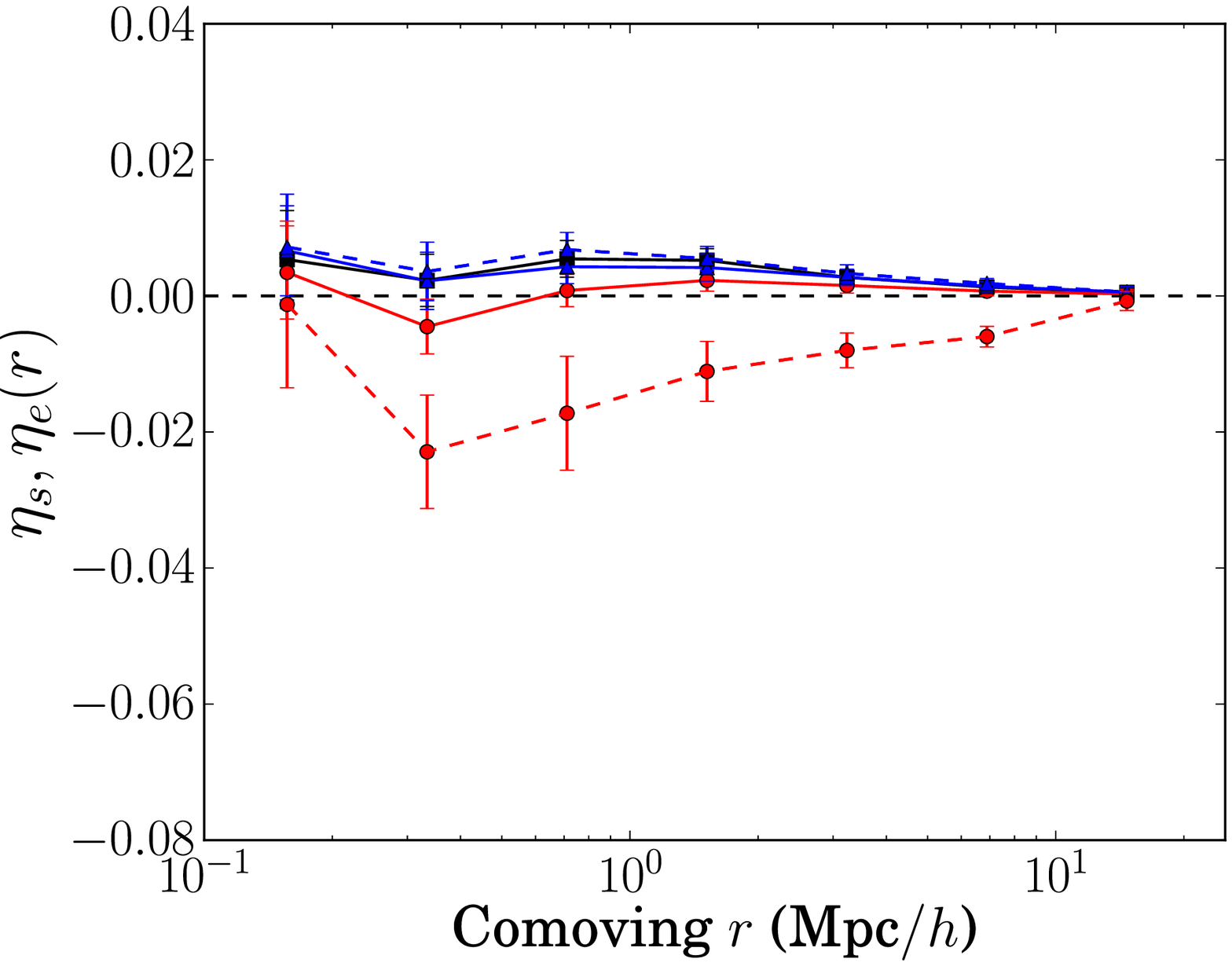}
  \includegraphics[width=0.32\textwidth]{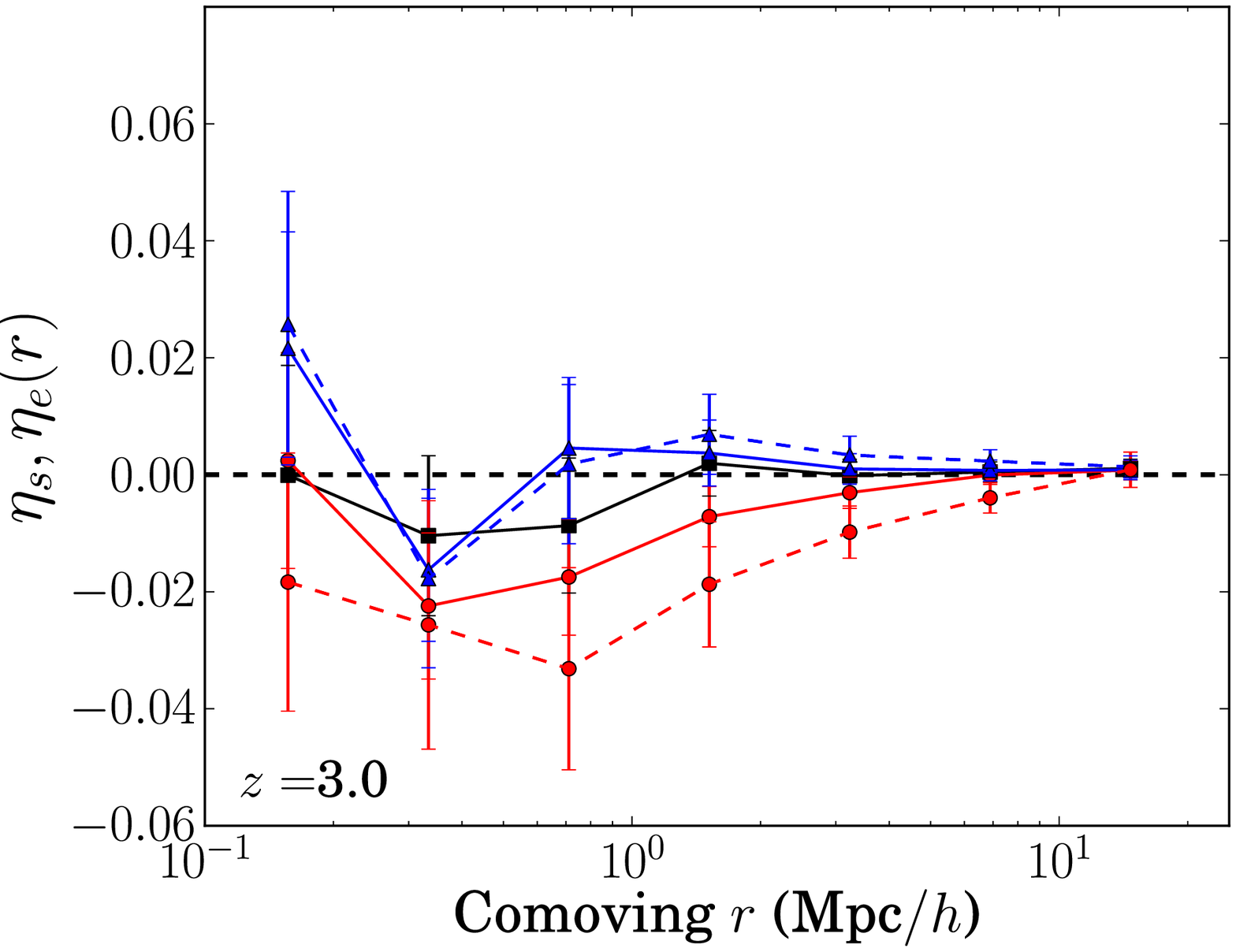}
  \includegraphics[width=0.32\textwidth]{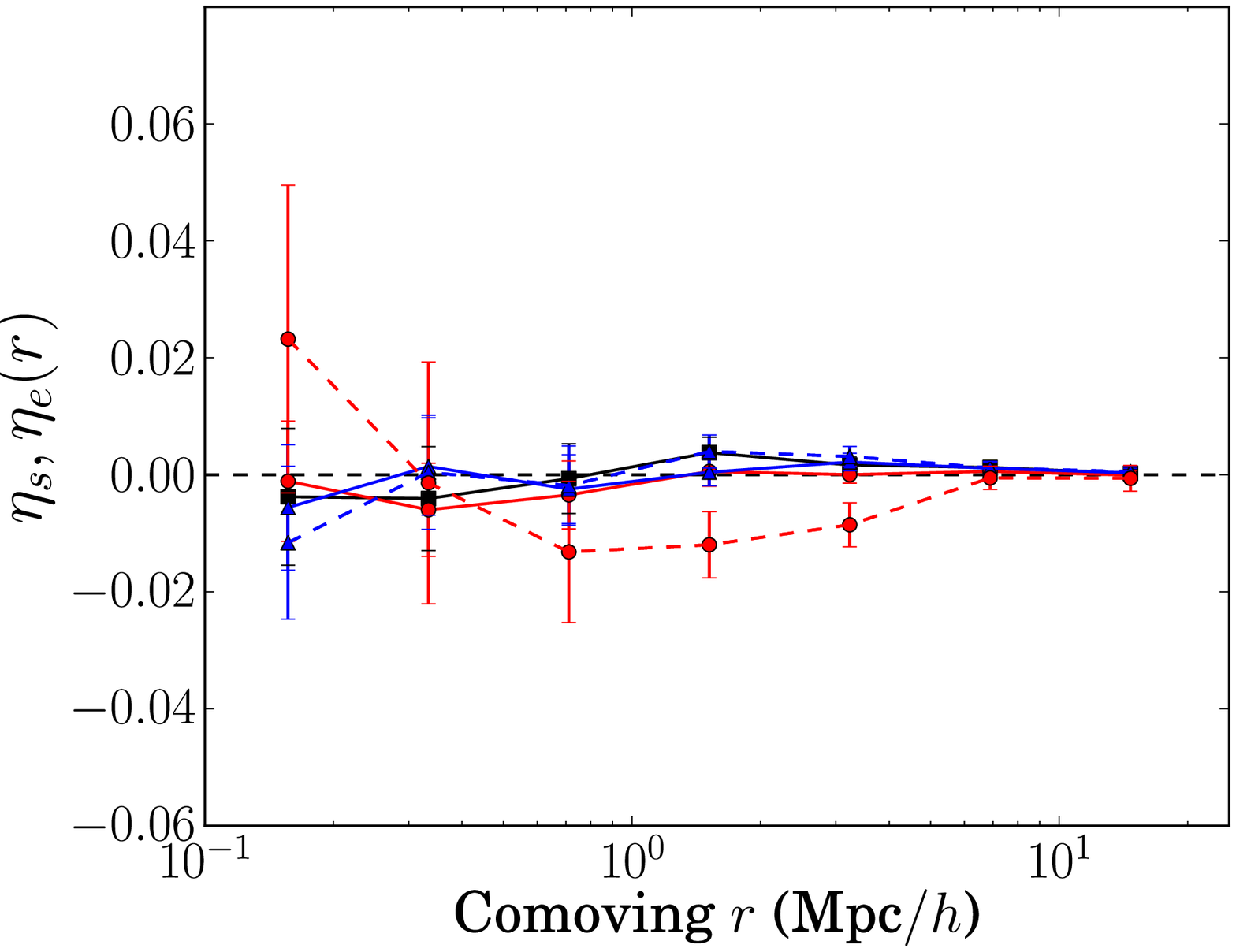}
  \includegraphics[width=0.32\textwidth]{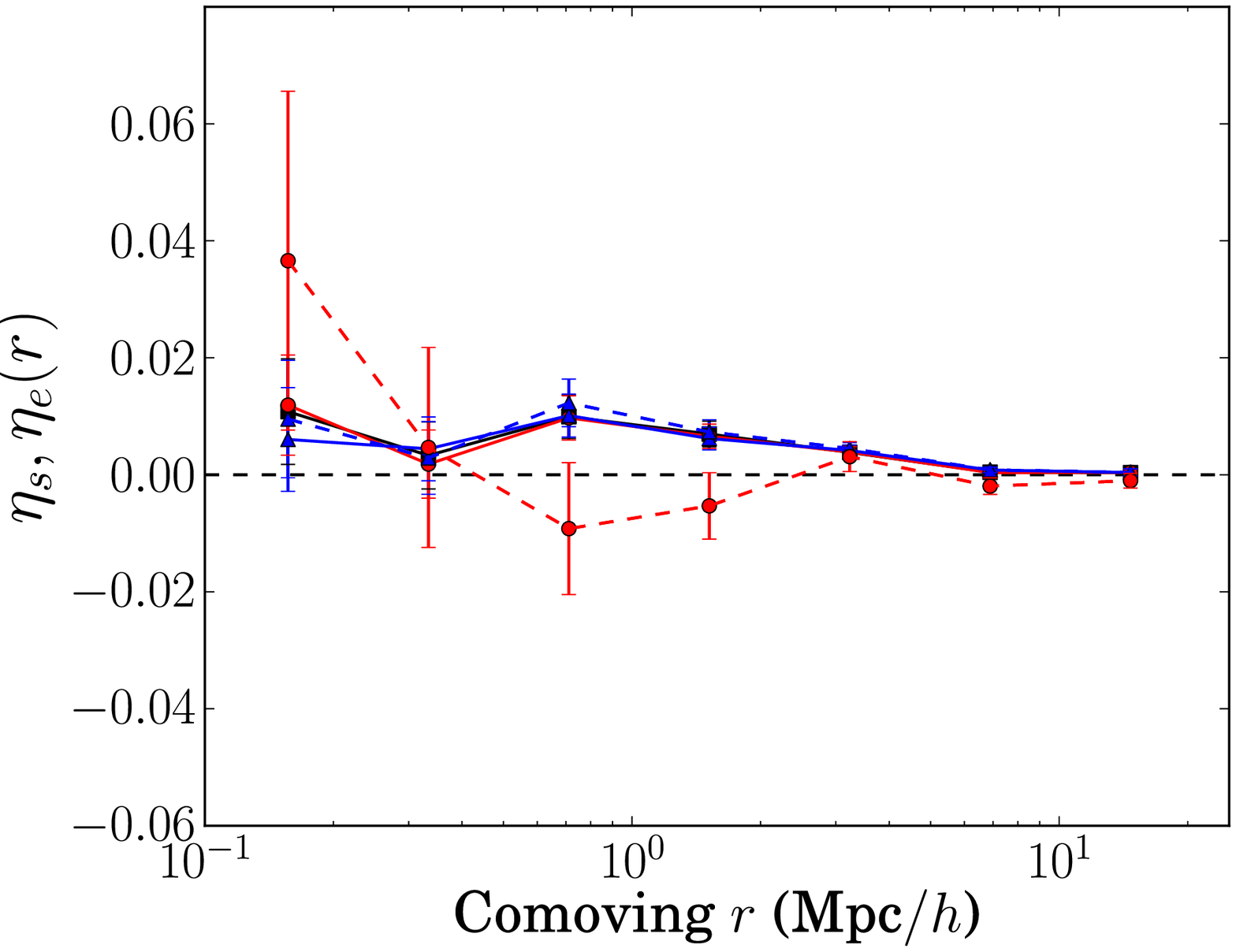}
  \caption{$\eta_e$ and $\eta_s$ around the subsampled dark matter field at redshifts $z=\{0.06,0.5,1.5,3\}$ from top to bottom. The first column corresponds to the most luminous galaxies, $M_r\leq -22$, in the simulation at each redshift. The middle column corresponds to $-22<M_r\leq-21$. The right column spans the range $-21<M_r\leq-20$. For the blue points, the orientation is determined from the spin. Red points correspond to orientations derived from the minor axis of the simple inertia tensor and black points, from the reduced inertia tensor. We also show $\eta_e$ for the subsample of ellipticals in each bin (red dashed, from the simple inertia tensor) and $\eta_s$ for the subsample of discs (blue dashed). Error bars are obtained through the jackknife.}
  \label{fig:align3d}
\end{figure*}

In Fig.~\ref{fig:align3d}, we show all $\eta_e(r)$ and $\eta_s(r)$ statistics as a function of redshift, luminosity and galaxy dynamics. These results discussed in depth in section~\ref{sec:align3d}; they are hereby presented for completeness.

The left column of Fig.~\ref{fig:align3d} shows the alignment correlations at $z=0$; the top left panel corresponds to Fig.~\ref{fig:jack3d761} for high luminosity galaxies at that redshift. In comparison to $M_r\leq-22$ measurements, alignment trends are reduced at intermediate luminosities. The measurement performed using the simple inertia tensor still indicates a significant radial alignment trend ($>99.99\%$ C.L.), while the reduced inertia tensor and the spin suggest no alignment at the $<2\sigma$ C.L. The significance is similarly high for the simple and reduced inertia tensor at low luminosities, and the spin alignment is consistent with being null at the $<3\sigma$ C.L. The strength of radial alignment is decreasing from the high luminosity to the intermediate luminosity sample; but it is comparable for the intermediate and low luminosity samples. However, the radial alignment of ellipticals alone clearly decreases with luminosity for the simple inertia tensor; this suggests there is an interplay between the strength of alignment and the fraction of ellipticals as a function of luminosity which yields the total signal observed in the left column of Fig.~\ref{fig:align3d}. In fact, the bottom right panel of Fig.~\ref{fig:align3d} shows a very significant trend for both shape and spin {\it tangential} alignments at low luminosity and high redshift, at $>99.9\%$ C.L. In Paper I, we similarly found a tangential alignment of discs around ellipticals. The results shown in the bottom right panel of Fig.~\ref{fig:align3d} show the same trend, but for low luminosity galaxies oriented around DM overdensities. It is clear from Figs.~\ref{fig:vsigmr} and \ref{fig:rfrac} that discs dominate the population of galaxies in this luminosity bin; and their relative fraction increases towards high redshift.

\begin{figure*}
  \includegraphics[width=0.32\textwidth]{figs_paper2/wdmplus_magr22_allz.eps}
  \includegraphics[width=0.32\textwidth]{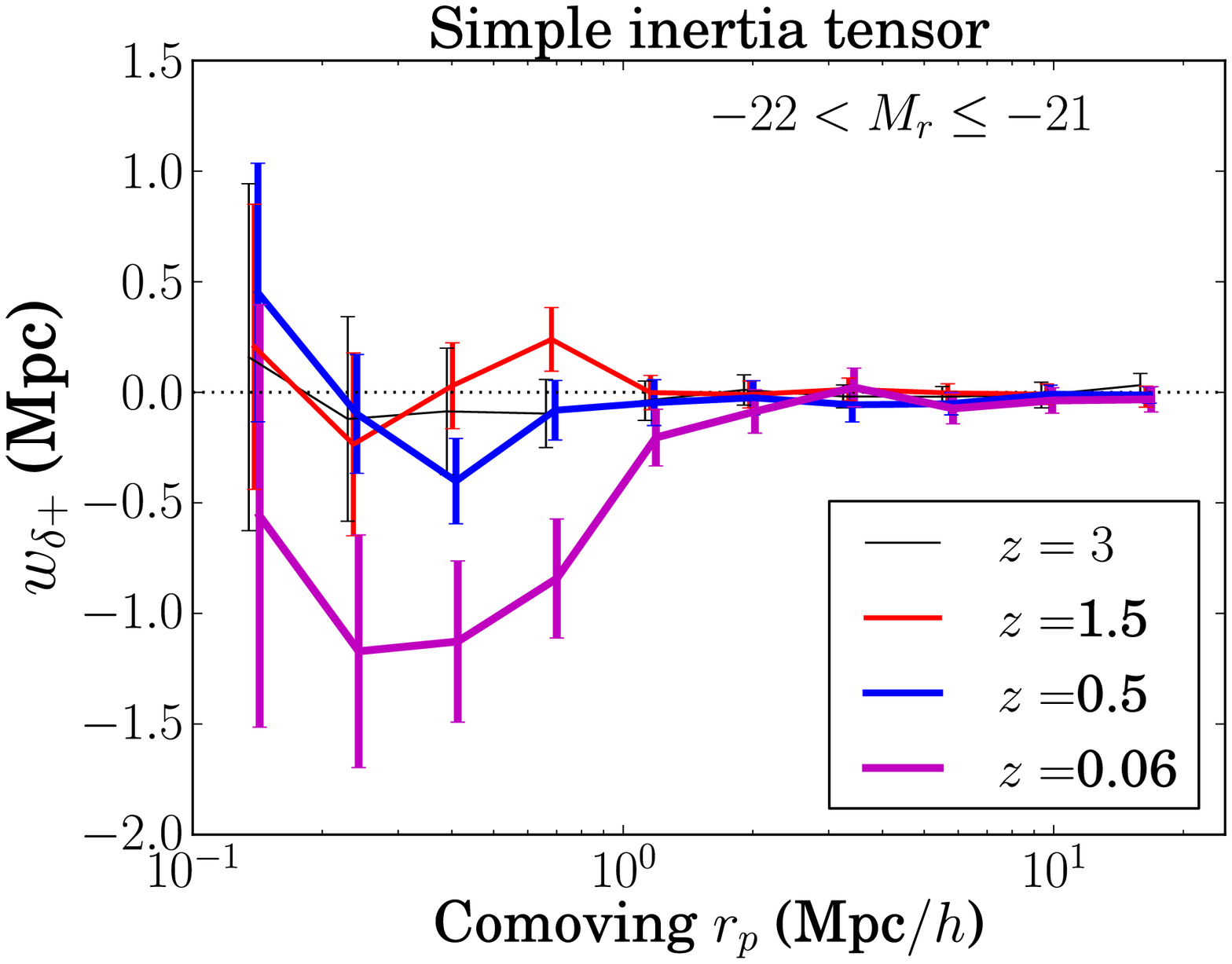}
  \includegraphics[width=0.32\textwidth]{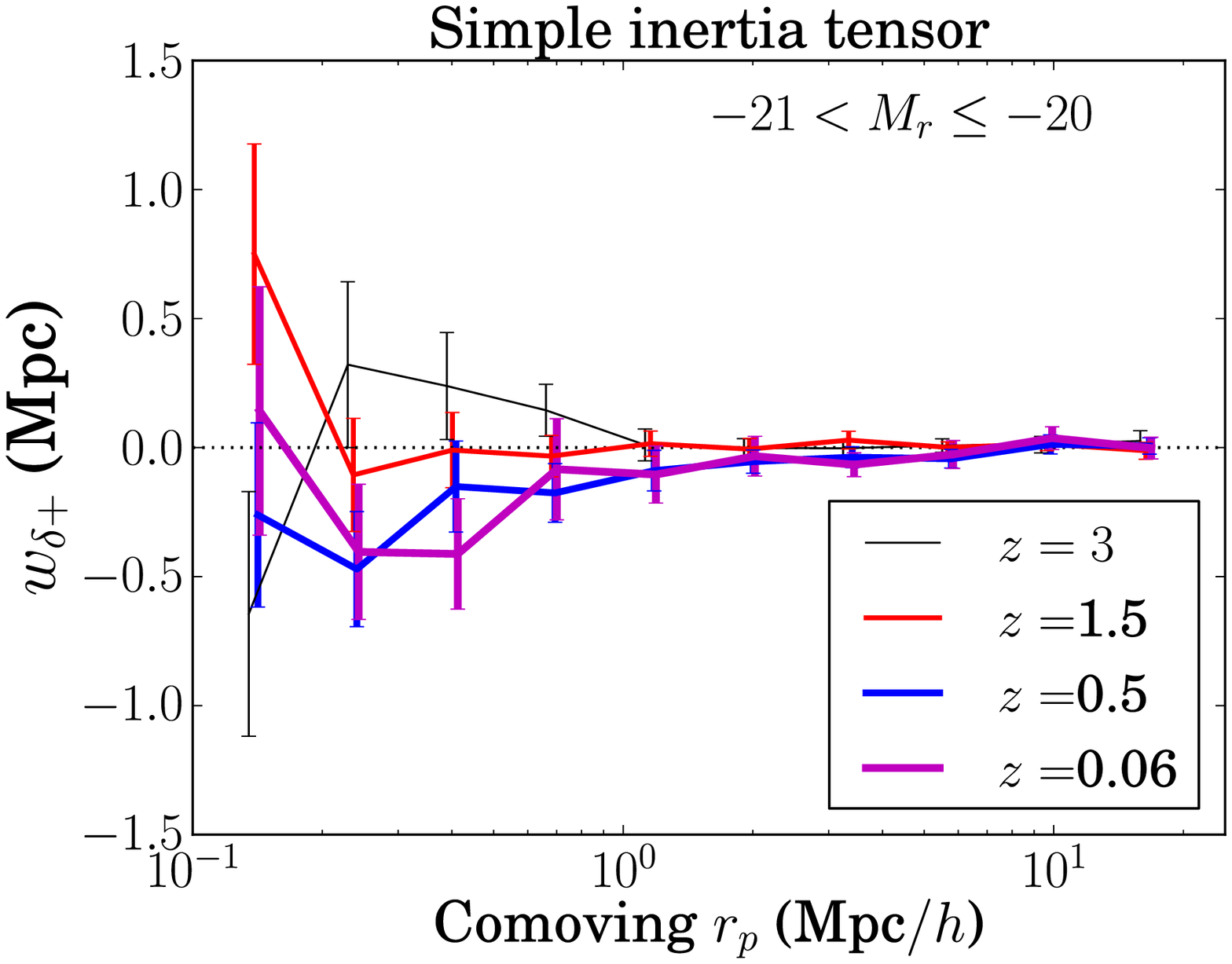}
  \includegraphics[width=0.32\textwidth]{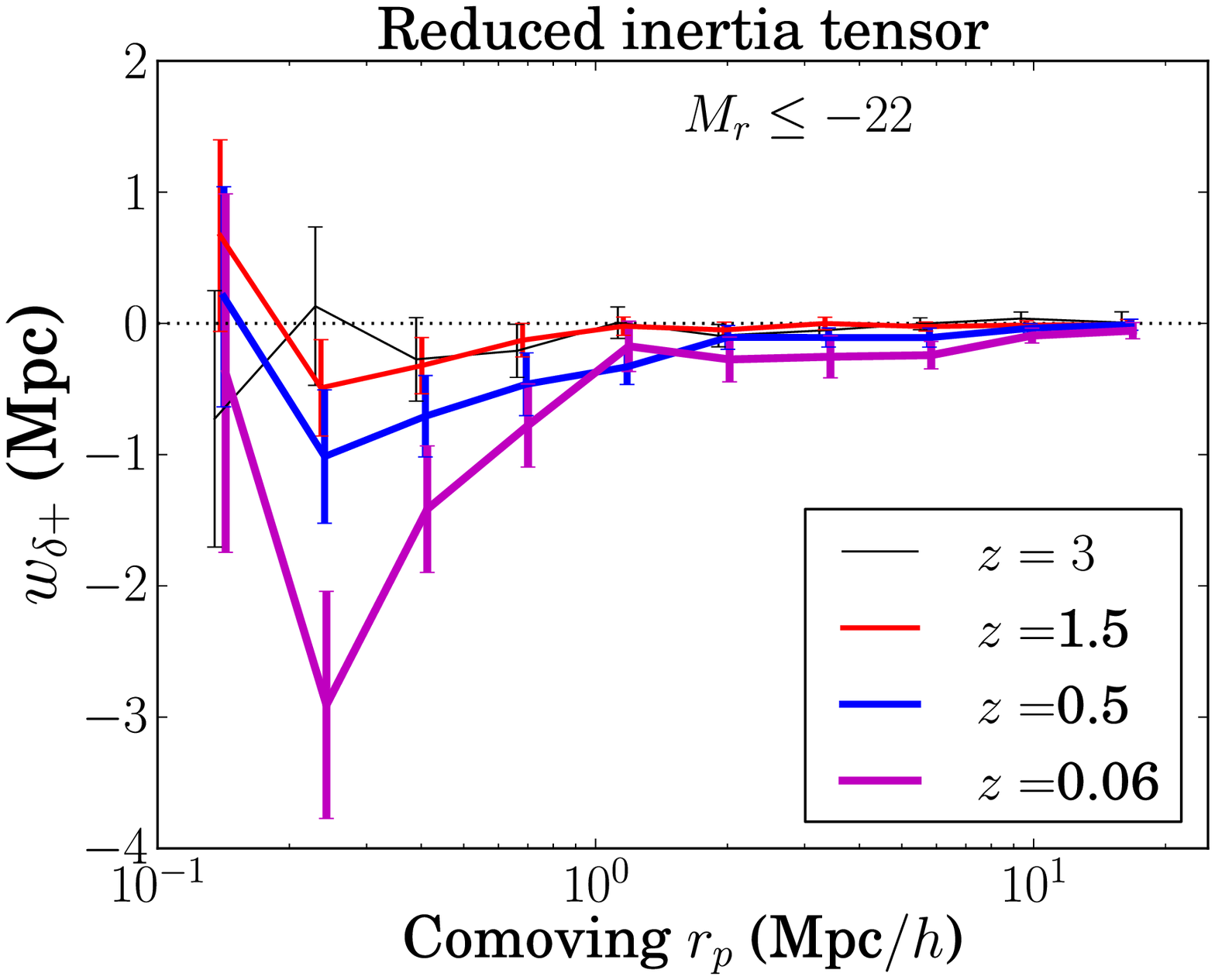}
  \includegraphics[width=0.32\textwidth]{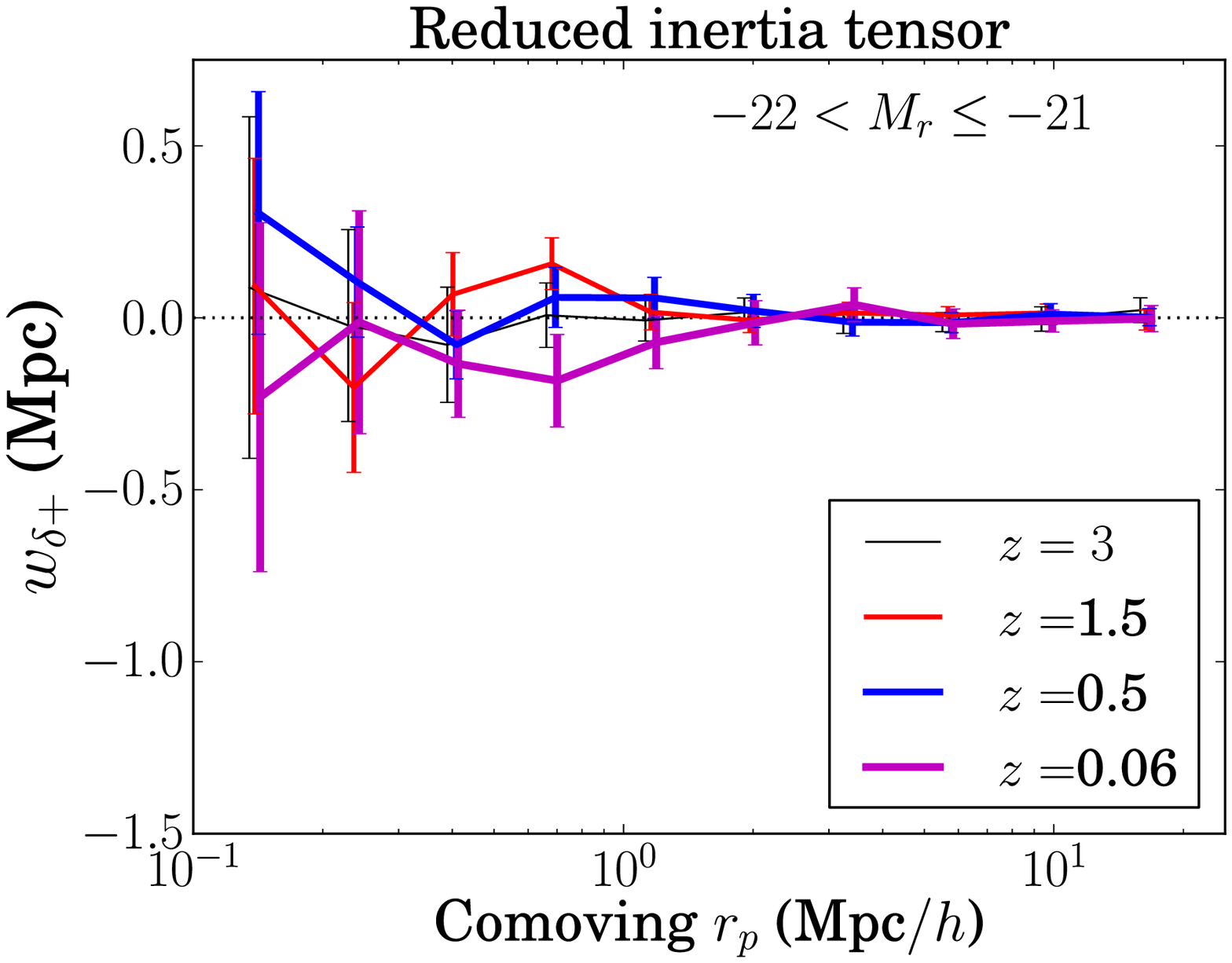}
  \includegraphics[width=0.32\textwidth]{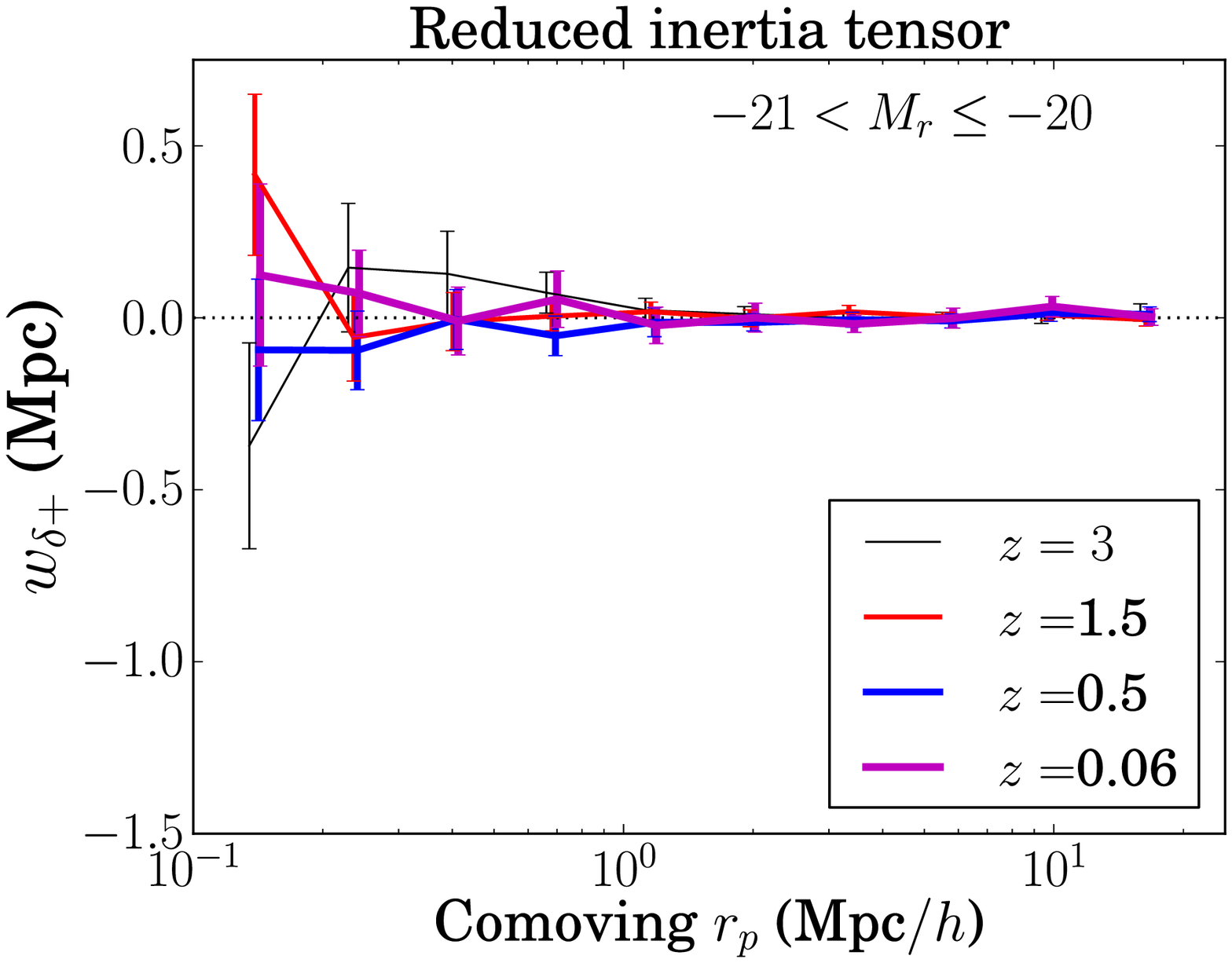}
  \includegraphics[width=0.32\textwidth]{figs_paper2/wgplus_magr22_allz.eps}
  \includegraphics[width=0.32\textwidth]{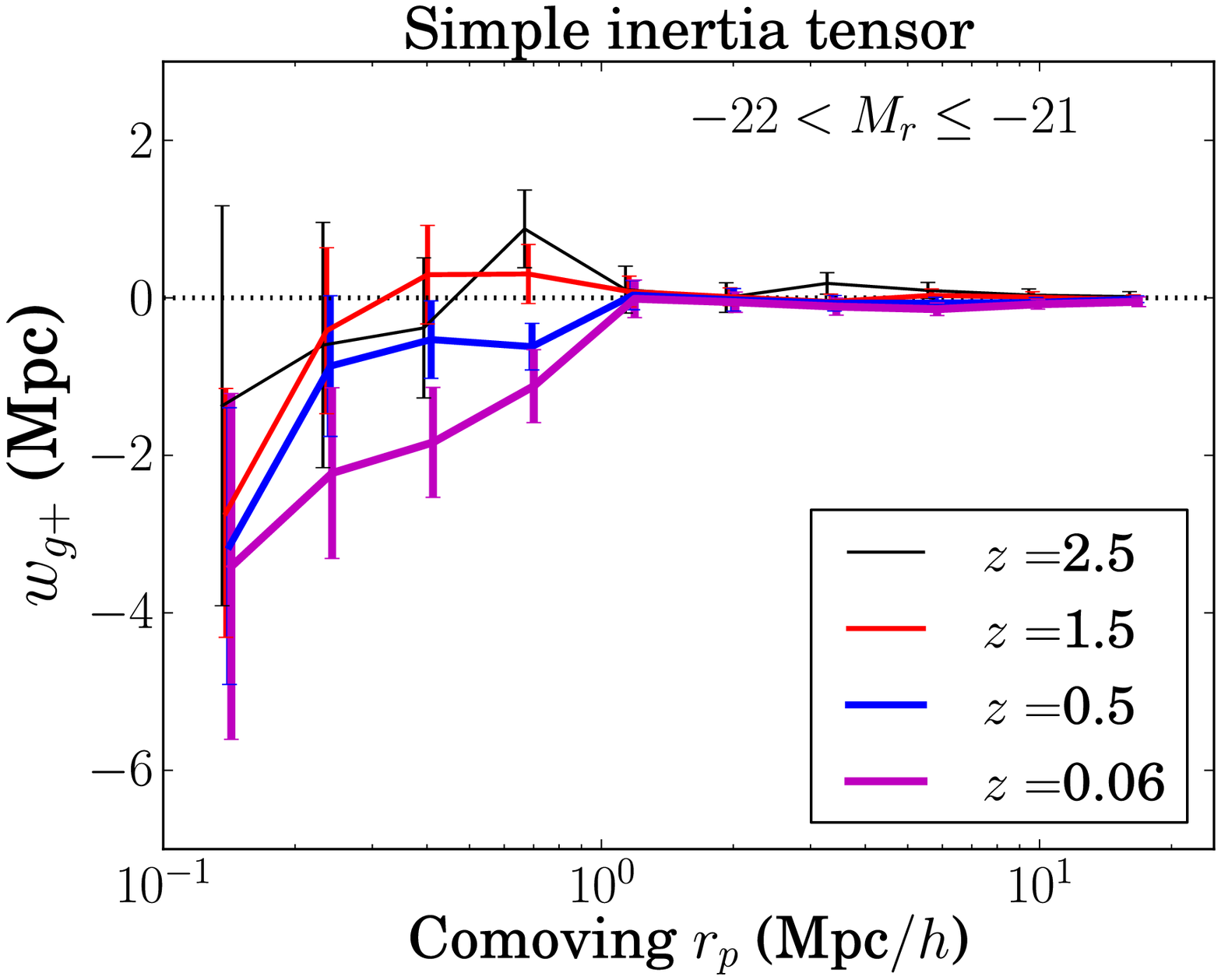}
  \includegraphics[width=0.32\textwidth]{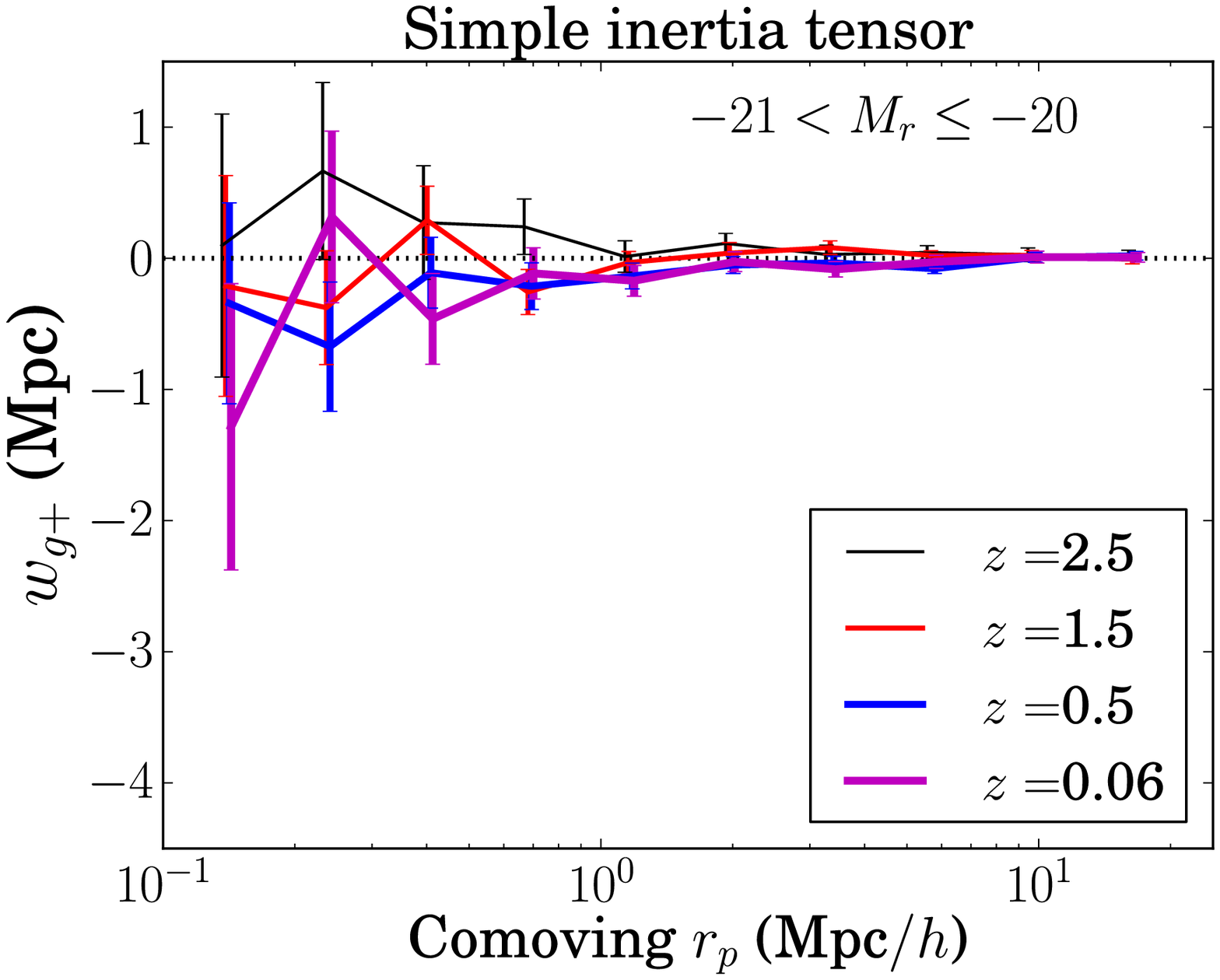}
  \includegraphics[width=0.32\textwidth]{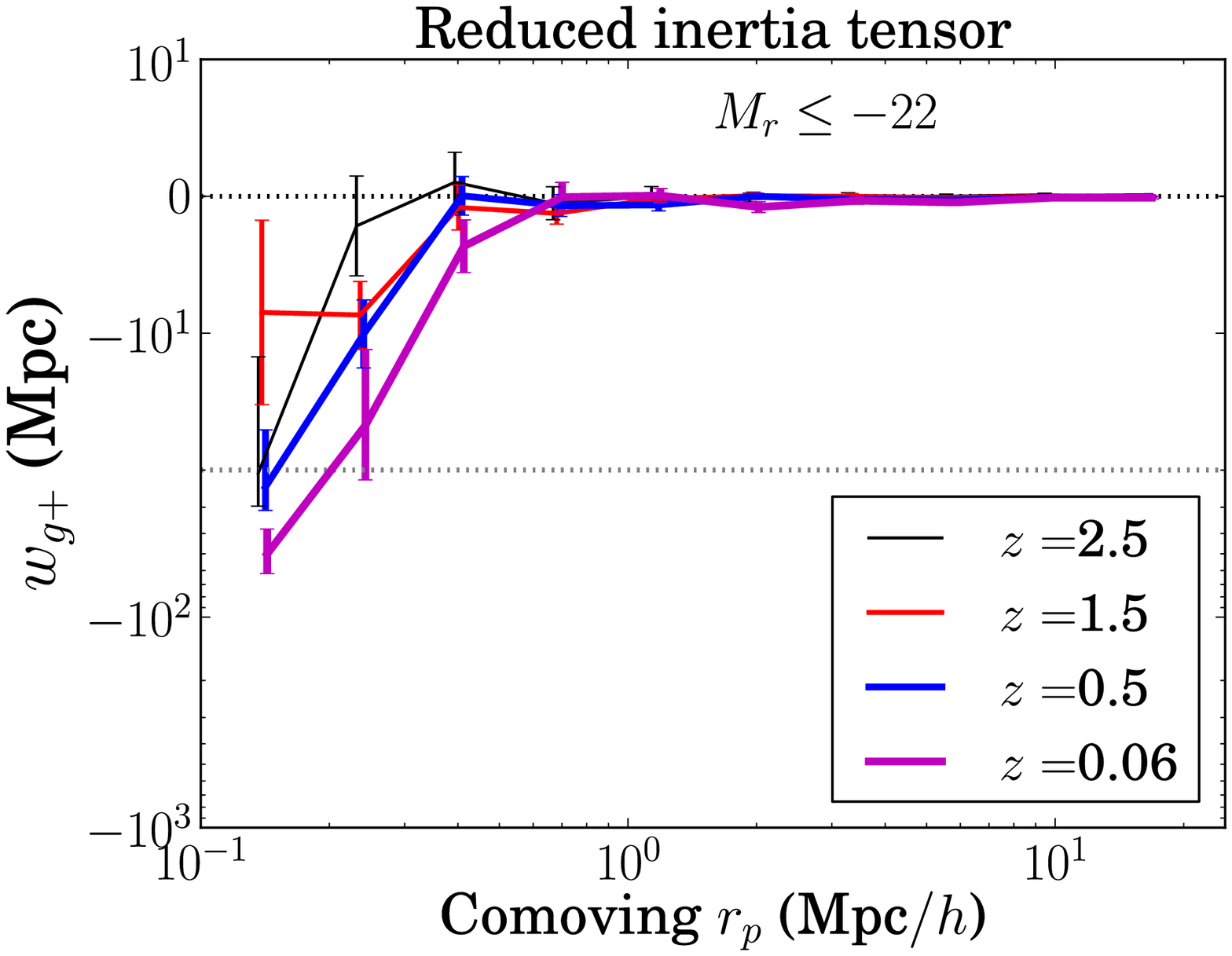}
  \includegraphics[width=0.32\textwidth]{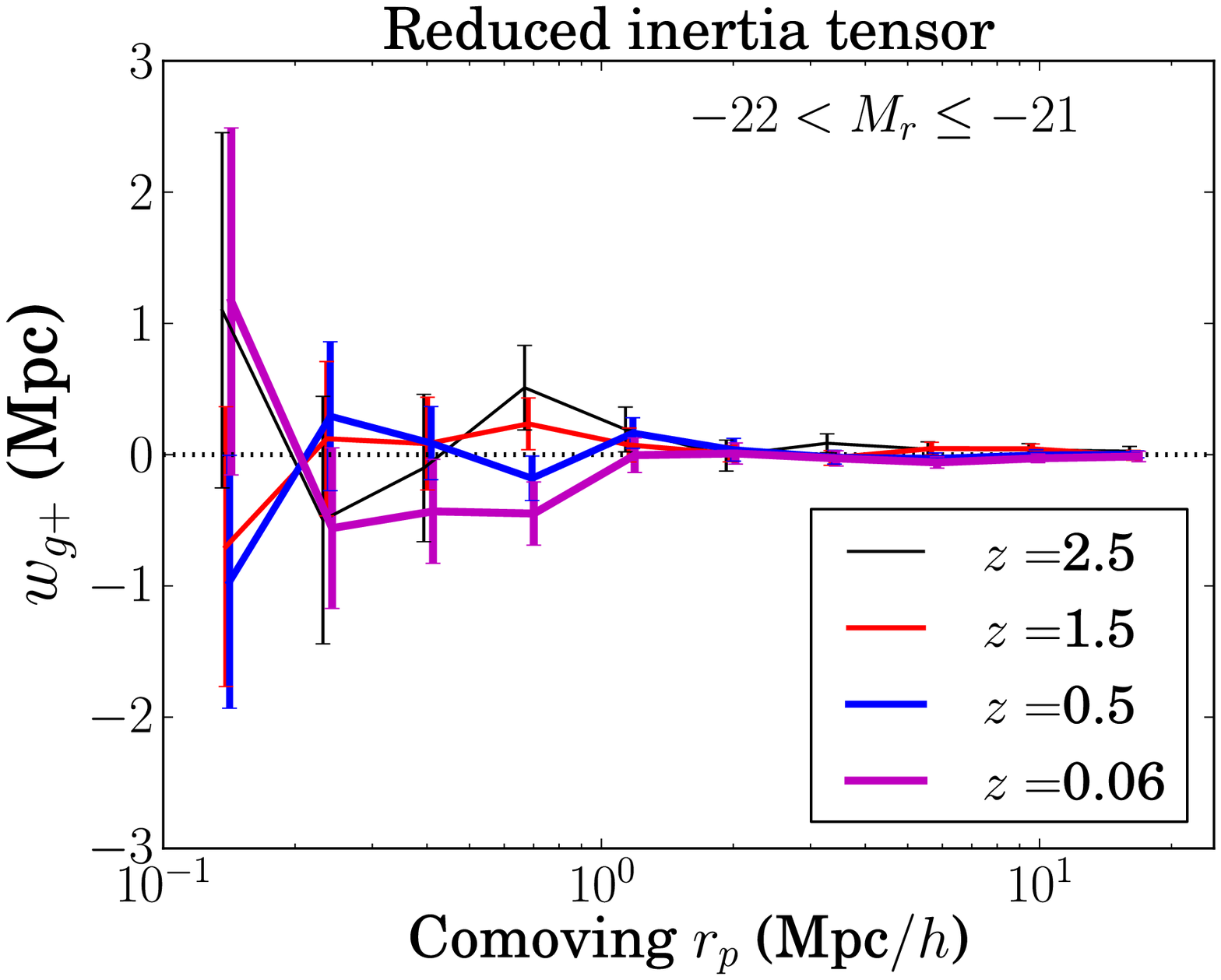}
  \includegraphics[width=0.32\textwidth]{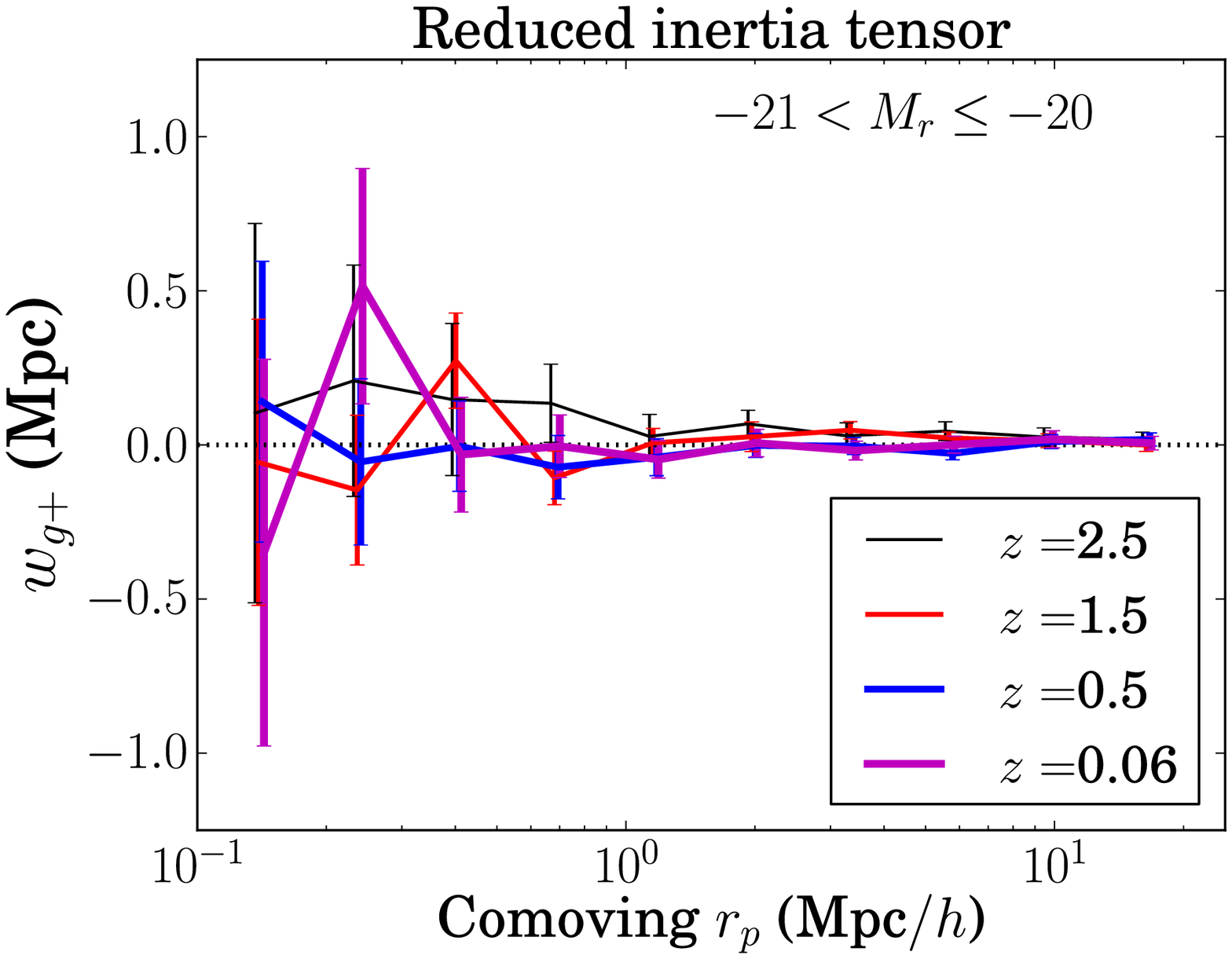}
  \caption{$w_{\delta +}$ (top two rows) and $w_{g+}$ (bottom two rows) for all luminosity bins as a function of redshift. We show results for both the simple and reduced inertia tensor, as indicated in the title of each panel. The alignment signal decreases with redshift and is stronger in the highest luminosity bin. Note that the black dotted line at $w_{g+}=-20$ Mpc indicates a change from logarithmic to linear scale in the $y$-axis.}
  \label{fig:allproj}
\end{figure*}

\begin{figure*}
  \includegraphics[width=0.32\textwidth]{figs_paper2/wdmplus_magr22_allz_vsig1.eps}
  \includegraphics[width=0.32\textwidth]{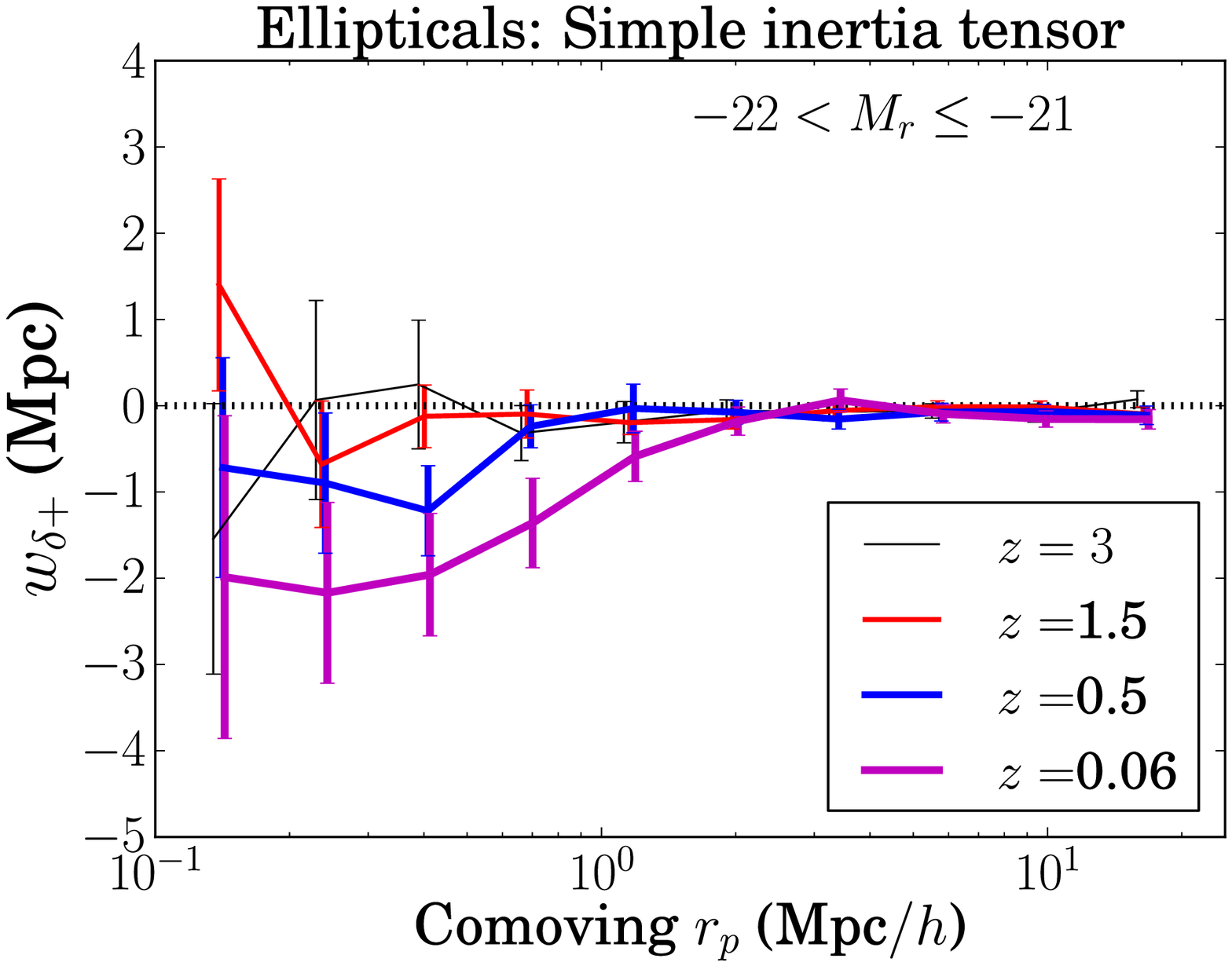}
  \includegraphics[width=0.32\textwidth]{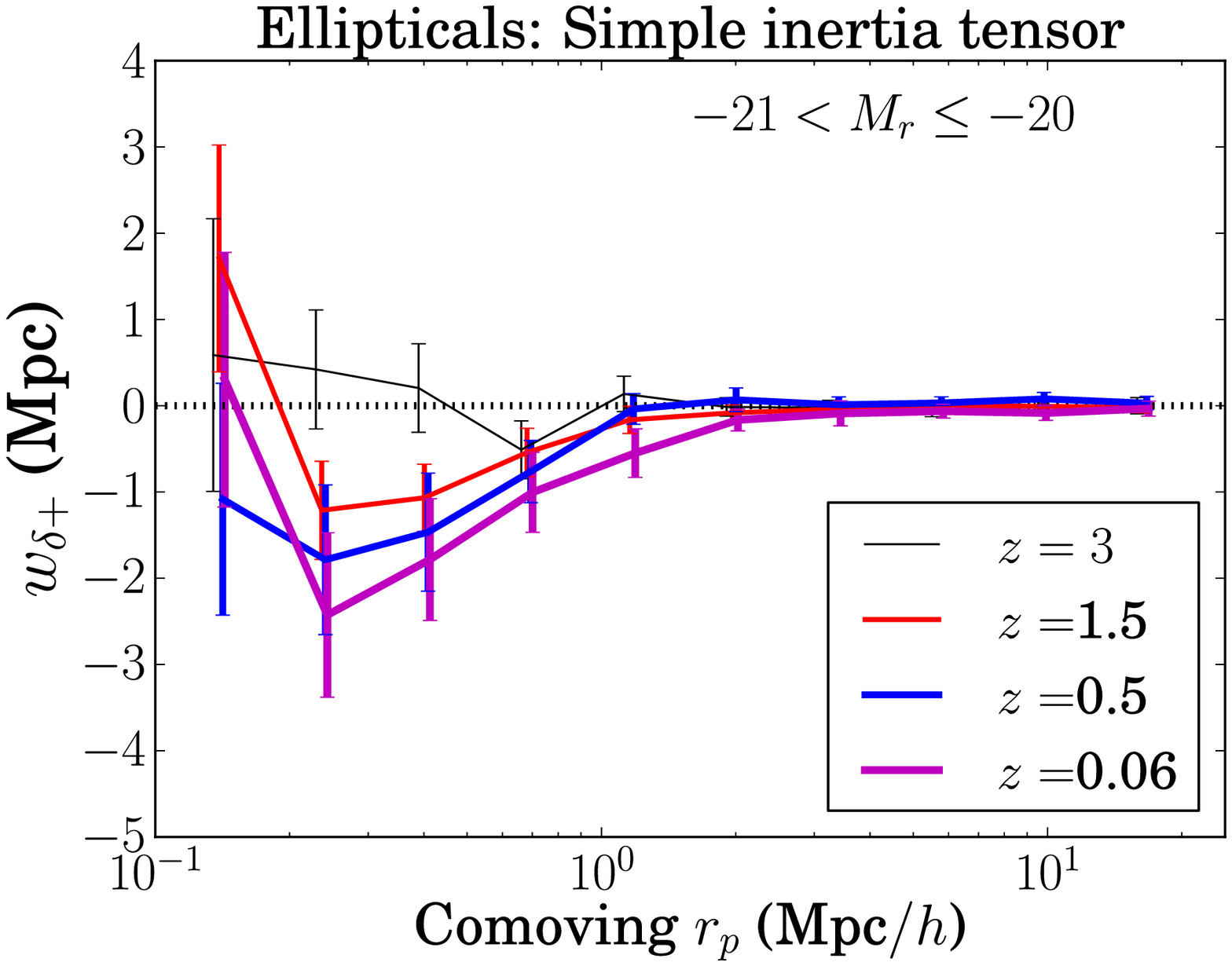}
  \includegraphics[width=0.32\textwidth]{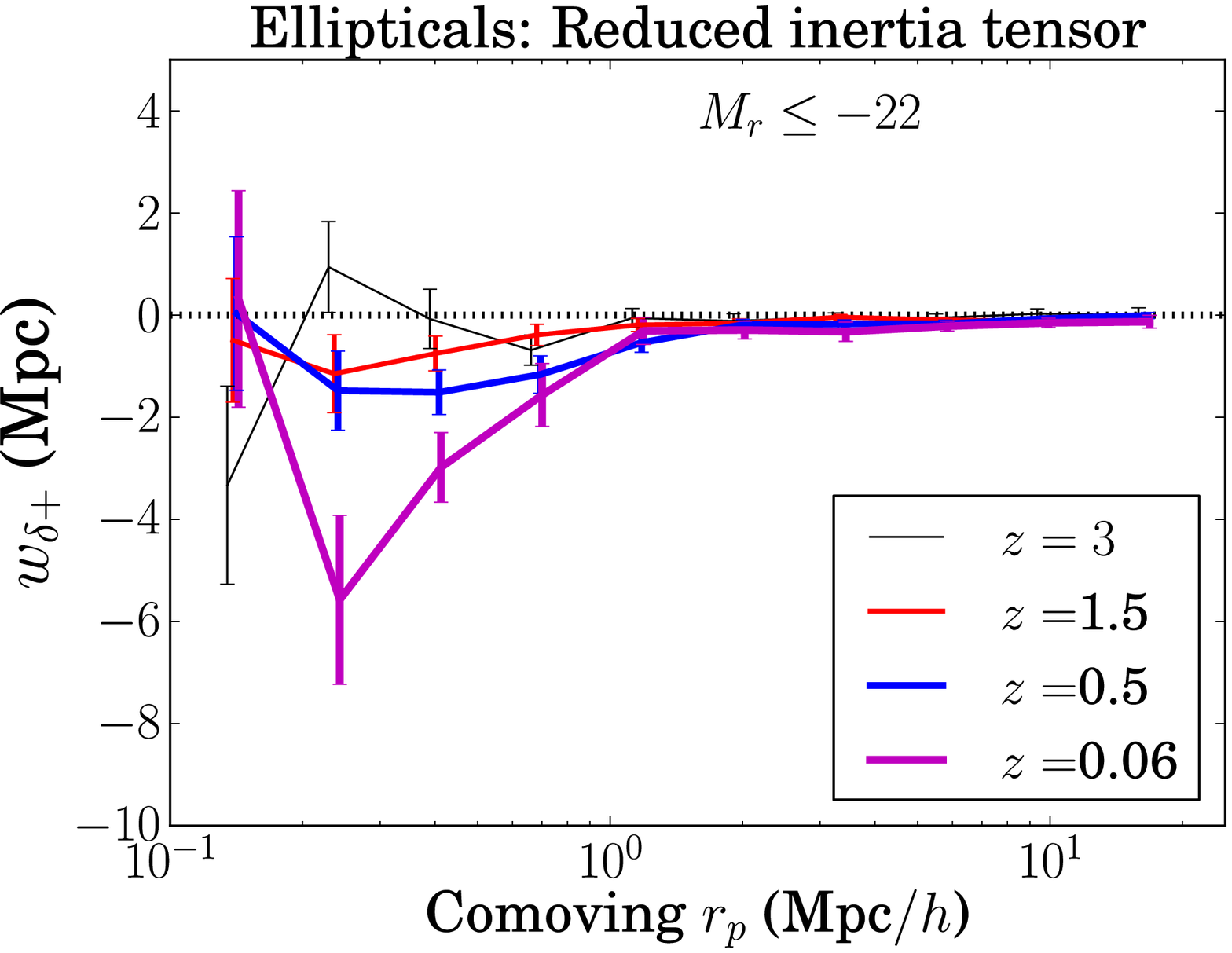}
  \includegraphics[width=0.32\textwidth]{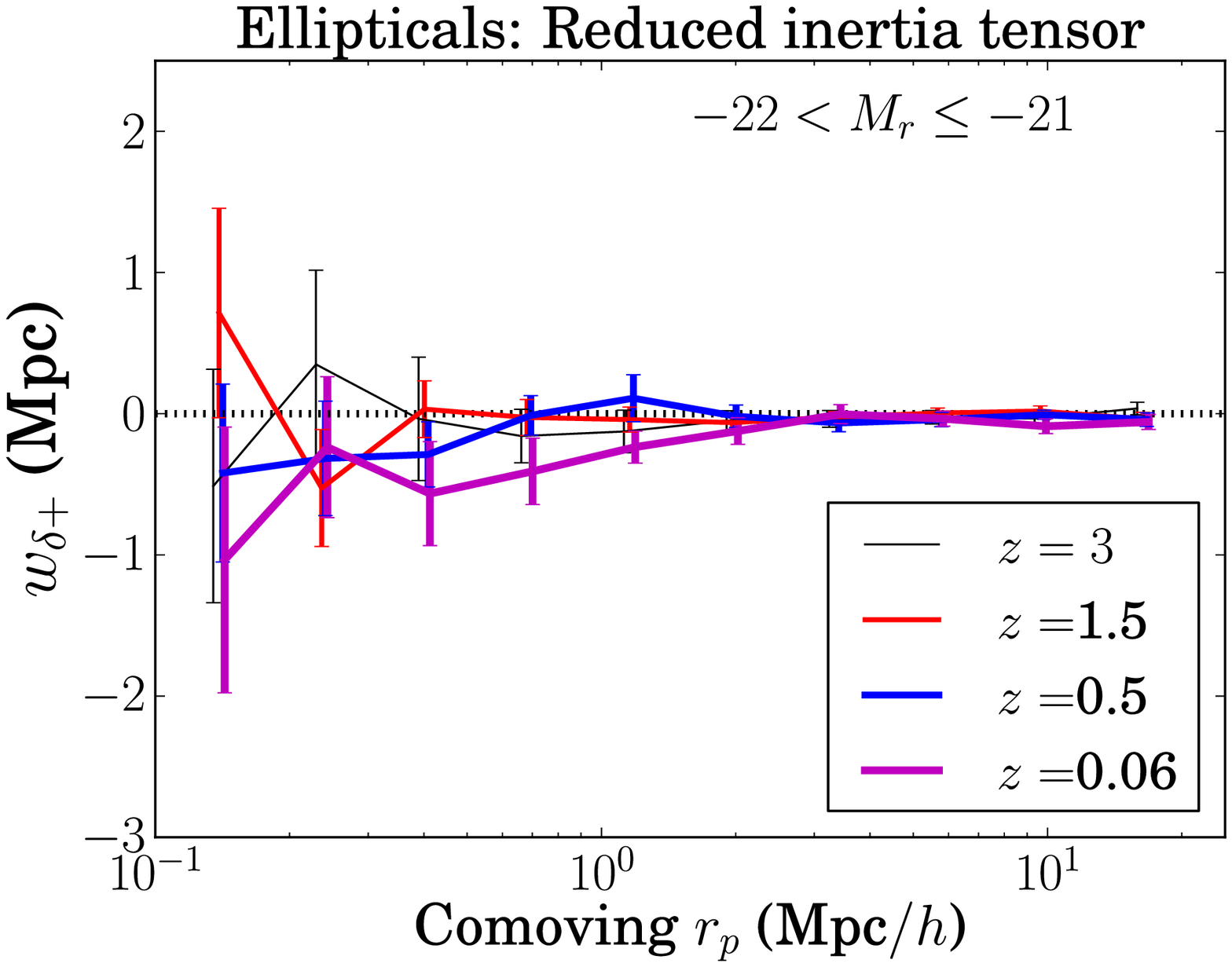}
  \includegraphics[width=0.32\textwidth]{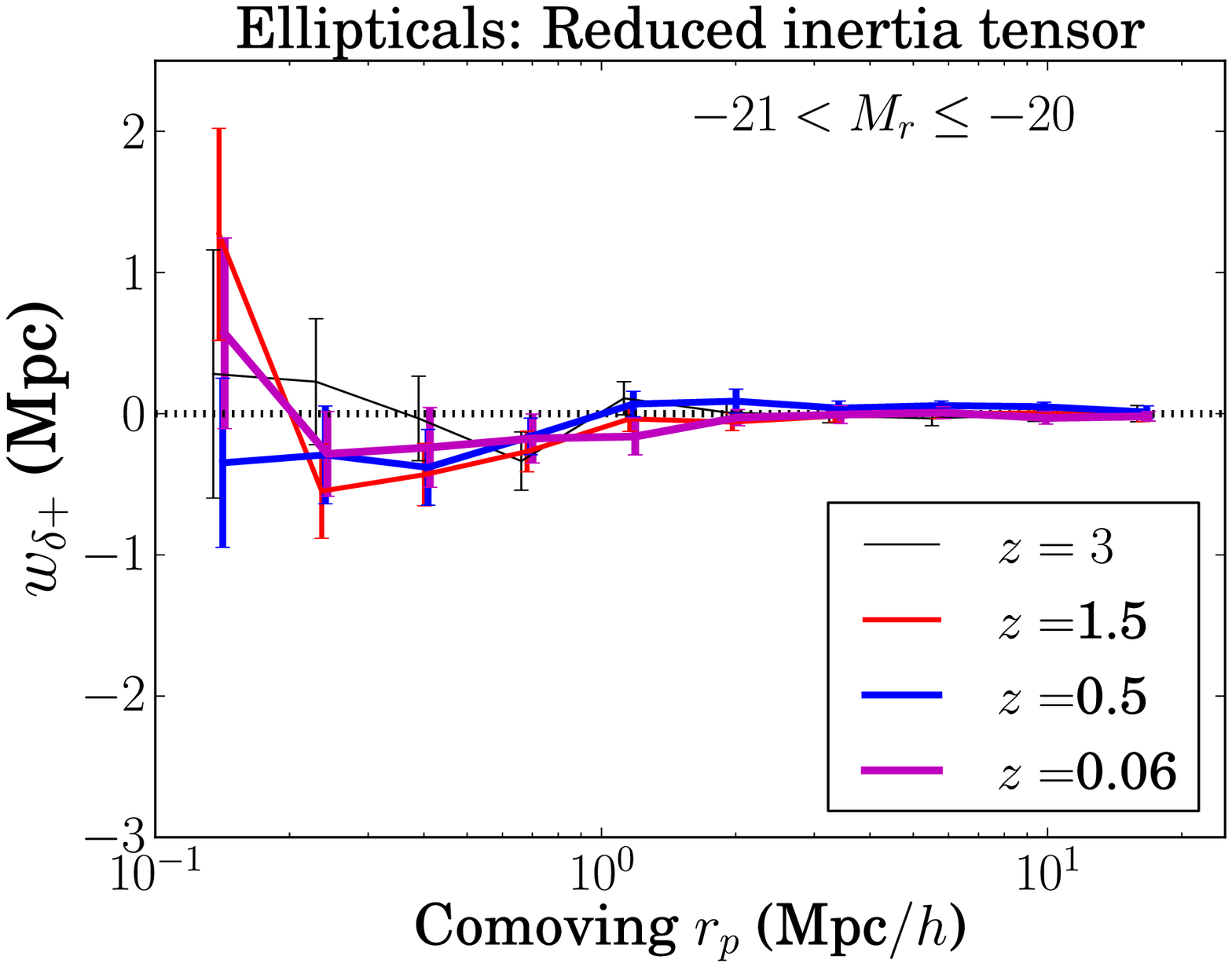}
  \caption{$w_{\delta +}$ for elliptical galaxies in all luminosity bins as a function of redshift. We show results for both the simple (top row) and reduced (bottom row) inertia tensor, as indicated in the title of each panel. The alignment signal decreases with redshift and is stronger for the highest luminosty bin. The $w_{\delta +}$ results are fit with the NLA model in section~\ref{sec:results} to obtain constraints on $A_I$ for ellipticals alone.}
  \label{fig:allprojvsig1}
\end{figure*}

\begin{figure*}
  \includegraphics[width=0.32\textwidth]{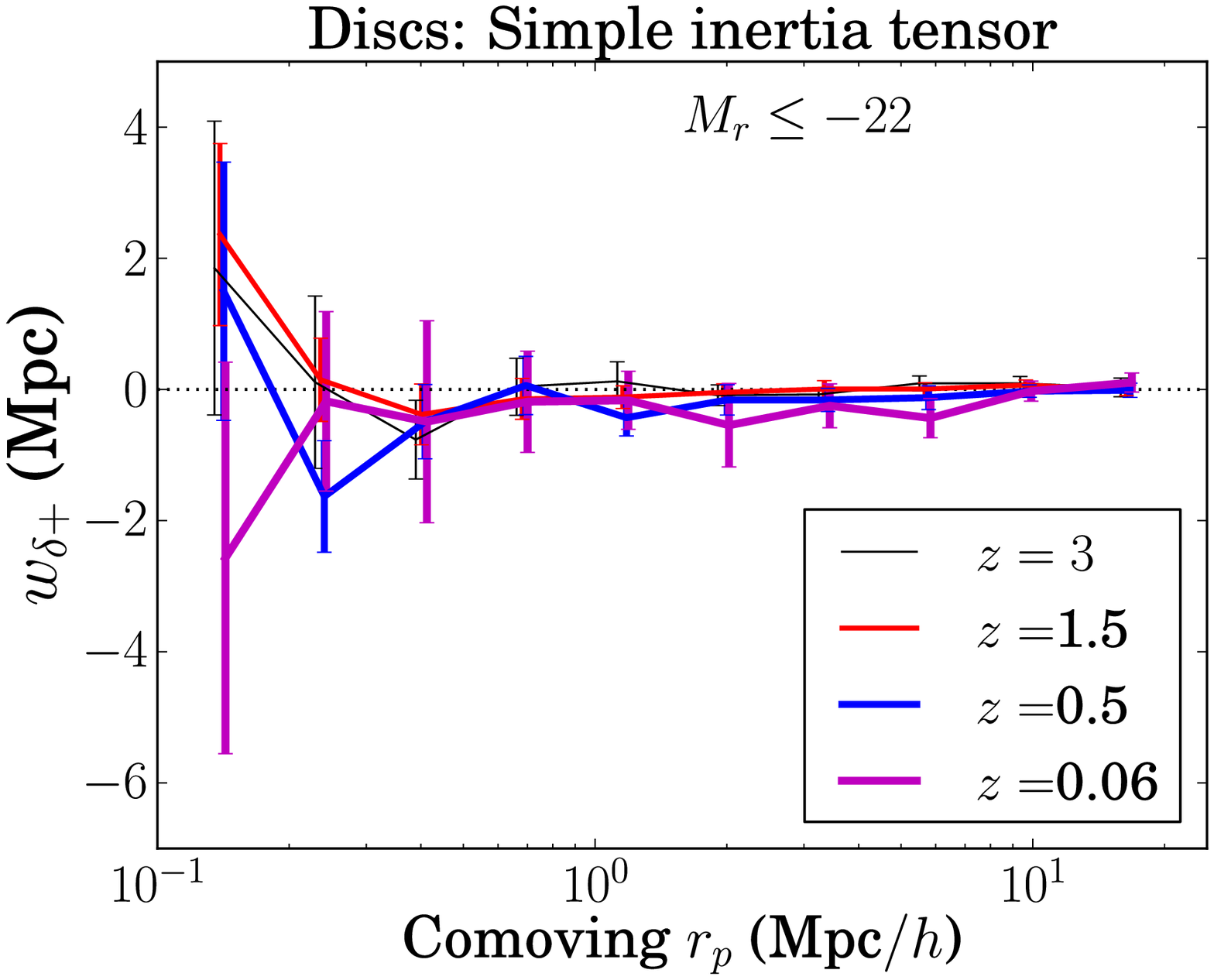}
  \includegraphics[width=0.32\textwidth]{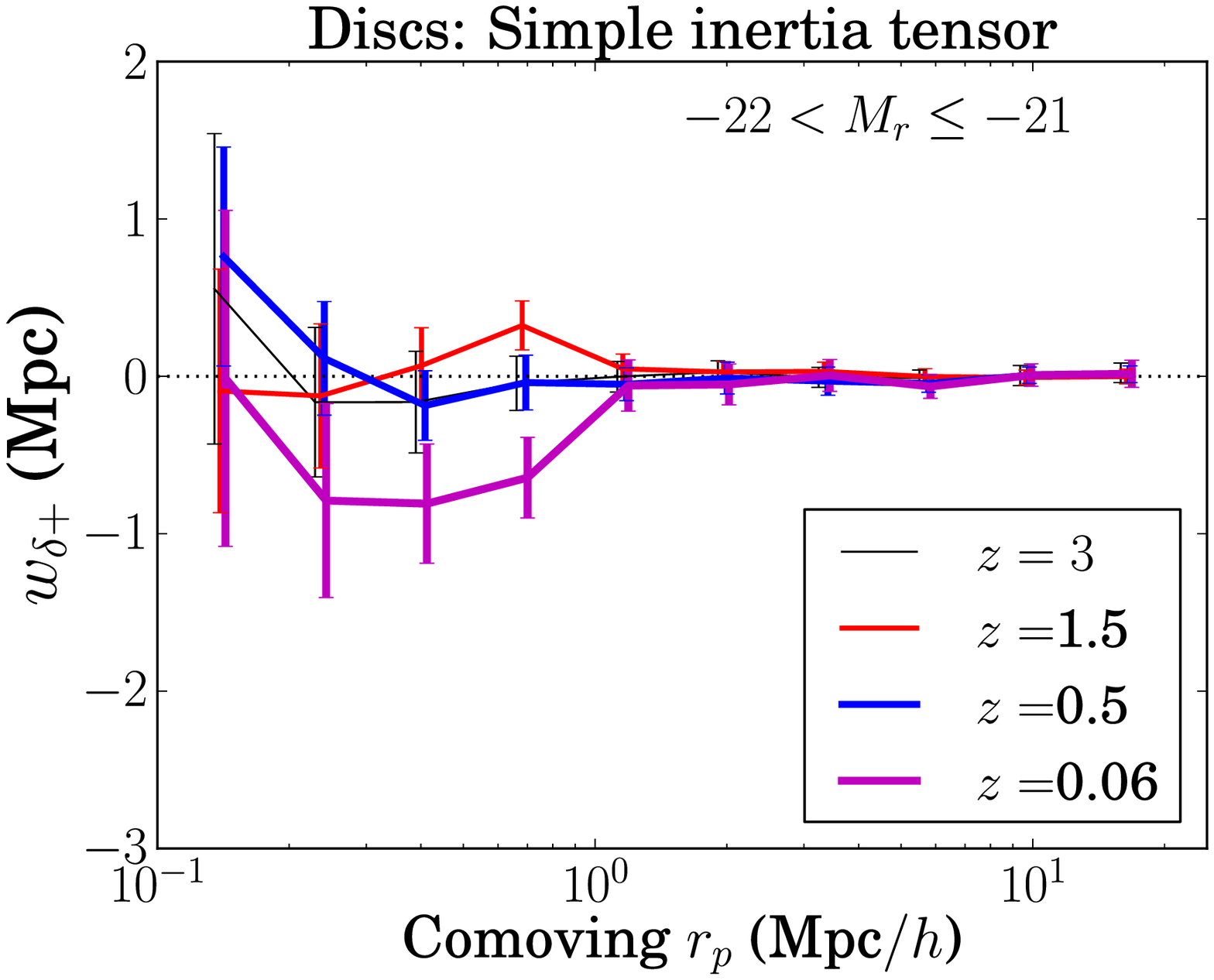}
  \includegraphics[width=0.32\textwidth]{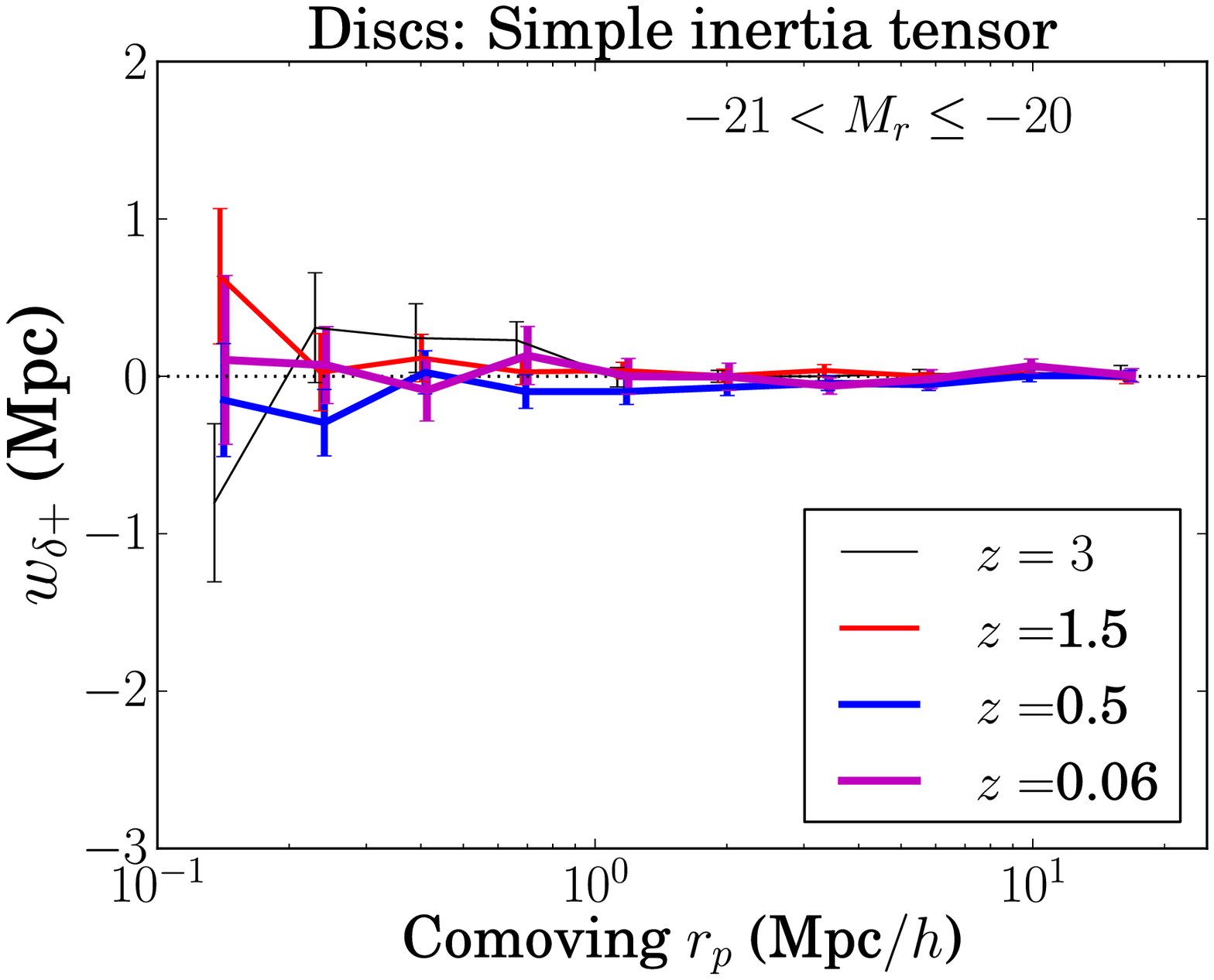}
  \includegraphics[width=0.32\textwidth]{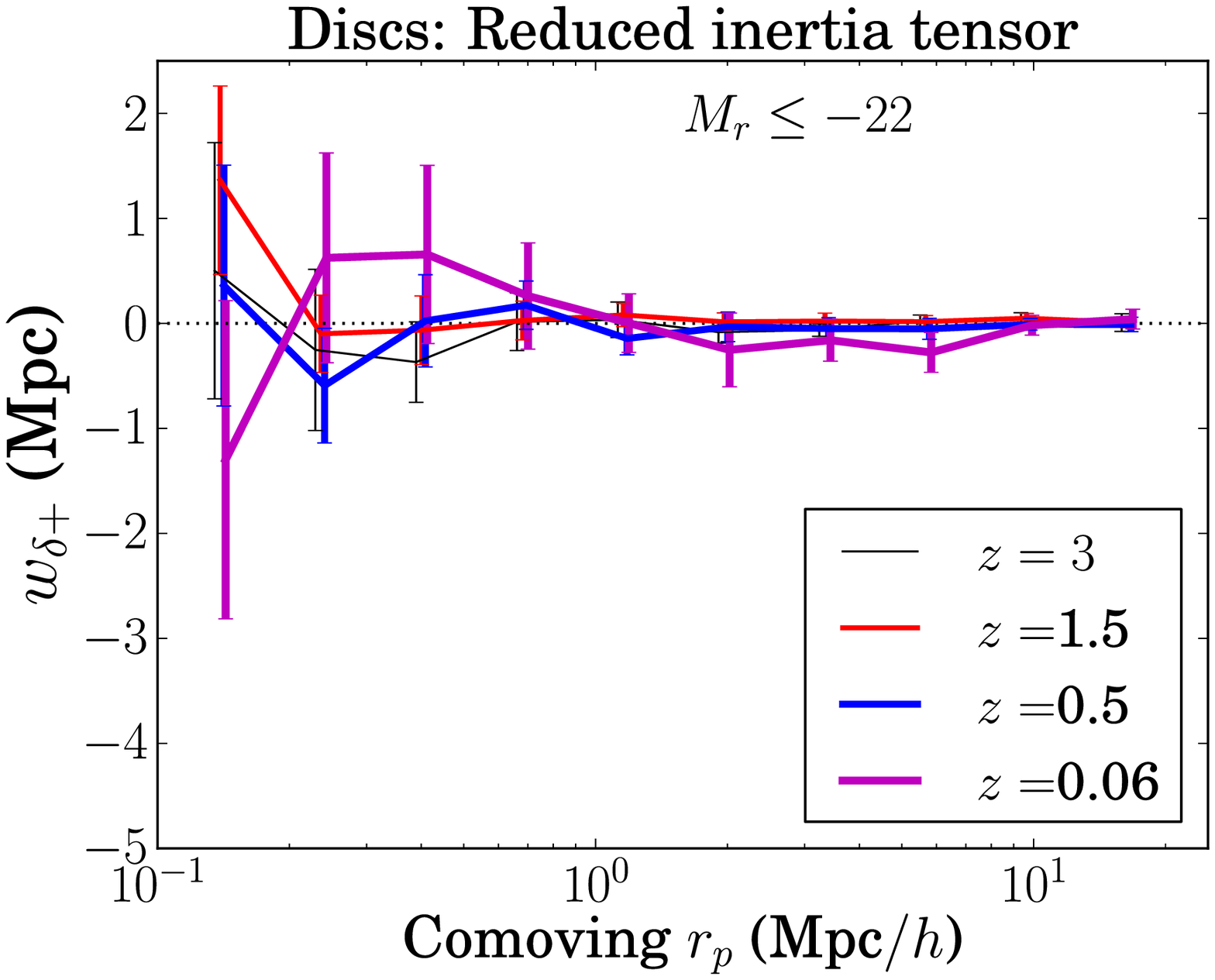}
  \includegraphics[width=0.32\textwidth]{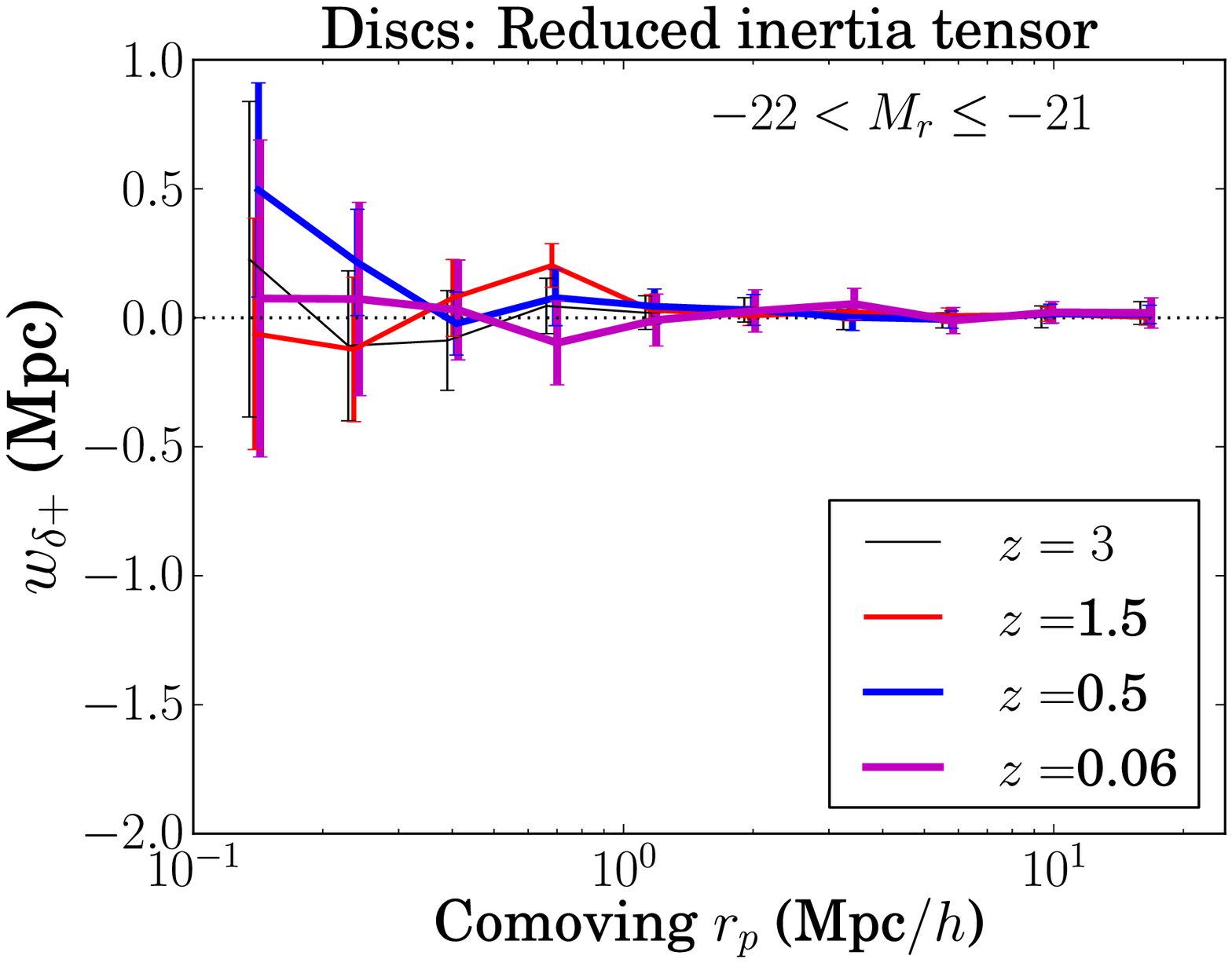}
  \includegraphics[width=0.32\textwidth]{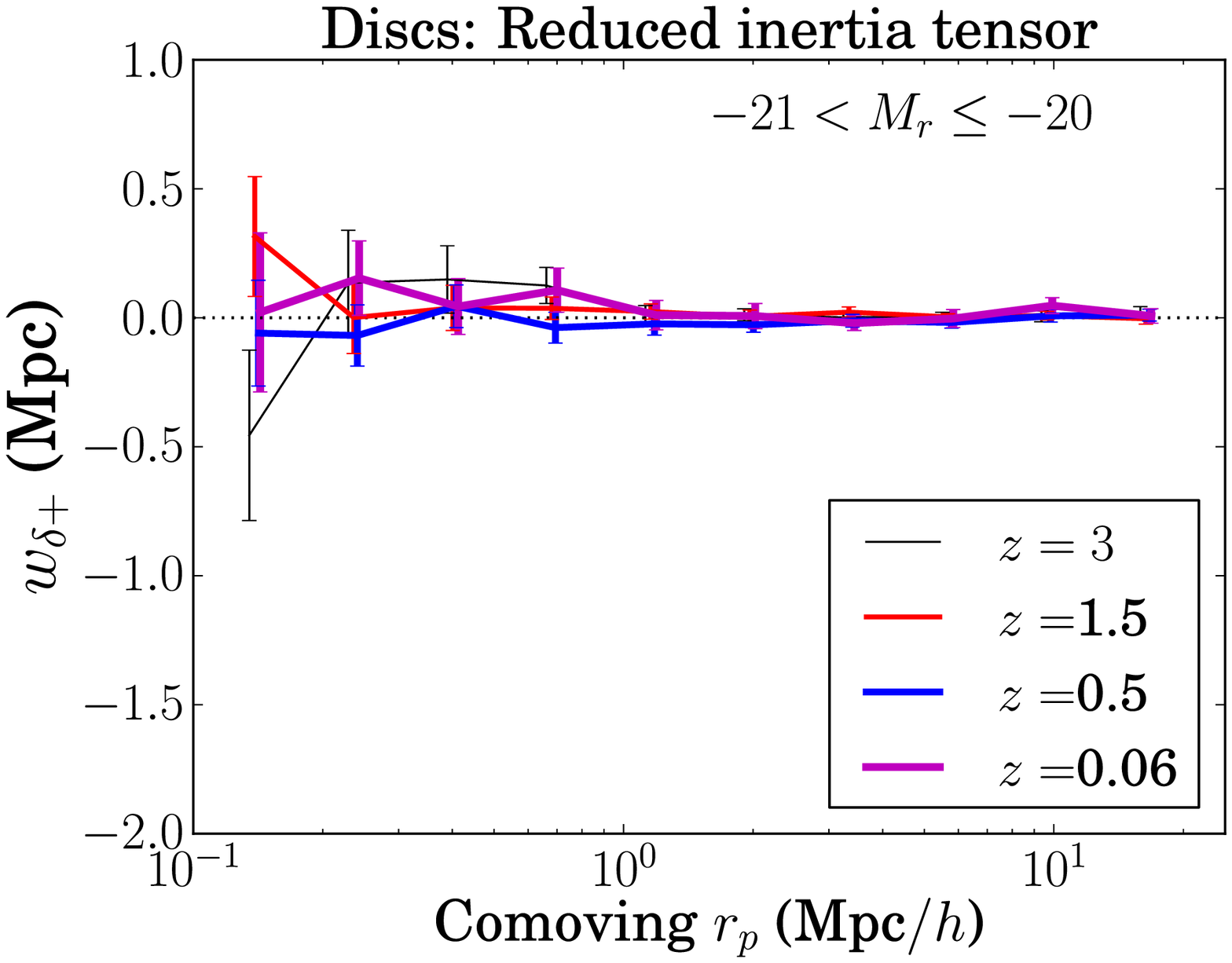}
  \caption{$w_{\delta +}$ for disc galaxies in all luminosity bins as a function of redshift. We show results for both the simple (top row) and reduced (bottom row) inertia tensor, as indicated in the title of each panel.}
  \label{fig:allprojvsig23}
\end{figure*}

Fig.~\ref{fig:allproj} shows the evolution of the projected correlation functions $w_{\delta +}$ and $w_{g+}$ for both the simple an reduced inertia tensor. Only selected redshifts are shown in each panel. The alignment signal is clearly detected at low redshifts and high luminosities in both $w_{\delta +}$ and $w_{g+}$ for the simple and reduced inertia tensors. Due to the rounder galaxy shapes obtained with the reduced inertia tensor, the amplitude of the signal is lower than with the simple inertia tensor. Also consistently with the results of Fig.~\ref{fig:align3d}, the alignment signal decreases with redshift and is higher of the $M_r\leq -22$ luminosity range. In section~\ref{sec:results}, we quoted the results for NLA model fits to $w_{\delta +}$ at large scales. Similarly, Fig.~\ref{fig:allprojvsig1} shows $w_{\delta +}$ for ellipticals alone, from which $A_I$ constraints are obtained in section~\ref{sec:results}. Fig.~\ref{fig:allprojvsig23} shows the corresponding results for discs galaxies.

\begin{figure*}
  \includegraphics[width=0.95\textwidth]{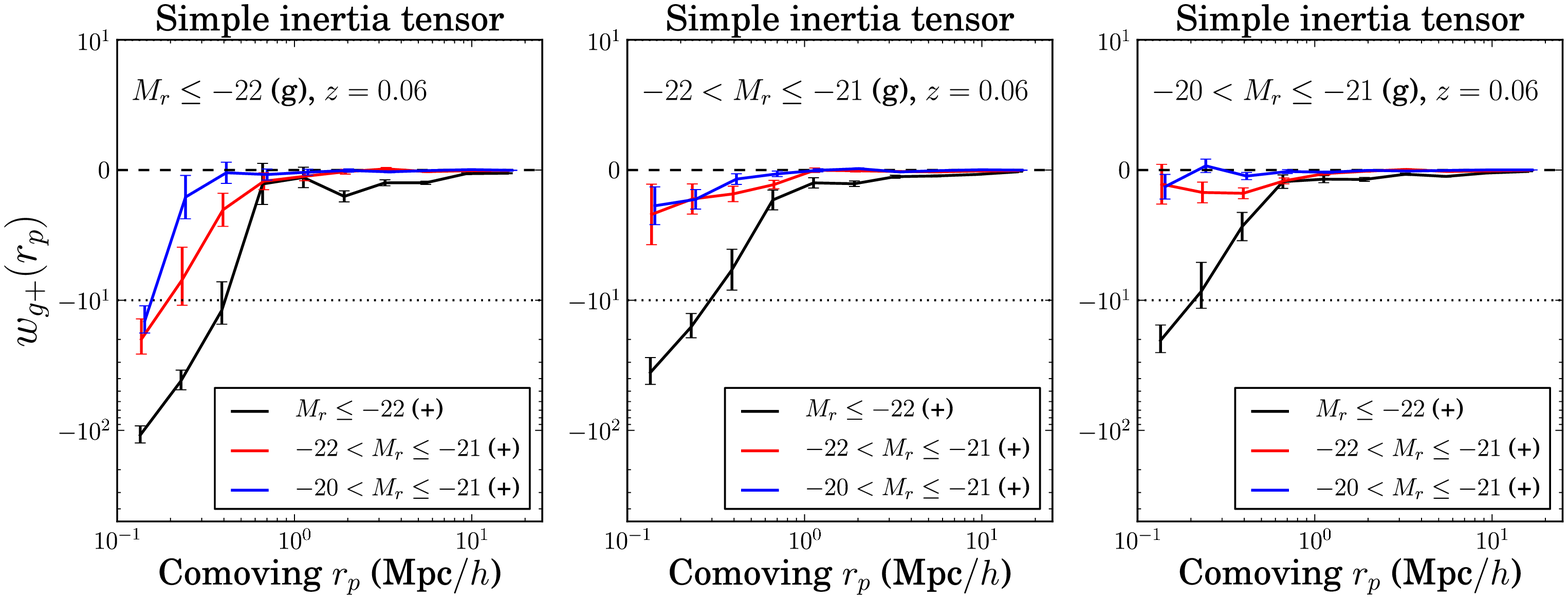}
  \includegraphics[width=0.95\textwidth]{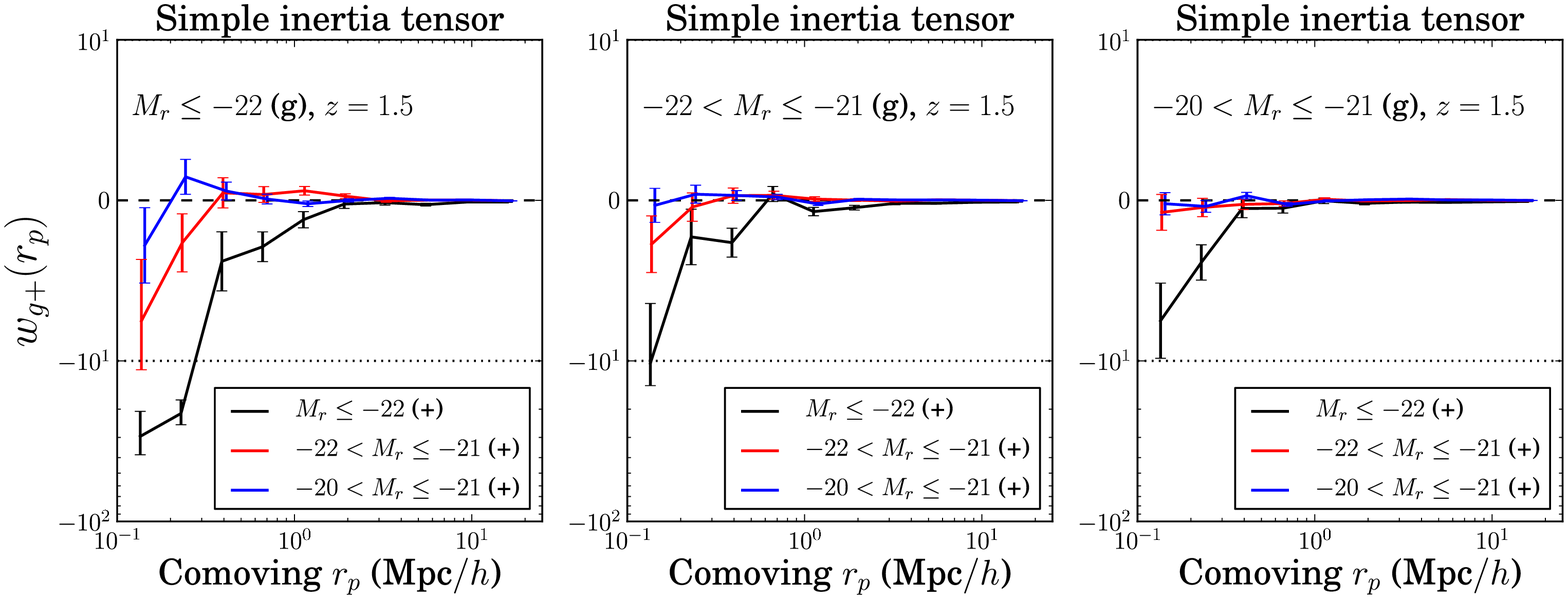}
  \includegraphics[width=0.95\textwidth]{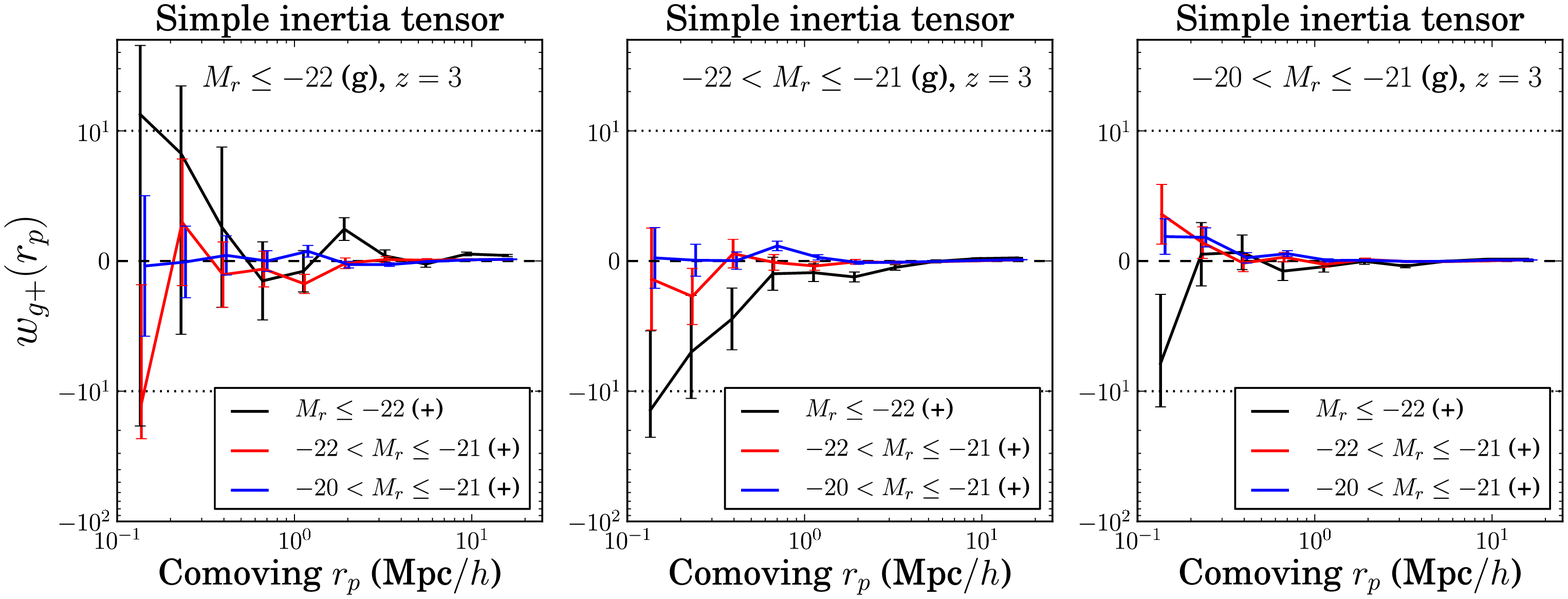}
  \caption{$w_{g+}$ for the simple inertia tensor when different luminosity galaxies are used as density tracers ($g$) and shape tracers ($+$). Solid black lines correspond to shape tracers with $M_r\leq-22$; red lines, to shape tracers with $-22<M_r\leq-21$, and blue lines, to shape tracers with $-21<M_r\leq-20$. The top row corresponds to $z=0.06$; the middle row, to $z=1.5$ and the bottom row, to $z=3$. Note that the black dotted line indicates a change from logarithmic to linear scale in the $y$-axis.}
  \label{fig:crossproj}
\end{figure*}

Finally, Fig.~\ref{fig:crossproj} shows $w_{g+}$ for the simple inertia tensor when different luminosity galaxies are cross-correlated as position and shape tracers. Similar conclusions were derived from Figure \ref{fig:threemodes} in section~\ref{sec:results}. The amplitude of alignment increases with both the luminosity of the density tracer, and of the shape tracer, similarly as for $\eta_e$. These correlations allow us to decompose the intrinsic alignment signal into several contributions that can be connected to previous work in the literature. It is interesting to note that very luminous galaxies are radially oriented towards other luminous galaxies, but the correlation persists even when low luminosity galaxies are used to trace the density field. This indicates that low luminosity galaxies are not randomly distributed around higher luminosity ones. Instead, they seem to cluster preferentially in the direction of the semimajor axes, as suggested by \citet{Mandelbaum06}, due to anisotropic infall or torquing from the central galaxy \citep{Welker15}. This is related to a well-known effect in the literature: \citet{Binggeli82} observed that the elongation of clusters is correlated with the elongation of the central galaxy. Of course, correlations between the shapes of centrals across large scales also give rise to correlations between the shapes of different clusters, an effect also measured by \citet{Binggeli82}, as well as more recently by \citet{Smargon12}. On the other hand, there is a small scale component to the alignment signal that comes from the elongation of low luminosity galaxies around high luminosity centrals. Hence, satellites are not only preferentially located in the direction of the projected major axis of the central, they are also radially elongated towards the central.


\end{document}